\documentclass{aa}  

\usepackage{graphicx}
\usepackage{longtable}

\usepackage{lscape}
\usepackage{eqnarray,amsmath}
\usepackage{txfonts}
\usepackage{xcolor}
\begin{document} 

\title{Self-gravitating collapsing star and black hole spin-up in long gamma ray bursts}
\titlerunning{Self-gravitating collapsar model for long GRBs}

   \author{Agnieszka Janiuk
          \inst{1}\fnmsep\thanks{agnes@cft.edu.pl}
          \and
          Narjes Shahamat Dehsorkh \inst{2}
           \and
          Dominika Krol\inst{3}
          }

              \institute{Center for Theoretical Physics, Polish Academy of Sciences, Al. Lotnik\'ow 32/46, 02-668 Warsaw, Poland\\
                \and Department of Physics, School of Science, Ferdowsi University of Mashhad, Mashhad, PO Box 91775-1436, Iran
                \\
                \and
                Astronomical Observatory of the Jagiellonian University 
Orla 171, 30-244, Krak{\'o}w, Poland \\
              }


 
  \abstract
   {Long Gamma Ray Bursts (GRBs) originate from the collapse of massive, rotating stars. Some of the GRBs exhibit much stronger variability patterns in the prompt GRB emission
 than the usual stochastic variations. We discuss the mechanisms which could account for this effect. }
   {We aim to model the process of stellar collapse in the scenario of a self-gravitating
     collapsing star. We account for the changes in Kerr metric induced by the growth of the black hole, accretion of angular  momentum, as well as the self-gravity effect due to a large mass of the collapsing stellar core falling onto black hole in a very short time. We also investigate the existence of accretion shocks in the collapsar, and role of magnetic field in their propagation. }
   {We compute the time-dependent axially-symmetric General Relativistic magnetohydrodynamic model of a collapsing stellar core in the dynamical Kerr metric. We explore the influence of self-gravity
     in such star, where the newly formed black hole is increasing the mass, and changing its spin. The Kerr metric evolves according to the mass and angular momentum changes during the collapse. 
We parameterize the rotation inside the star, and
account for the presence of large-scale poloidal magnetic field. For the set of the global parameters, such as the initial black hole spin, and initial content of specific
angular momentum in the stellar envelope, we determine the evolution of black hole parameters (mass and spin) and
we quantify the strength of the gravitational instability.
Then we estimate the variability timescales and amplitudes.
}
   {We found that the role of the gravitational instability measured by the value of the Toomre parameter
     is relatively important in the innermost regions of the collapsing star. The character of accretion rate variability strongly depends on the assumption of self-gravity in the model, and is also affected by the magnetic field. Additional factors are initial spin and rotation of the stellar core. We find that for sub-critical rotation of the pre-collapsed star, a centrifugally supported mini-disk is present at the equatorial plane, and it may be subject to fragmentation due to self-gravitating instability.
     We also find that self-gravity may play a role in the angular momentm transport and it
     generally lowers the final mass and spin of the black hole,
     while the accretion rate variability amplitude is much larger in self-gravitating objects. 
     The effect of magnetic field is rather weak, while it seems to decrease the strength of
     accretion shocks.
     The magnetisation affects the global properties of the flow in a non-linear way,
     and is manifested mostly in models with moderate initial black hole spins, but for super-critial rotation of the collapsing star.
      }
   {Our computations confirm that the gravitational instability can account for flaring activity in GRBs and the variations in their prompt emission. Rapid variability detected in case of the brightest GRBs (most likely powered by rapidly spinning black holes) is consistent with the self-gravitating collapsar model where the transonic shocks are formed. The effect should be weakened by magnetic field. 
   }


   \maketitle
%

\section{Introduction}

Massive stars are born, live and die collapsing under their gravitational force, and eject their outer hydrogen-rich envelopes, after billions of years of transforming light elements into heavier ones, via nuclear fusion. In the last stage of their evolution they are called collapsars if the star rotation velocity was large enough to enable formation of an accretion disk in the core. 
In contrast to the low-mass stars which leave white dwarfs as their compact remnants, the more massive ones with masses $\gtrsim 8M_{\odot}$, die violently in supernova explosions that inject freshly synthesized elements, enriching the interstellar medium. In this case, the iron core of the progenitor collapses to a neutron star or black hole.
These types of explosions are called core-collapse Supernova (CCSN) and the particular type of remnant after at the final stage of the massive stars' evolution depends on its mass, metallicity, and rotation rate \citep{Janaka2007, Woosley2015}. More precisely, in the simplest case of no rotation and no mass loss, for the stars of mass $8-30M_{\odot}$ their iron cores collapse to neutron stars, leading to supernova. However, some of these stars may either not explode or explode incompletely, leaving black holes as their remnants. This occurs especially for stars with massive helium cores from  $7M_{\odot}$ up to $10M_{\odot}$. In cases of stars with mass $30-80M_{\odot}$ (helium core mass $10-35M_{\odot}$ ), the black hole formation is quite likely. Rotation generally shifts the main sequence mass ranges (but not the helium core masses) downwards for each outcome. Mass loss complicates the relation between initial main sequence mass and final helium core mass \citep{Woosley2015}.

Gamma-ray bursts (GRBs) may accompany some of the type I b/c supernovae explosions. These transient events are manifested in a sudden release of about $10^{51}-10^{54}$ ergs of energy in a volume with a radius of less than 100 km, which last from 0.01 to 100 s (for reviews, see e.g. \cite{Piran2004, Kumar2015}). According to the duration time $T_{90}$, which is defined as the time interval over which $90\%$ of the total background-subtracted counts are observed, GRBs are usually separated into two classes: long GRBs (LGRBs; $T_{90}>2$ s), whose existence emanates from the core collapse of massive stars \citep{Woosley1993, Hjorth2003}, and short GRBs (SGRBs; $T_{90}<2$ s), whose origins are thought to be the coalescence of neutron stars (NSs) or NS-black hole binary systems \citep{Eichler1989, Narayan1992}.
Most observed GRBs ($70 \%$) have a duration greater than two seconds and are classified as LGRBs. Since these events constitute the majority of the population, and as they tend to have the brightest afterglows,
they have been observed in much greater detail than their short counterparts.
Almost every well-studied long GRB has been linked to a galaxy with rapid star formation, and in many cases to a CCSN as well (\cite{Woosley2006}). Long GRB afterglow observations which reflect their association with high redshift ($z \gtrsim 5$), are also consistent with the GRB having originated in star forming regions \citep{Pontzen2010}. However, not all the collapsing stellar cores give birth to GRBs. It is about being of a sufficiently large angular momentum for the progenitor to have an accretion disk at the equatorial plane, in order to be capable of producing LGRBs (see e.g., \cite{JP2008, JMP2008}, and \cite{Dominika2021}). If this condition is not satisfied, the collapse will proceed without an electromagnetic transient and lead to a disappearance of the star from the filed of view of our telescopes \citep{Murguia2020}. Otherwise, the creation of a jet 
via the process of accretion, is a key factor to account for the formation of a GRB. The strong magnetic field that can be sustained during the process of collapse, and amplified by the differential rotation and dynamo effects, helps launching relativistic jets from the accretion disk. 
These jets further scrape off through the stellar surface and produce emission in gamma rays (\cite{Zhang2003}).

In the previous work, we have build a numerical model that accounts for a dynamical change of black hole parameters, and related Kerr metric, during the collapse onto a newly formed black hole. We used general relativistic  hydrodynamical \citep{Janiuk2018} or magnetohydrodynamical simulations \citep{Dominika2021}, to probe the amount of the angular momentum in the collapsing star envelope and conditions which are sufficient for producing either a GRB, or just a massive, moderately rotating black hole with no electromagnetic transient. In that work, we evolved the Kerr metric of the space-time surrounding the rotating black hole with changing mass and spin. However, we neglected the gravitational force of the massive star itself, which acts on the collapsing gas. This simplification was justified by the fact that the massive core is very compact in comparison with diluted stellar envelope residing in a much larger volume.
In this work, we release this simplifying assumption and we 
extend our model with a new numerical scheme where we account for the self gravity of the star. We are doing this via a perturbative approach, so still not by solving the full set of Einstein equations.
 This is done by integration of mass and angular momentum in the flow
around the center at any given radius, and adding this component as a small term to the
mass and spin of the black hole which are constituents of the Kerr metric defined locally. In this way we represent more correctly the influence of self-gravitating mass enclosed in the volume on the orbiting material. We analyze the possible instabilities in the self-gravitating collapsing stellar core, and we also find regions where shock discontinuities are formed.
We also study the process of stellar collapse by allowing the initial core radius to slightly vary and hence define different initial conditions for the subsequent formation of a rotationally supported accretion disk. Finally, we account for the magnetic field component to perturb the accretion rate
and we estimate the time scale of variability which may be reflected in the observed properties of GRBs, if the jest are formed. 

Similarly to our previous studies, we explore the properties of an exemplary model, assuming the $25 M_\odot$ envelope of a collapsing star and  the $3 M_\odot$ initial
black hole mass formed from the core.
We probe the parameter space similar to our previous study, i.e. we vary the value of initial black hole spin, magnitude of angular momentum inside the envelope, and strength of magnetic fields. However, we extend this parameter space to larger specific angular momentum endowed in the star, and we implement several alternative magnetic field configurations. 
We compare the results of new models, i.e. those with self-gravity of the collapsing core, to those without self-gravity force.

The article is organized as follows. In Section \ref{sect:model_general} we present the general framework of our model, which has been developed upon the general relativistic MHD code and extended to work in time-dependent Kerr spacetime metric in previous papers. In Section \ref{sect:model}, we describe the current advancement of the previous model, which is the implementation of self-gravity of the collapsing star in the spacetime dynamics. In Section \ref{sect:rin} we describe a modification in the inner boundary condition, in Section \ref{sect:selfgrav} we describe the perturbative method used to compute the change of mass and angular momentum of the black hole due to self-gravity, and in Section \ref{sec:magnetic_field} we define the magnetic field configurations used in our testing models.  
In Section \ref{sect:results} we present the results of our calculations. In particular, Section \ref{sect:global_evolution} describes time evolution of the self-gravitating, non-magnetized stellar cores, and compares them to our previous, non-selfgravitating models, Section \ref{sec:toomre} presents a detailed analysis of the gravitational instability, which is a new feature found in the newly developed models, and Section \ref{sect:magnetized} presents evolution of magnetized, self-gravitating collapsing cores. 
In Section \ref{sec:diss} we discuss our results, in the context of GRB phenomenology, and in Section \ref{sec:conclusions} we summarize our conclusions.

\section{Time evolution with accreting black hole mass and spin update}
\label{sect:model_general}

Apart from the modifications described in Sect.~\ref{sect:rin} and \ref{sect:selfgrav},
we follow the time evolution of the collapsing core as described in \citet{Dominika2021}.
We use the general relativistic magneto-hydrodynamic code, called High Accuracy Relativistic Magnetohydrodynamics (HARM) which has been originally established by \cite{Gammie2003} (see also \cite{Noble2006}). The code introduces a conservative, shock-capturing scheme with low numerical viscosity, to solve the hyperbolic system of partial differential equations of GR MHD.
The numerical scheme uses the plasma energy-momentum tensor, $T_{\mu \nu}$, with contributions from matter (gas) and electromagnetic field. For the GR MHD  evolution, we solve two fundamental equations: the equation of mass and energy-momentum conservation
which are as follows:

\begin{equation}
(\rho u^{\mu})_{;\mu}=0;~~~~~~~~T_{\nu;\mu}^{\mu}=0.
\end{equation}

The energy stress tensor is a sum of two parts, gas and electromagnetic:

\begin{equation}
T_{(m)}^{\mu \nu}=\rho hu^{\mu}u^{\nu}+pg^{\mu \nu}
\end{equation}

\begin{equation}
T_{(em)}^{\mu \nu}=b^{k}b_{k} hu^{\mu}u^{\nu}+\frac{1}{2} b^{k}b_{k}g^{\mu \nu}-b^{\mu}b^{\nu}
\end{equation}

\begin{equation}
T^{\mu \nu}=T_{(m)}^{\mu \nu}+T_{(em)}^{\mu \nu}
\end{equation}
where $u^{\mu}$ denotes the four-velocity of gas, $u$ represents internal energy density, $b^{\mu}$ and $h$ are magnetic four-vector and the fluid specific enthalpy, respectively. 
The MHD scheme is brought in conservative form, by implementing a Harten-Lax-van Leer (HLL) solver
\citep{Harten1983} to calculate numerically the corresponding fluxes. 

The fluid equation of state is that of a polytrope with a pressure $P=K\rho^\gamma$, where $\rho$ is the density,  $\gamma=4/3$ is the adiabatic index, and $K$ is the specific entropy, in this case taken to be that of a relativistic fluid with inefficient cooling.

We have been developing a new version of the HARM code which was already implemented by \cite{Janiuk2018}, where we
considered a dynamically evolving space-time owing to the changes in the central black hole's parameters. The simulations are started after the black hole has already formed and it is assumed that its gravitational field controls the subsequent space-time evolution. Then, the matter gets to accrete onto it, and the code applies a sequence of quasi-stationary solutions with black hole's mass and spin updated by a very small value in each time step. The Kerr black hole's line element (metric) in the Boyer-Linquidst coordinates is given by:

\begin{eqnarray}
ds^{2}=(1-\frac{2M_{BH}r}{\Sigma})dt^{2}+\frac{4M_{BH}arsin^{2}\theta}{\Sigma}dtd\phi- \\ \nonumber
- \frac{\Sigma}{\Delta}dr^{2}-\theta^{2}-sin^{2}\theta (r^{2}+a^{2}+\frac{2M_{BH}a^{2}rsin^{2}\theta}{\Sigma})d\phi^{2}
\end{eqnarray}
where $\Delta=r^{2}-2M_{BH}r+a^{2}$, $\Sigma=r^{2}+a^{2}cos^{2}\theta$, and $a=\frac{J}{M_{BH}}$, with $M_{BH}$ and $J$ are the mass and angular momentum of the black hole. To put the evolution of the black hole's parameters into effect, one can consider the metric is changed discretely between the consecutive time steps, according to mass and spin small changes, $\Delta M=(M^{i}_{BH}/M_{BH}^{0}-1)$, and $\Delta a=(\dot{J}/M^{i}_{BH}-a^{i-1}/M^{i}_{BH}\dot{E})\Delta t$, where $M^{i}_{BH}$ reflects the current black hole mass at time $t>0$, $M^{0}_{BH}$ denotes initial mass of the black hole at $t=0$, and $\dot{J}$ and $\dot{E}$ are the flux of angular momentum and energy flux transmitted through the black hole event horizon. The six non-trivial Kerr metric components are then updated at every time step to get their new values. 

Our gride size is that or $R_{out} = 1000~r_{g}$, and the resolution is 256x256 points in the radial and polar directions. The outer boundary conditions in radial direction are free outflow (variables are copied to the two ghost zones). In the polar direction, reflecting boundary conditions are assumed (velocity and magnetic field components change their signs on the polar axis).

\section{Self-gravitating collapsar model}
\label{sect:model}

Now, we calculate the time evolution of the collapsing massive star using the GR MHD scheme, using the new version, upgraded upon \cite{Janiuk2018}. The evolution of the space-time Kerr metric is again accounted for by the increasing mass and changing spin of the black hole in the collapsing stelar core. However, new terms due to self-gravity of the star are computed and volume-integrated, at every time-step during dynamical simulation.

\subsection{Initial conditions}

We adopt the initial conditions similar to those used in \cite{Dominika2021}, namely a slowly rotating, transonic, quasi-spherical accretion flow with small angular momentum. The initial distribution of density and radial velocity in the accreting sphere is given by numerically integrated Bondi solution, which we 
parameterize with the location of the sonic radius.
In our models, it is fixed at $r_{s}=80 r_{g}$. Below this radius the matter falls into black hole supersonically, reaching the speed of light at the black hole horizon.

Once the critical point is determined, the velocity at this critical point is \citep{1986bhwd.book.....S}:
\begin{equation}
 (u^r_{\rm s})^2=\frac{GM_{\rm BH}}{2r_{\rm s}}, 
\end{equation}
 where $r$ is the radial coordinate and  $u^r$ is the radial component of the four-velocity. 
The radial velocity can be obtained by numerically solving the relativistic Bernoulli equation:
\begin{equation}
\bigg{(} 1+\frac{\gamma}{\gamma-1}\frac{P}{\rho}\bigg{)} ^2 \bigg{(}  1-\frac{2GM_{\rm BH}}{r}+(u^r)^2 \bigg{)} = \rm{constant},
\end{equation}
and the density is set by the mass accretion rate $\dot{M}$: 
\begin{equation}
\rho={\frac{\dot{M}}{4\pi r^2u^r}}.
\end{equation}

The specific entropy value, $K$, depends on the radial velocity and is taken to be \citep{Sukova2015,2017MNRAS.472.4327S,2019MNRAS.487..755P}:
\begin{equation}
K=\bigg{(} u^r 4\pi r^2\frac{c_{\rm s}^{\frac{2}{\gamma-1}}}{\gamma^{\frac{1}{\gamma-1}}\dot{M}}\bigg{)}^{\gamma-1},
\end{equation}
where $c_{\rm s}^2=\frac{\gamma P}{\rho}$ is the local sound speed.

The small angular momentum is imposed on the spherically distributed gas,  similarly to \cite{Dominika2021}. The specific angular momentum is normalized with the parameter $S$, such that the flow is circularized at the innermost stable circular orbit (ISCO). In addition, the rotation velocity scales with the polar angle, so that at the equator, $\theta=\pi/2$, the rotation of the star is maximal.

   \begin{equation}
l=S l_{\rm isco}r^2\sin^2{\theta} 
   \end{equation}
  with
   \begin{equation}
l_{\rm isco}=u_{\phi, \rm isco}=\frac{r_{\rm isco}^{1/2}-2a/r_{\rm isco}+a^2/r_{\rm isco}^{3/2}}{\sqrt{1-3/r_{\rm isco}+2a/r_{\rm isco}^{3/2}}}.
   \end{equation}
   
Here the radius $r_{\rm ISCO}$ in Kerr geometry depends on the black hole spin. Our black hole spin parameter is set in the initial setup, and from now on, it is denoted as $A_{0}$. We use both sub-critical and super-critical rotation speeds, parameterized with $S<1$ and $S>1$, respectively. 
Our model parameter space is therefore defined by $S$ and $A_{0}$.

\begin{figure}
    \centering
    \includegraphics[scale=0.5]{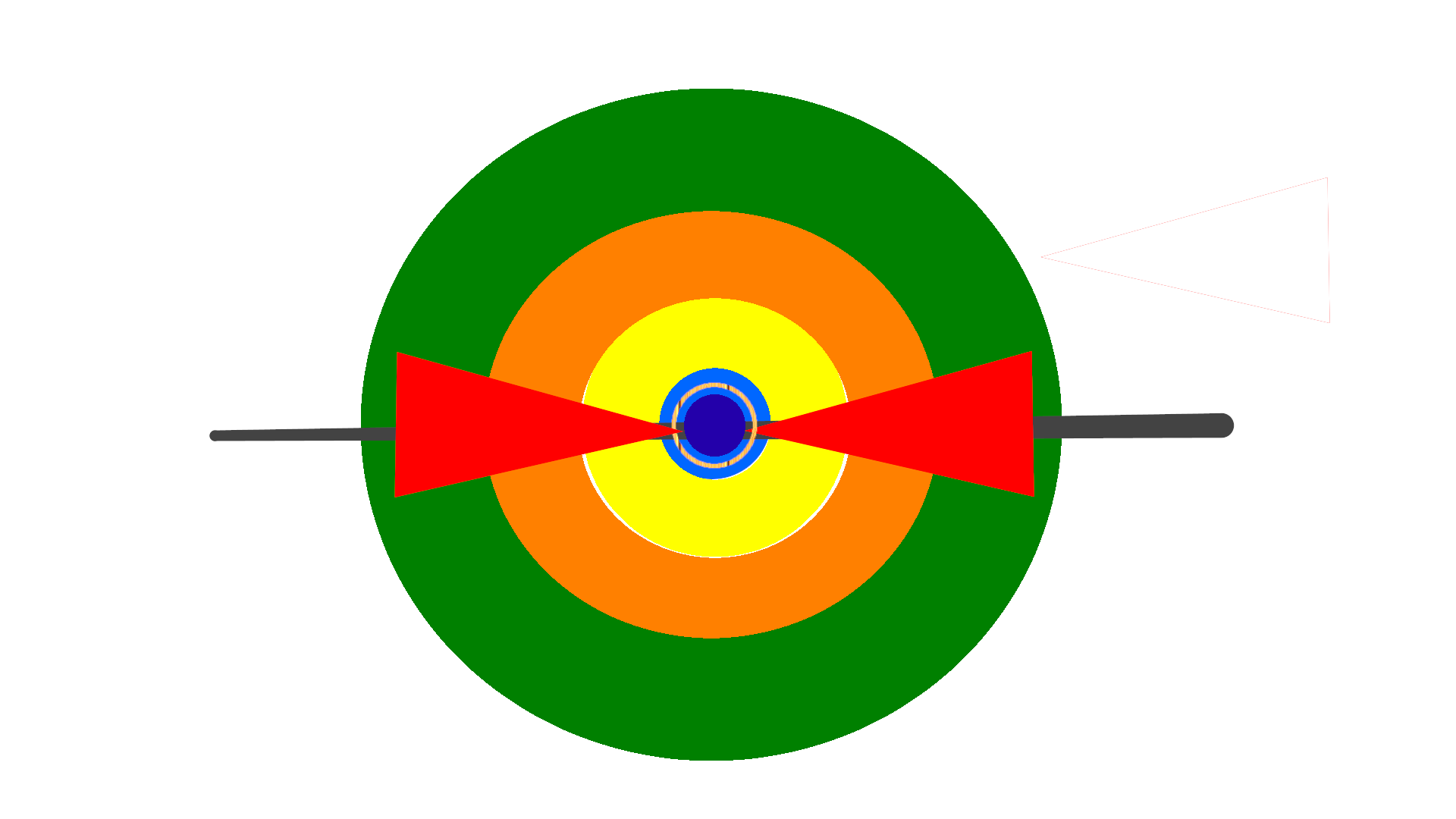}
    \caption{Schematic view of the collapsar at the onset of the GRB: stellar core (dark blue), stellar envelope composed of
      subsequently accreting
      shells with decreasing density (light blue-yellow-orange-green), and
      rotationally supported accretion disk formed at the equatorial region (red).
      Horizontal black line represents the equatorial plane. The circle marked with a
      stripped line represents 
      an exemplary chosen radius above the horizon, at which the gas feels the perturbative force due to self gravity of matter enclosed within this radius (see Eq. 7.).
      } 
    \label{fig:our_model}
\end{figure}

We notice here that in this formulation, the black hole in the center of collapsing stellar core is already as massive as $M_{BH}=3 M_{\odot}$, which scales the length unit of $r_{g}=4.45\times 10^{5} ~ cm$. This means that our computational grid size in cgs units is only $4.45\times 10^8 cm$. Therefore, if a compact C-O core of a Wolf-Rayet star, or a
  pre-supernova of $25 M_{\odot}$ \citep{WoosleyHeger2006} is assumed as our initial model, it is now squeezed into a much smaller volume and reaching order of magnitude larger densities in the center. 
  Our model is compact enough to address the problem of self-gravitating gas close to the horizon of a newly formed black hole. We do not address here any prior or ongoing supernova explosion. 
  Depending on the rotation parameter, $S$, the ultimate outcome might be either a direct collapse, or a formation of a mini-disk inside the stellar core, i.e. collapsar (as depicted in Figure \ref{fig:our_model}). The latter may lead to an electromagnetic transient.

\subsection{Boundary condition between the initial stellar core and outer flow}
\label{sect:rin}

For numerical reasons, we need to ensure that enough number of grid cells are located below the black hole horizon. The inflow of matter proceeds then smoothly through the horizon, thanks to change of coordinates to the Kerr-Schild ones, which are non singular.
In the initial setup, we assume that the transition between the newly formed black hole and the accreting stellar core is shifted by 
a certain factor. We choose a shift radius, on the order of 1.2 $r_{g}$, to account for a smooth free-fall of the stellar shells onto the black hole, that has been already 
shielded by the event horizon.
Therefore, we introduce an initial offset between the black hole horizon radius, 
$r_{h}=1 + \sqrt{1-a^2}$, in $r_{g}$ units, and the dense surrounding gas.
The inner radius of the 
accreting stellar core is now placed at $R_{in}=R_{shift} \cdot r_{h}$.
Our model parameter $R_{shift}$, represents therefore the initial inner radius of the stellar core before the collapse starts. Its minimum value, 1.0, would imply that the collapse starts immediately when the black hole has formed. Otherwise, values larger than 1, imply a slight delay of the black hole growth.
The shift
provides numerical stability of the initial phase of the simulation.

\subsection{Modification of the stellar core structure due to the self-gravity}
\label{sect:selfgrav}

To describe the self-force acting in the collapsing gas, the Teukolsky equation has been chosen. This equation describes gravitational, electromagnetic, and scalar field perturbations of a rotating Kerr black hole \citep{Teukolsky1972}. The global vacuum solution of the Teukolsky equation is given by the CCK method (see \cite{Chrzanowski1975}; \cite{Cohen1974}, \cite{Wald1978}) which reconstructs the metric perturbation and shows that only perturbations of the mass and angular momentum ($\delta M$ and $\delta J$, defined below) are to be concluded within the Kerr metric. 
 We notice that in the weak field limit, the Einstein gravitational field equation will be reduced to the Poisson equation leaving us with the only non-zero component of the perturbed metric potential (cf. \cite{Ryu2020}).

Recently, \cite{Meent2017}
showed that the perturbation due to a particle on a bound
orbit around black hole described by CCK metric affects the Kerr parameters
describing the mass and angular momentum of the black hole for the metric ’outside’ the
particle’s orbit and vanishes ’inside’ the orbit.
Our work proposes numerical implementation of this method, and follows the assumptions of \cite{Meent2017} for fluid dynamics, instead of particles, so we compute volume integrals of the corresponding stress-energy tensor components. By definition, the problem does not assume spherical symmetry. The potential wells may therefore appear off-axis, in the whole region 'outside' the orbit of a given fluid element.

We have developed a new
version of GRMHD numerical code HARM. As the initial results have shown
\citep{Janiukproc}, we expect to see more mass and spin growth of the black hole after incorporating the perturbation effects of the accreting disk into the updating metric.

In our new simulations, both the mass and angular momentum accreted onto black hole horizon, and used to update the Kerr metric coefficients, are now 
modified with perturbation acting on the metric in the region above the horizon, due to the self-gravity force that the gas feels at a given distance from the horizon (a schematic view of the calculation is depicted in Fig. \ref{fig:our_model}).
These perturbative terms are calculated from stress-energy tensor. Hence, 
in addition to the two Equations governing the growth of black hole mass and spin via the mass and angular momentum transfer through the horizon:
\begin{equation}
 \dot M_{BH} = \int d\theta d\phi\, \sqrt{-g}\, {T^{r}}_t,
\end{equation}
and
\begin{equation}
    \dot{J} = \int d\theta d\phi\, \sqrt{-g}\, {T^{r}}_\phi,
\end{equation}
(see \cite{Janiuk2018} for more details), 

we now calculate:
\begin{equation}
    \delta M_{BH} (t,r) = 2\pi \int_{r_{hor}}^{r} T^{r}_{t}\sqrt{-g} d\theta 
\end{equation}
computed at every radius above horizon.
Analogously, the angular momentum of the black hole will change by adding the perturbation:
\begin{equation}
    \delta J(t,r) = 2\pi \int_{r_{hor}}^{r}T^{r}_{\phi}\sqrt{-g} d\theta 
    \end{equation}
The dimensionless spin of the black hole, as a result, will change by
\begin{equation}
    \delta a = {{J + \delta J(r)} \over {M_{BH}+\delta M_{BH}(r)}} - a^{i} 
\end{equation}
Here $a^{i}=a^{i-1}+\Delta a$, according to eq. (7) in \cite{Janiuk2018}.

Having different $\delta M$ and $\delta J$ at each grid point in radial direction at each time, affects the metric coefficients which are sensitive to the mass and spin update. 
Our main change of the \cite{Janiuk2018} model is developing a new module in the time-dependent code, that accounts for self-gravity force. In the rest of the paper, we present the results computed with this module enabled in the simulation setup. We also compare them with the runs without self-gravity (the new modules are switched off), to emphasize the difference and investigate the role of self-gravity in the collapsar physics.

\subsection{Post-collapse magnetic field and its chosen structure}
\label{sec:magnetic_field}

We assume that  evolution of the collapsing stellar core prior to the simulated phase was unaffected by the particular configuration of magnetic field.
This is justified, as we address the class of massive stars with relatively weak fields. The surface magnetic fields can be constrained by observations and suggest that there might be a bimodal distribution \citep{Petit2019}. The magnetic fields in the stellar core are unknown and all scenarios are uncertain.
Currently, various pre-supernova models and scenarios adopt only general scaling of magnetic fields, such as a large scale dipole \citep{Reichert2023}. Some models have been using the fields inherited from the stellar evolution models \citep{WoosleyHeger2006}, possibly supplemented by a toroidal component and amplified via a dynamo mechanism in a differentially rotating star \citep{Spruit2002}, which may lead to ultimate formation of a proto-magnetar in the core \citep{Obergaulinger2009}.

In our simulations, we consider some possible effective modification of the collapsing zone due to the action of the weak magnetic field which is dynamically unimportant. 
The application of the realistic evolved star field geometry to our quasi-spherical mass distribution is not unique, and we introduce two specific prescriptions: (i) uniform magnetic field, and (ii) dipole magnetic field.

Therefore we introduce the magnetic field in our initial conditions, by defining the shape of the magnetic vector potential, and setting up a proper normalisation.
First, we  assume that the initial accreting gas is embedded in a simple poloidal configuration of the magnetic field, as discussed already in \cite{Dominika2021}. In this case, the only non-vanishing component of the magnetic field vector potential is given by:
\begin{equation}
A_{\varphi} \propto {1 \over 2} r \sin(\theta)
    \label{eq:vertical}
\end{equation}

Furthermore, as a variation of the uniform field derived for the Kerr black hole, we adopt the formula by \cite{Wald1974}
\begin{equation}
    A_{\varphi} \propto \big[ {1 \over 2} (r^{2}+a^{2}) - a^{2} r \Sigma^{-1} (1 + \cos^{2}(\theta) \big] \sin^{2}(\theta)
    \label{eq:Wald}
\end{equation}

We normalize this uniform field, assuming an initial maximum gas-to-magnetic pressure ratio, $\beta = (\gamma -1) u /(0.5 b^{2})$.

The second scenario assumes a dipole field, where vector potential is given by
\begin{equation}
A_{\varphi} \propto {\sin (\theta) \over r}
\label{eq:dipole}
    \end{equation}

We normalize the magnetic field to a chosen initial value of maximum gas-to-magnetic pressure ratio, $\beta_{0}=p_{gas}/p_{mag}$, which in case of the Bondi gas distribution is reached at the black hole horizon.
By implementing these various field strengths and setups, we aim to verify its astrophysical implications for the collapsar scenario as the central engine for LGRBs.

To sum up, in comparison with previous code applications \citep{Janiuk2018, Dominika2021}, we allow for three major modifications:
\begin{itemize}
\item the offset of the initial stellar core,
\item the modification of the metric evolution to account for self-gravity effects,
\item several different setups and strength of magnetic fields, aimed to model realistic stellar cores. 
\end{itemize}

\section{Results}
\label{sect:results}

The HARM code works in dimensionless units of G = c = 1. Conversion coefficients can be found in Table \ref{tab:units}.

\begin{table}[h]
    {\scriptsize
    \begin{center}
        \caption{Geometric units to cgs units conversion. We adopted M=$3M_{\odot}$ (the initial central black hole mass).}
        \label{tab:units}
        \begin{tabular}{ccc}
            \hline\hline
            Physical quantity & Geometrical units & cgs units  \\ 
            \hline \hline
            Length & $r_g=\frac{GM}{c^2}$ & $4.44 \times 10^{5}$cm \\ 
            Time & $T_{unit} = \frac{r_g}{c}$ & $1.38 \times 10^{-5}$s \\ \hline
        \end{tabular}
    \end{center}
    }
\end{table}

We calculated several set of models, which differ with respect to the chosen initial parameters, 
and with respect to presence or absence of self-gravity effects, and magnetic fields.
All models and their input parameters are listed in Table ~\ref{tab:modele}, where we give the values of initial black hole spin $A_{0}$ in dimensionless units, the inner radius of the collapsing stellar core, $R_{shift}$ (equal to the stellar core radius, or no pre-collapsed core and then the inner radius is located at ISCO), and the initial specific angular momentum, $S$, in the accreting gas.
The self-gravity effect is denoted as either "yes" or "$-$" in the Table.
For magnetized models, we provide the type of field initial configuration (vertical, dipole) and we give the initial strength of the field, normalized either with the ratio of gas to magnetic pressure at
$\beta_{0}$, or with the initial magnetisation, $\sigma_{0}$. 

Below, we show the results for the time evolution of the disk under the self-force, and we compare them with trends previously observed,
i.e. in calculations where the self-gravity was neglected.
We consider for now only the non-magnetized models.
In the next subsection, we will present a few characteristic properties of the low angular momentum, quasi-spherical accretion 
models,
analyzed with respect to their gravitational stability. The magnetized models will be presented in more detail in Sect. \ref{sect:magnetized}.

\subsection{Time-dependent evolution}
\label{sect:global_evolution} 

From the point of view of the evolutionary timescales, the black hole mass is the key parameter as it determines the object's scale size. Here we concentrate on the results for the initial black hole mass of $M_{BH} = 3 M_{\odot}$ and we check whether the whole star collapsed to the black hole, contributing to its final mass.

We check how the evolution proceeds for various initial spin parameters, and several values of rotation parameter.
We compare the time profile, and time-averaged value of the accretion rate during the collapse, for SG and non-SG models.
We also check the evolution of black hole spin, and check what was the maximum spin value reached during the collapse. We then compare the
final black hole spin value, which saturates when the collapse is ended.
These results are given in the Table.

\begin{figure*}
    \centering
    \includegraphics[scale=0.28]{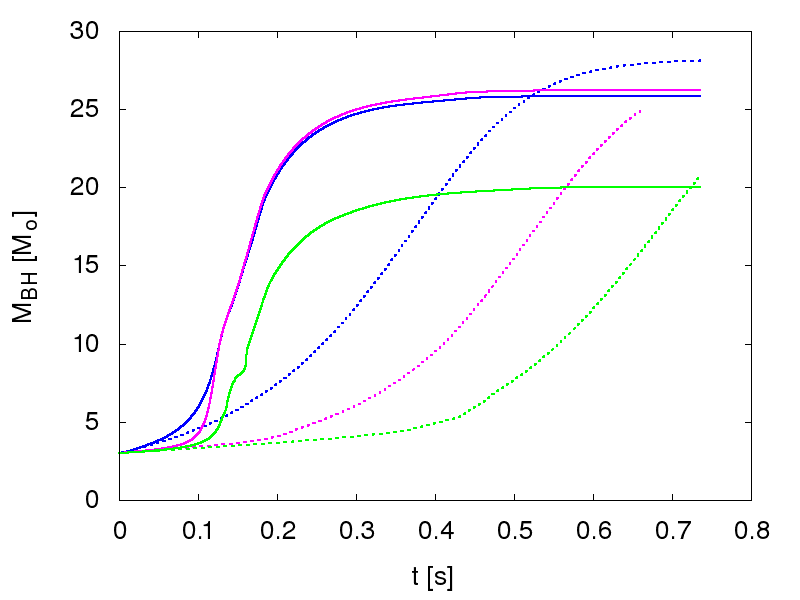}
    \includegraphics[scale=0.28]{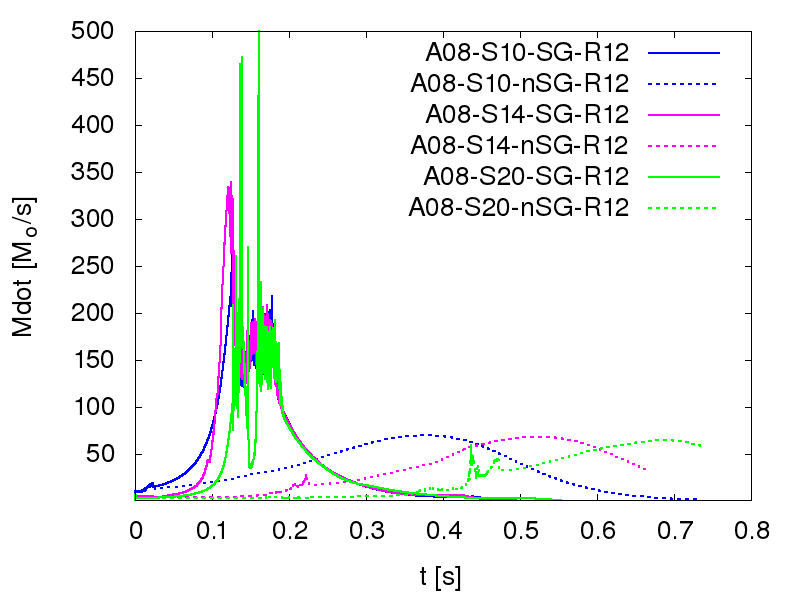}
    \includegraphics[scale=0.28]{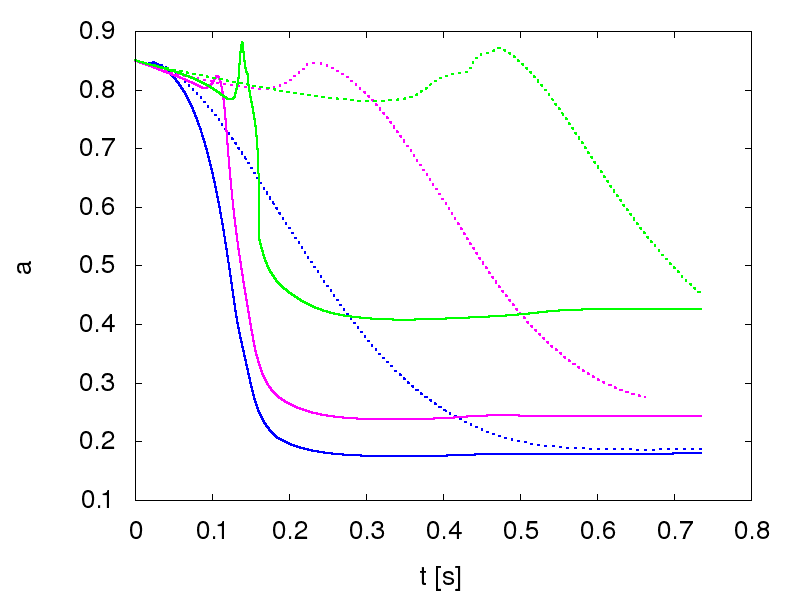}
    \includegraphics[scale=0.28]{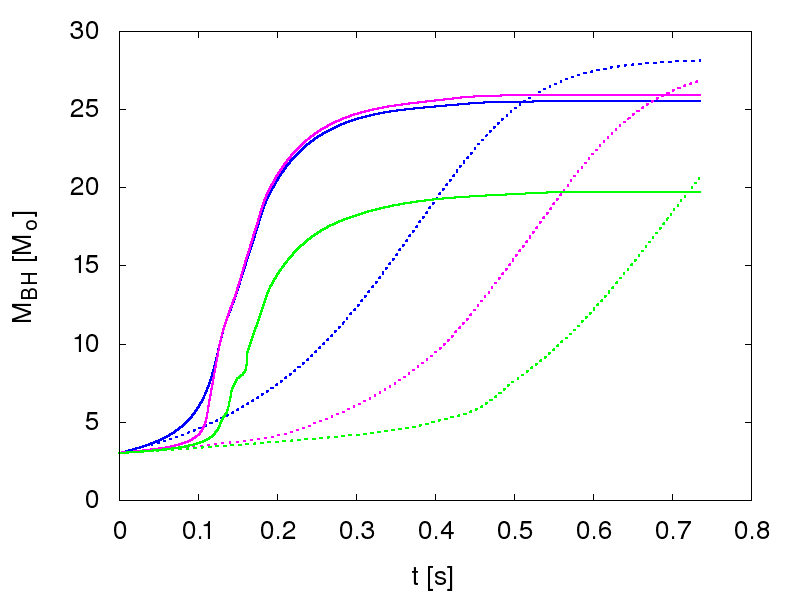}
    \includegraphics[scale=0.28]{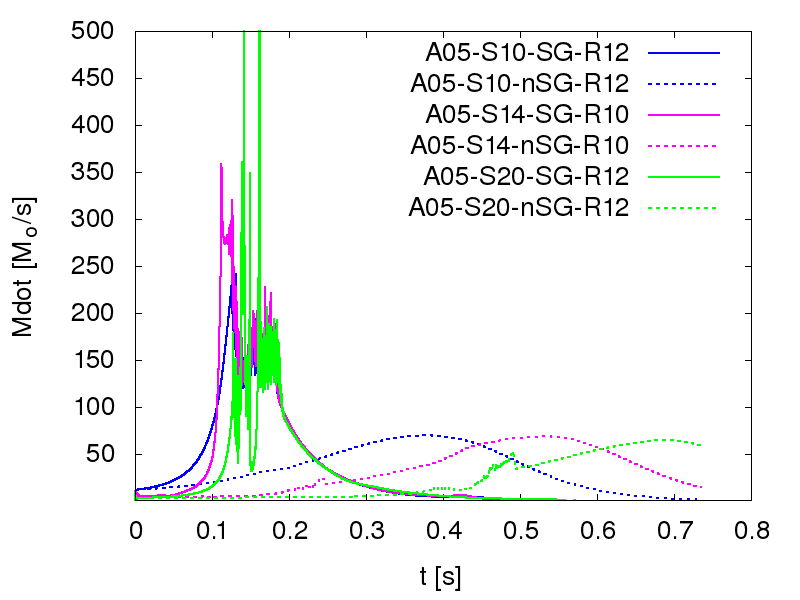}
    \includegraphics[scale=0.28]{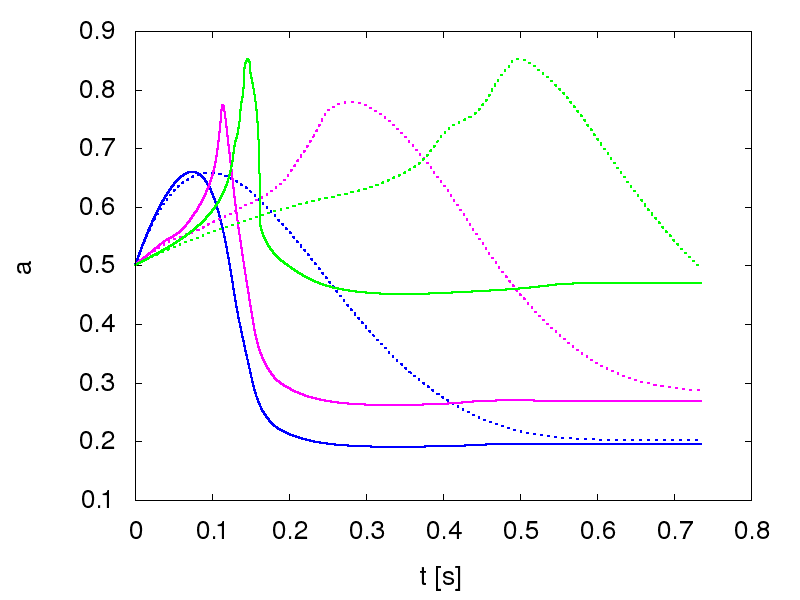}
    \caption{ Evolution of the accretion rate and black hole parameters including and excluding self-gravity plotted by the thick and thin curves, respectively. Top panels refer to the initial black hole spin parameter $A_{0}=0.85$ while the bottom panels correspond to $A_{0}=0.5$. Three different values for the initial rotation parameter, i.e. $S=1, 1.4, 2$ related to the blue, pink and green curves, have been considered as well. The left plots refer to the black hole mass evolution, the middle one show the accretion rate temporal behavior and the panels at the right demonstrate the evolution of black hole’s spin parameter. Models are labeled in the middle panels with symbols referring to Tab \ref{tab:modele}. }
    \label{M_S_A}
\end{figure*}

Assigning three different values to the initial rotation parameter, $S=1,1.4,2$, and two values of $0.5$ and $0.85$ to the initial black hole spin parameter $A_{0}$, we analyze the evolution of our model features, i.e. accretion rate, black hole's mass and dimensionless spin parameter. The time profiles of these quantities are plotted in Figure \ref{M_S_A}. Both cases including and excluding self-gravity are compared. The three panels show the black hole mass profile (left), the accretion rate (middle) and black hole spin evolution.
(In the rest of the paper, we label the models with their acronyms, corresponding to these used in Table \ref{tab:modele}.)

Giving different initial black hole spins, one can find no remarkable changes in the $M_{BH}$ evolution.
The situation differs when it comes to various rotation parameters $S$, however. We notice that the larger the initial rotation magnitude, the longer it takes for the black hole mass to evolve. The non-SG simulations end with very different final black hole mass, depending on the $S$ parameter. This is because material kept by super-critical rotation is affected by the centrifugal force and remains in the accretion torus, while only the material from polar regions is able to reach the black hole quickly, in the free fall time-scale, as was already found in \cite{Janiuk2018}. 
In contrast, the self-gravity of the envelope can speed up the evolution of the collapsing stellar core significantly. We also observe no big difference in the black hole mass evolution in both cases where $S \ge 1$. Only in the $S=2$ run, the final black hole mass is significantly smaller than the total mass of the evolved stellar core, and it saturates at the value of about $20 M_{\odot}$ (see Table \ref{tab:modele} for details).

The middle plots in Fig \ref{M_S_A} confirm that with no self-gravity effects taken into account, a considerably less fluctuating behavior of the accretion rate during a longer time is deduced. In this case, there exist some oscillations in the accretion rate during some time intervals (around 0.2 sec. for $S=1.4$, and $0.4-0.5$ sec for $S=2$).
Before these periods of time, there seem to be some mass accumulation in the inner regions, 
observed in the density profiles, and also in specific angular momentum distributions in different time snapshots (see next Subsection).
This mass accumulation followed by fluctuations in the accretion rate is attributed to the generation of the rotationally supported torus in the inner stellar core, which is concentrated on the equatorial plane. 
On the other hand, in curves with self-gravity the mass accretion rate increases and drops much more sharply than in non-SG models, presenting a characteristic pattern of a sudden fluctuating rise in very short periods of time (less than $0.1s$). We notice here that those sharp peaks have very large magnitude, but their height may be to some extent affected by numerical resolution, and overestimated.
The mass accumulation prior to the oscillations can be seen in this case as well. We found the fluctuating reduction in the size of the rotational supported torus, followed by the formation of an inhomogeneous structure in density profiles that may account for this temporal behavior. We will present more details on these events in the next section.

Considering the black hole spin (dimensionless, so the angular momentum of black hole scaled with mass inverse), one can find that in self-gravitating models, a notable increase in the black hole's mass is corresponding to the spin parameter reaching its maximum. Further accretion after this time, brings more matter than angular momentum to the central object resulting in a considerable rise of $M_{BH}$ and later drop of spin. It coincides with the remarkable increase of accretion rate onto the black hole.
The non-SG models behave differently. Here, the black hole mass and spin rise more slowly than in SG models.
However, a maximal spin is reached earlier than maximum mass (for sub-critical rotation, $S=1$), or is reached at a rather late time, when still the black hole mass has not saturated on the final value (for $S=2$).

We also found that  black hole's spin is reaching its maximum value when the rotationally supported region close to the black hole is shrinking. Self-gravity effect seems to speed up this evolution. Moreover, higher value of the rotation parameter $S$, as well as the presence of disk at late times in the collapsar with $S=2$, leads to a less massive but more spinning black hole, as can be found in the green curves of the right panel in Figure \ref{M_S_A}. Final spin in this case is about $A=0.4-0.5$.

\begin{figure*}[ht]
      \centering
   \includegraphics[scale=0.4]{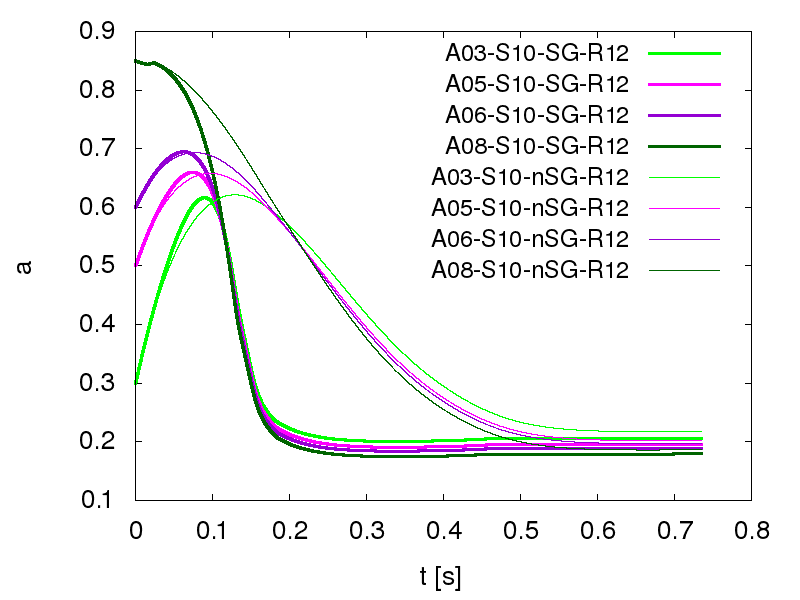}
   \includegraphics[scale=0.4]{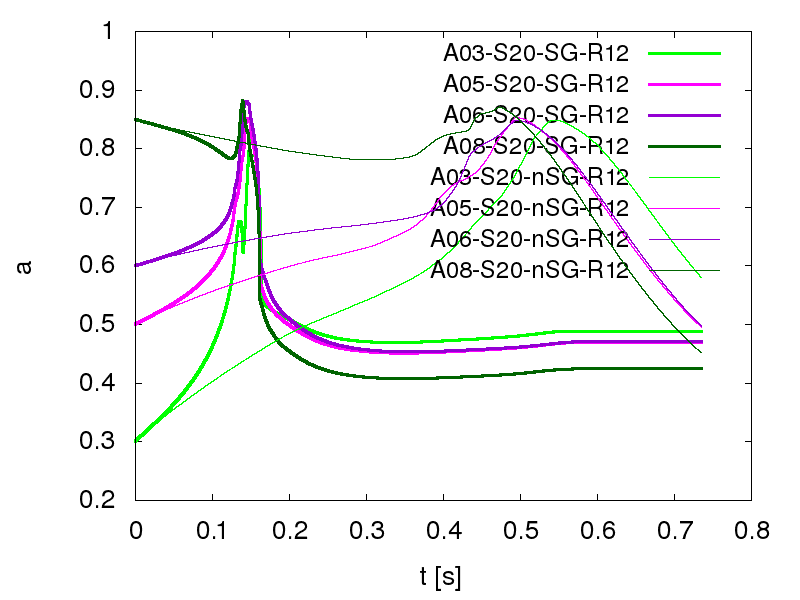}
             \caption{Time evolution of the
              black hole spin
             for the rotation normalized with specific angular momentum at ISCO $S=1.0$ (left) and $S=2.0$ (right). Thick lines represent self-gravitating models and thin lines are for self gravity neglected. Various initial black holes spin values are shown, as labeled in the plots and marked by dark green for $a_{0}=0.85$, violet for $a_{0}=0.6$, magenta for $a_{0}=0.5$, and light green for $a_{0}=0.83$. Models are labeled in the both panels with symbols referring to Tab \ref{tab:modele}.
             }
    \label{fig:models_rshift1x2_rcirc}
\end{figure*}

In Figure \ref{fig:models_rshift1x2_rcirc} we show the evolution of the black hole spin, where we compare the runs with various initial spin $A_{0}$. 
The thick lines represent self-gravitating stars, while the thin lines are shown for comparison, and represent calculations where self-gravity was neglected. The initial black hole spin was taken in the range between  $A_{0}=0.3-0.85$.  In the two panels, we show the simulations with critical and super-critical angular momentum content inside the collapsing star, namely $S=1$ and $S=2$.
As the figure shows, the models with $S=1$ tend to have smaller value of the maximum and final black hole spin. The maximum value reached during the collapse is reached at about $t\sim 0.1 s$ (around time 5000 $t_{g}$ in geometric units) and it correlates with the initial spin value. For smallest $A_{0}=0.3$, the net spin-up is largest, while for $A_{0}=0.85$ the black hole actually does not increase its spin, and only temporary flattening of the spin time evolution profile is observed. Similarly to the cases presented in Fig. \ref{M_S_A}, we notice that the evolution of the spin proceeds much faster in the self-gravitating star.
Nevertheless, the final black hole spin is almost the same for all models (about 0.2).
For super-critical rotation of the star, on the other hand, the value of maximum spin is almost the same for all models, regardless of the initial spin value, and it is about $A_{max}=0.87$ (see Table \ref{tab:modele}). In contrast, the final spin value, is systematically higher than for the case of $S=1$, and it somewhat depends on the initial black hole spin. For $A_{0}=0.85$, the net spin-down of the black hole during the whole simulation is largest.
The values of maximum and final spins for all models are reported in Table \ref{tab:modele}.

 We notice that all our models evolve very quickly, as a consequence of the small radius of the computational domain. Our collapsing stellar core is squeezed into a volume that is much reduced wtr. to a typical stellar progenitor. For a star of the radius about $10^{12}-10^{14} ~cm$ it would take hours to entirely collapse.
  On the ther hand, for a long gamma ray burst, it is sufficient to sustain the accretion disk in the collapsar for about 100 seconds. As was demonstrated with a toy model by \cite{JMP2008}, the duration of the GRB event between 20-40 s, or around 100-150 s, is expected
  for a collapsing star where the black hole is being fed with mass and angular momentum, while the rotationally-supported accretion disk is sustained (cf. Fig. 9 in their paper). The most violent changes of the black hole parameters occur at the very initial phase, corresponding to the shells collapsing from radii below $10^{9}-10^{10}$ cm, where most mass is enclosed. Our calculations, now beyond this old toy model and employing full GR hydrodynamics, confirm this qualitative picture.

\subsection{Effects of the self-force on the flow properties}

In this Section, we present in more detail the properties of the self-gravitating collapsing star for chosen models.
We analyze specific time snapshots, which correspond to the characteristic features found in the time profiles of mass accretion rate, and black hole spin. 

{\subsubsection{Density inhomogeneities and formation of the accretion shocks}}
\label{sect:self_shocks}

 \begin{figure*}[ht]
   \centering
    \includegraphics[scale=0.27]{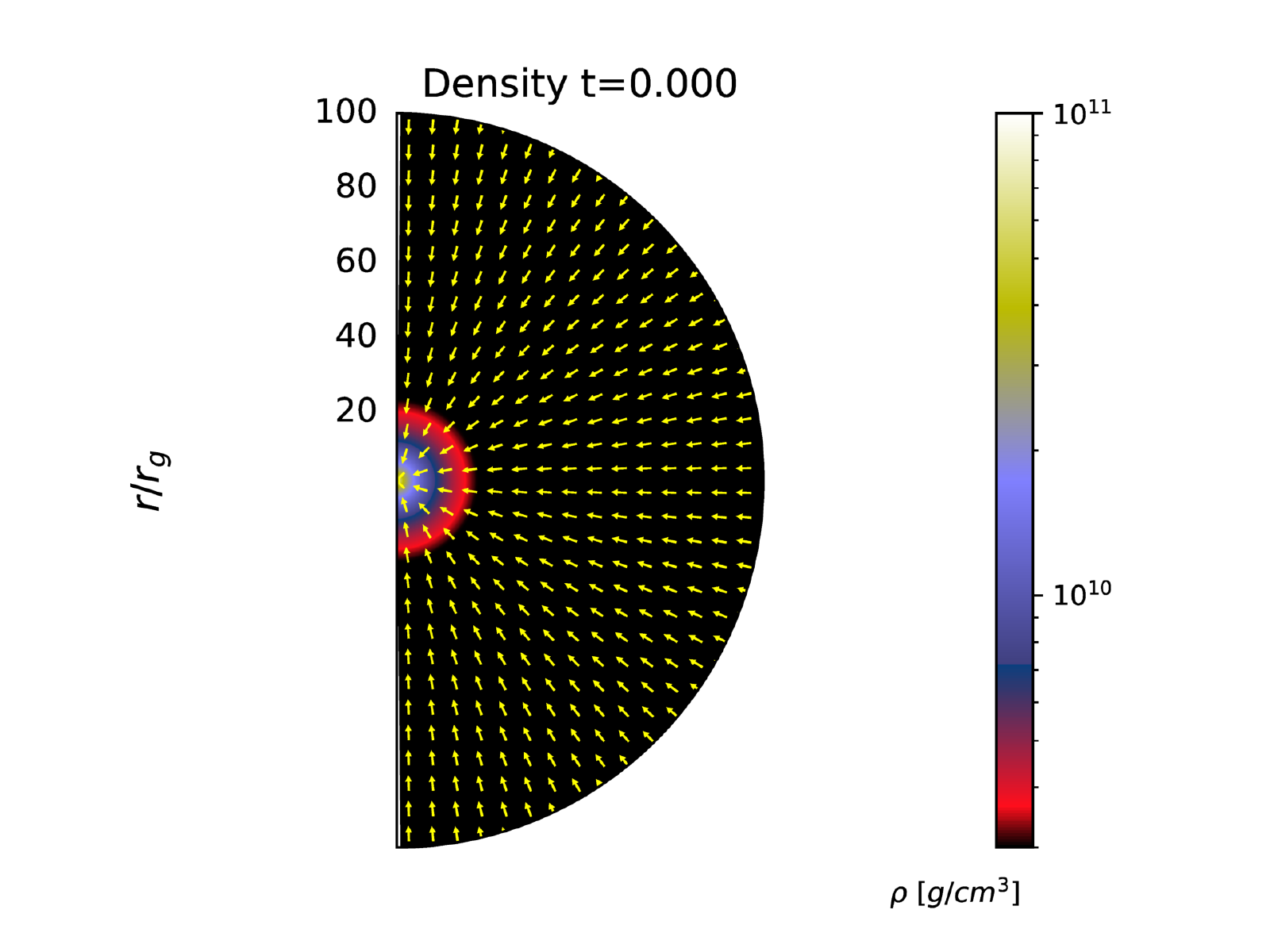}
    \includegraphics[scale=0.27]{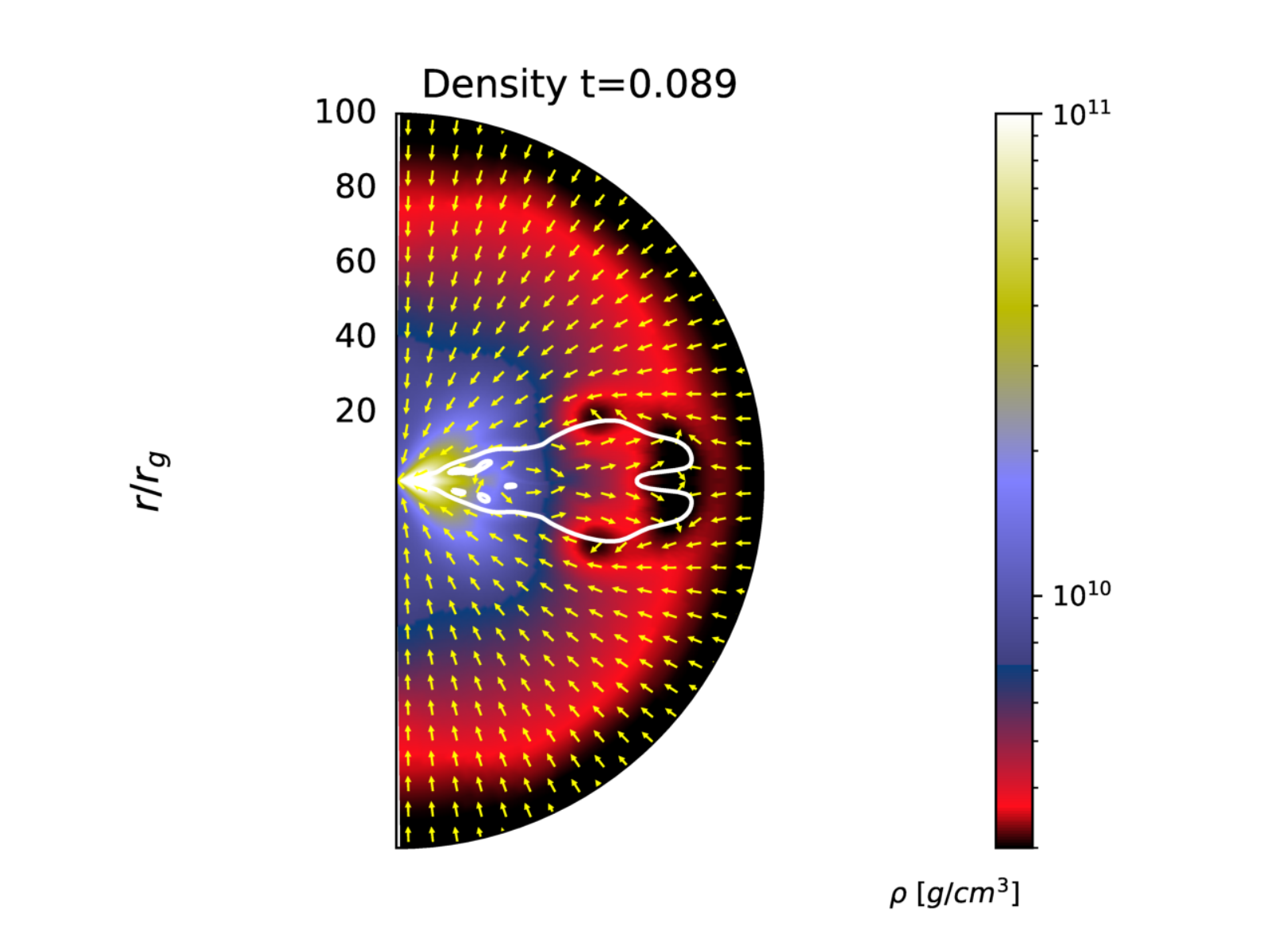}
    \includegraphics[scale=0.28]{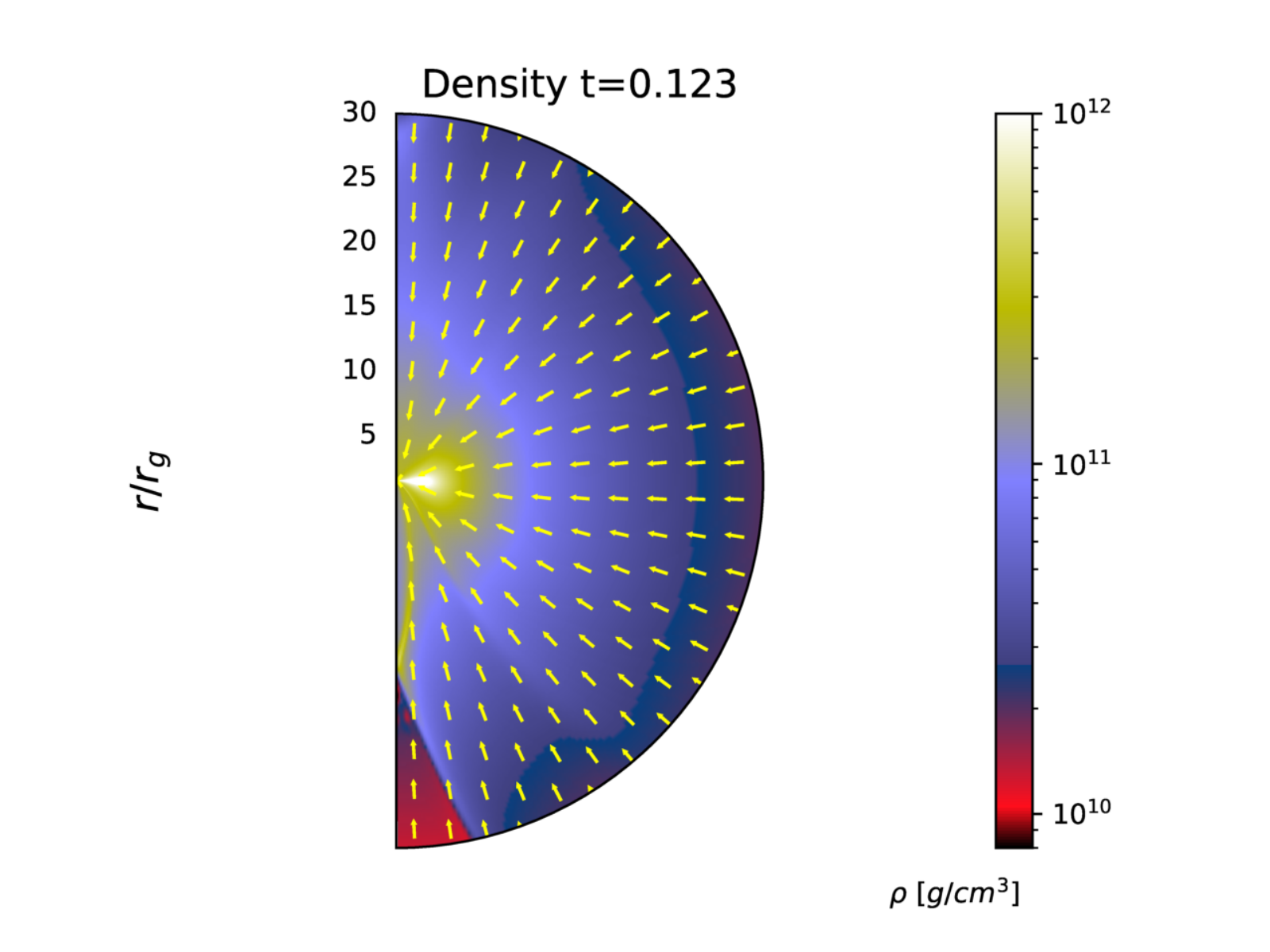}
    \includegraphics[scale=0.28]{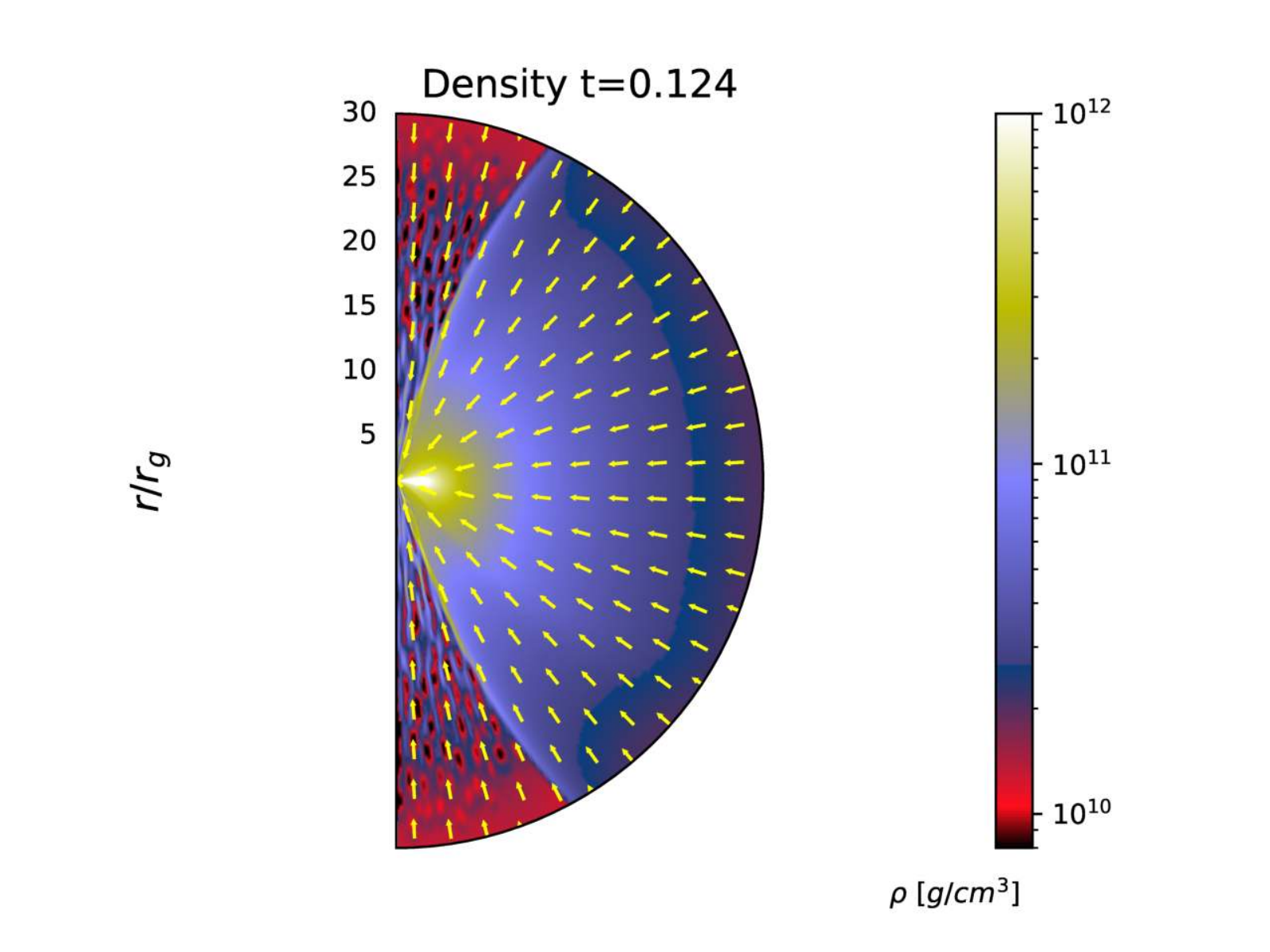}
    \includegraphics[scale=0.27]{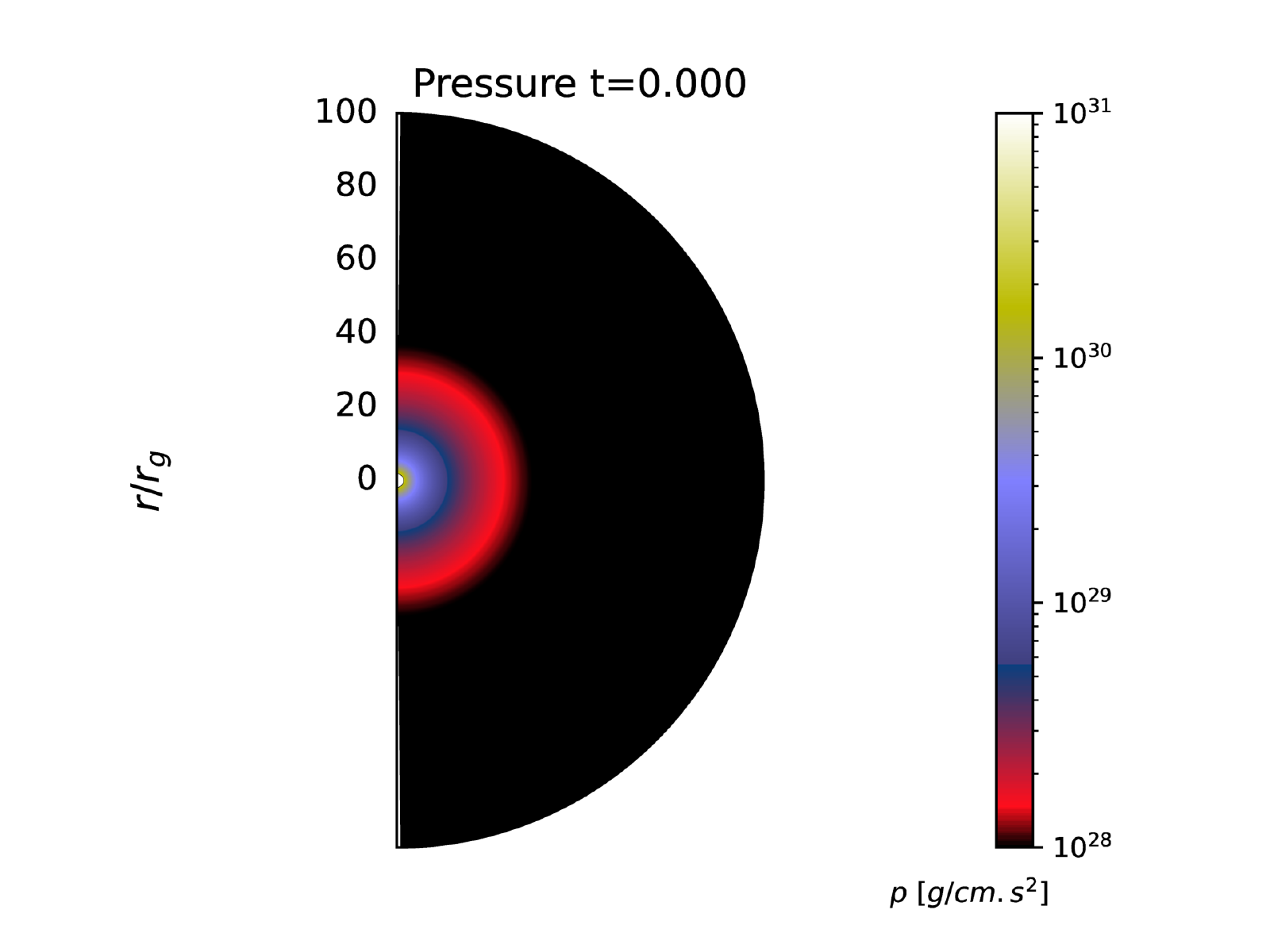}
    \includegraphics[scale=0.27]{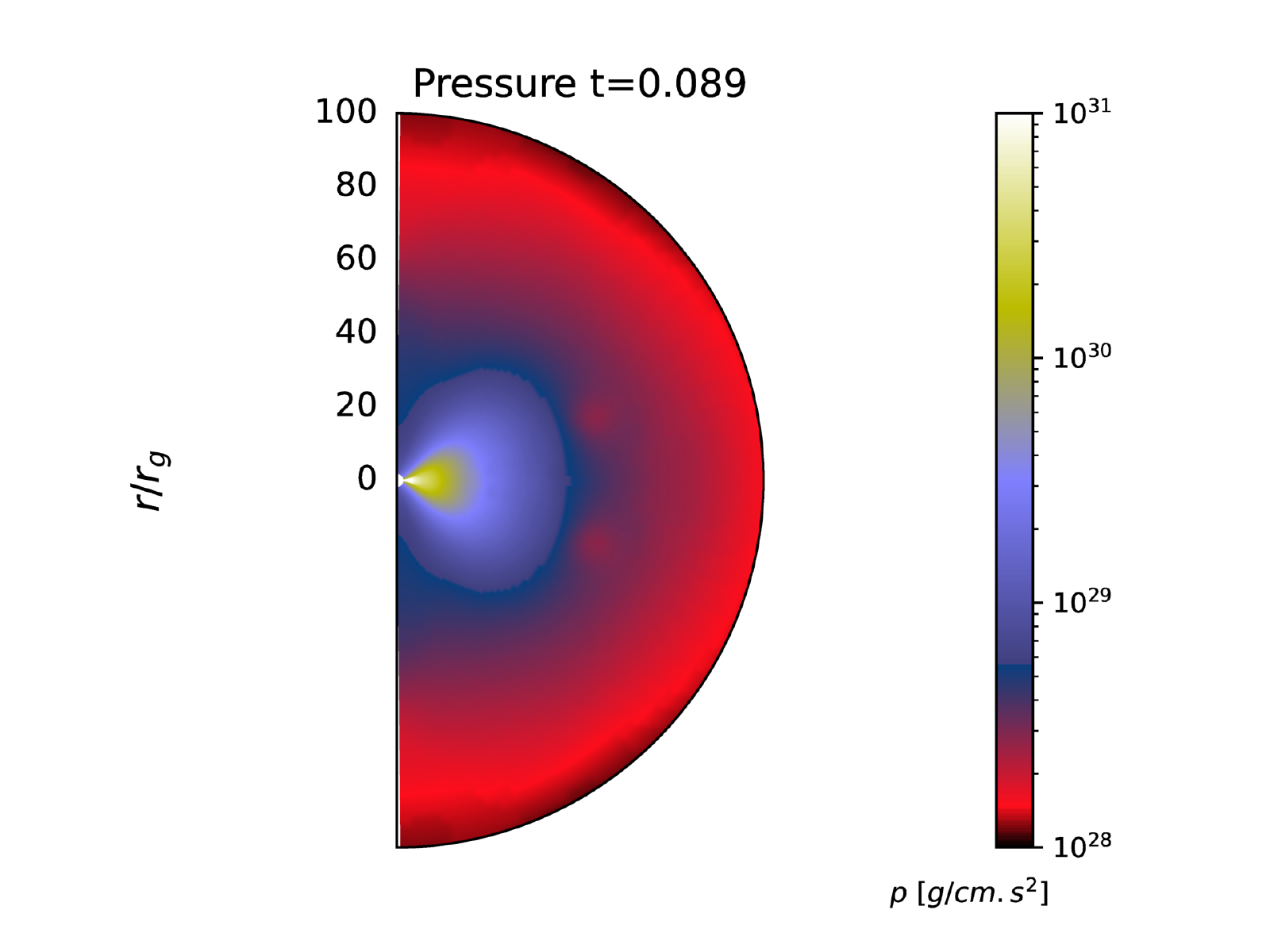}
    \includegraphics[scale=0.28]{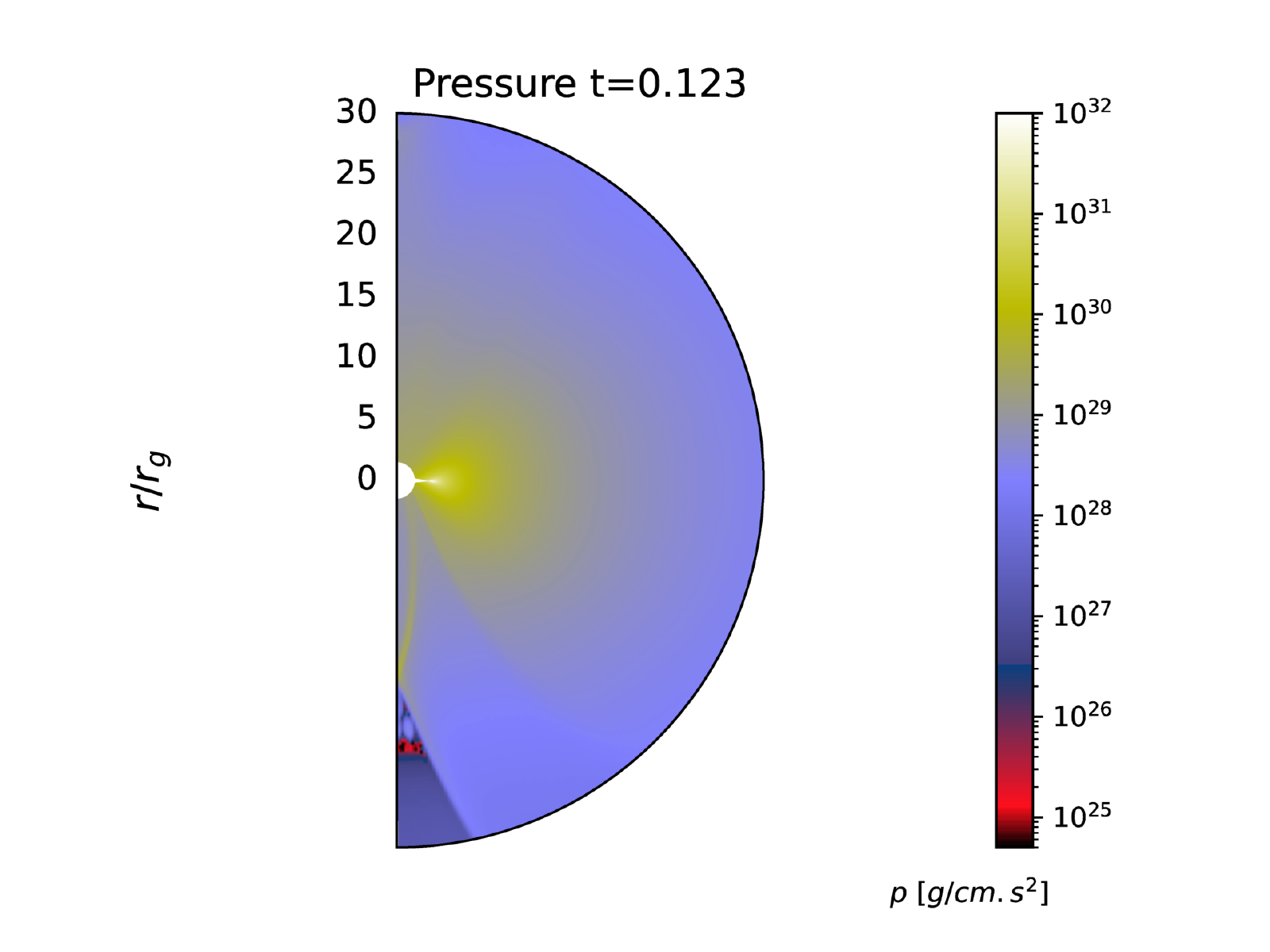}
    \includegraphics[scale=0.28]{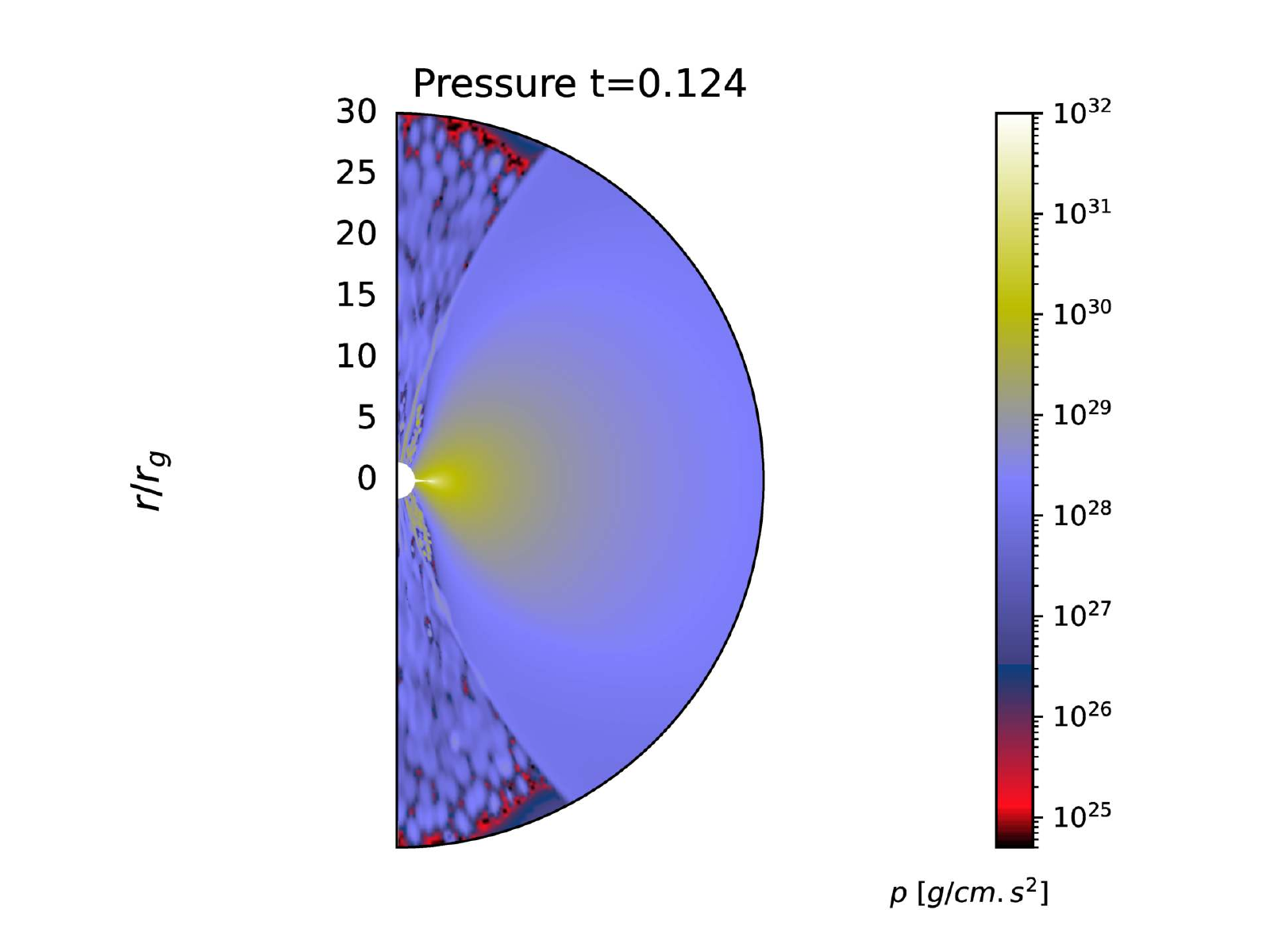}
    \caption{ Snapshots of density and pressure at different time steps, for self-gravitating model. The rotation parameter is $S=1.4$ and initial black hole spin is $A=0.5$. The top three profiles demonstrate the velocity vector field overlaid on the background of density profiles, with thick white curves as the contour plot of the sonic surface. The three bottom snapshots show to the pressure profiles and indicate how it varies through the inner regions, corresponding to density, as a hint to SGI/RT instability. Note that the first snapshot ($t=0.089s$) is zoomed to a larger area of about $100 r_{g}$ while the other two time snapshots illustrate the zoomed in profiles. This is done for a clear indication of shocked region.
     Model shown is A05-S14-SG-R10, as listed in Table \ref{tab:modele}.
    }
    \label{RT/SGI_s1.4}
 \end{figure*}
 
 \begin{table}[h]
    {\scriptsize
    \begin{center}
        \caption{ RT and SGI growth rates for the self-gravitating case of $S=1.4$ and $A_{0}=0.5$ (model A05-S14-SG-R10). Different radii located around the mixing boundary at $\theta=\frac{\pi}{40}$ are considered.}
        \label{growth_rate}
        \begin{tabular}{cccc}
            \hline\hline
            $r~(r_{g})$  & $\sigma_{RT} (s^{-1})$ & $\sigma_{SGI} (s^{-1})$ & $\frac{\partial ln \rho}{\partial r}\frac{\partial ln p}{\partial r}$  \\ 
            \hline \hline
            20.0 &  $2.18\times 10^{-4}$ & 1.1 & $>0$ \\ 
            21.0 &  $2.75\times 10^{-4}$ &  0.97 & $>0$ \\
            22.0 &  $3.64\times 10^{-4}$ &  0.82 & $>0$ \\
            23.0 &  $4.99\times 10^{-4}$ &  0.63 & $>0$ \\
            24.0 &  $7.13\times 10^{-4}$ &  0.4 & $>0$ \\
            25.0 &  $9.5\times 10^{-4}$ &  0.18 & $>0$ \\\hline
        \end{tabular}
    \end{center}
    }
\end{table}

\begin{figure*}[ht]
   \centering
    \includegraphics[scale=0.32]{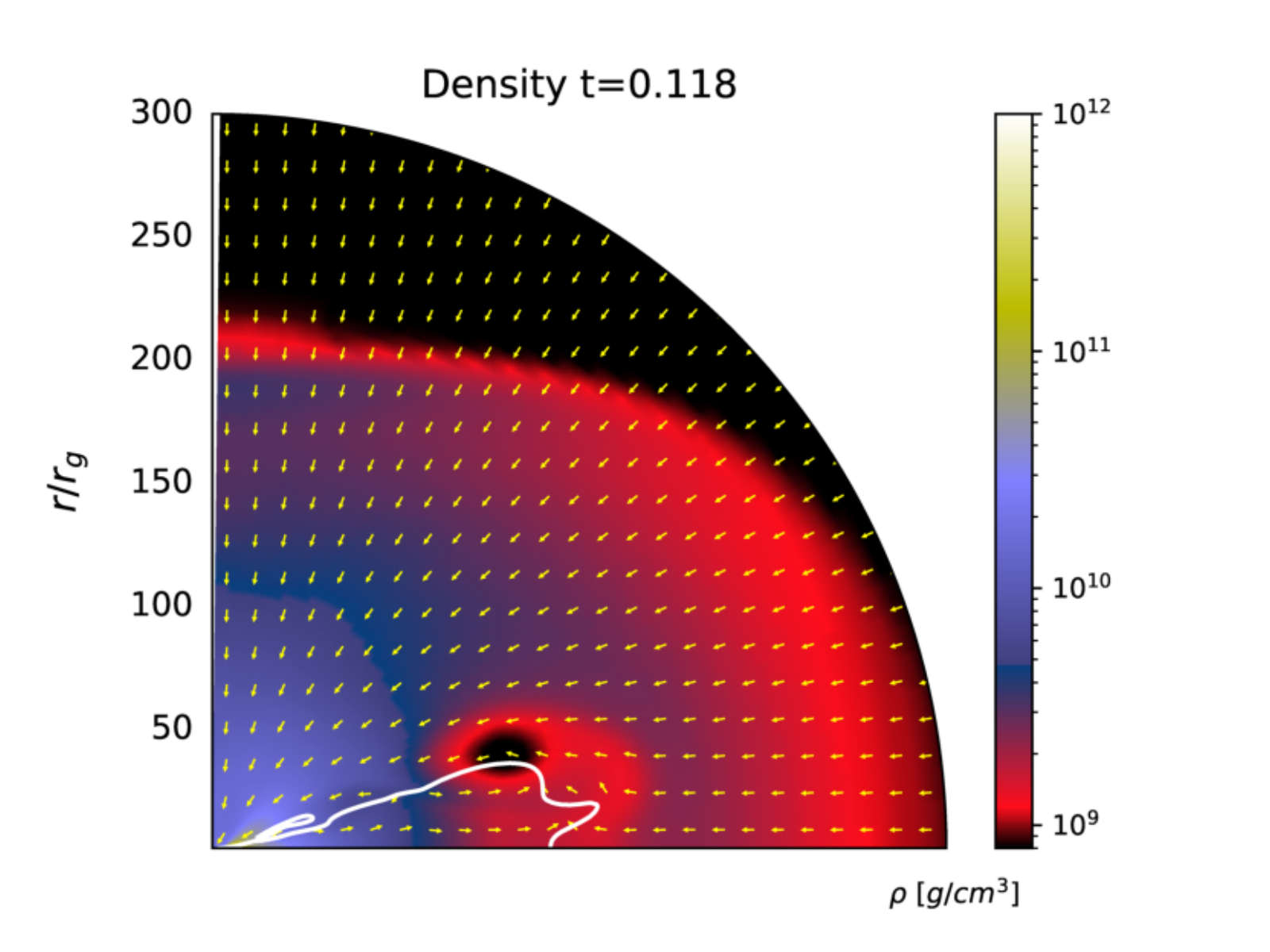}
    \includegraphics[scale=0.32]{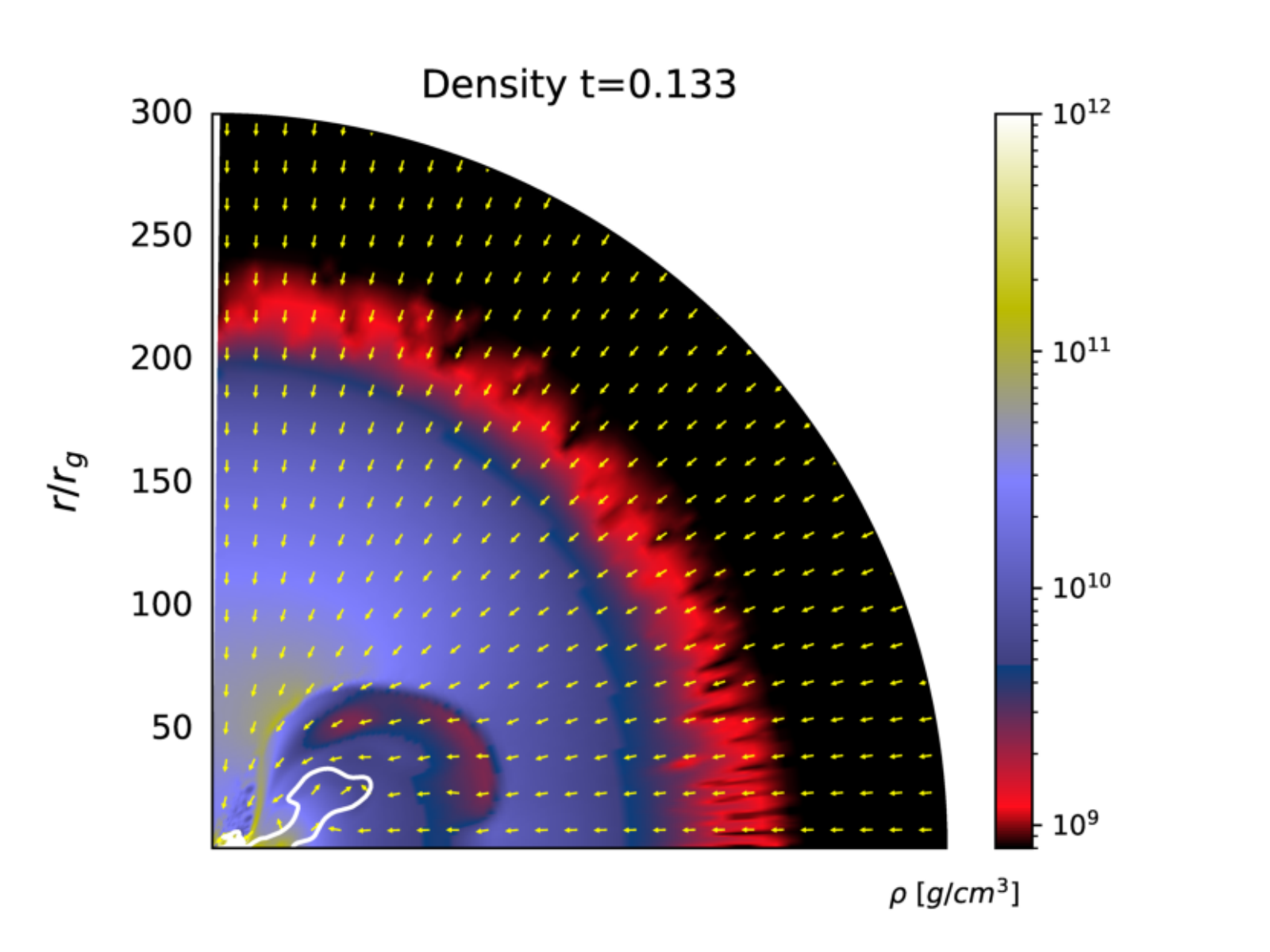}
    \includegraphics[scale=0.32]{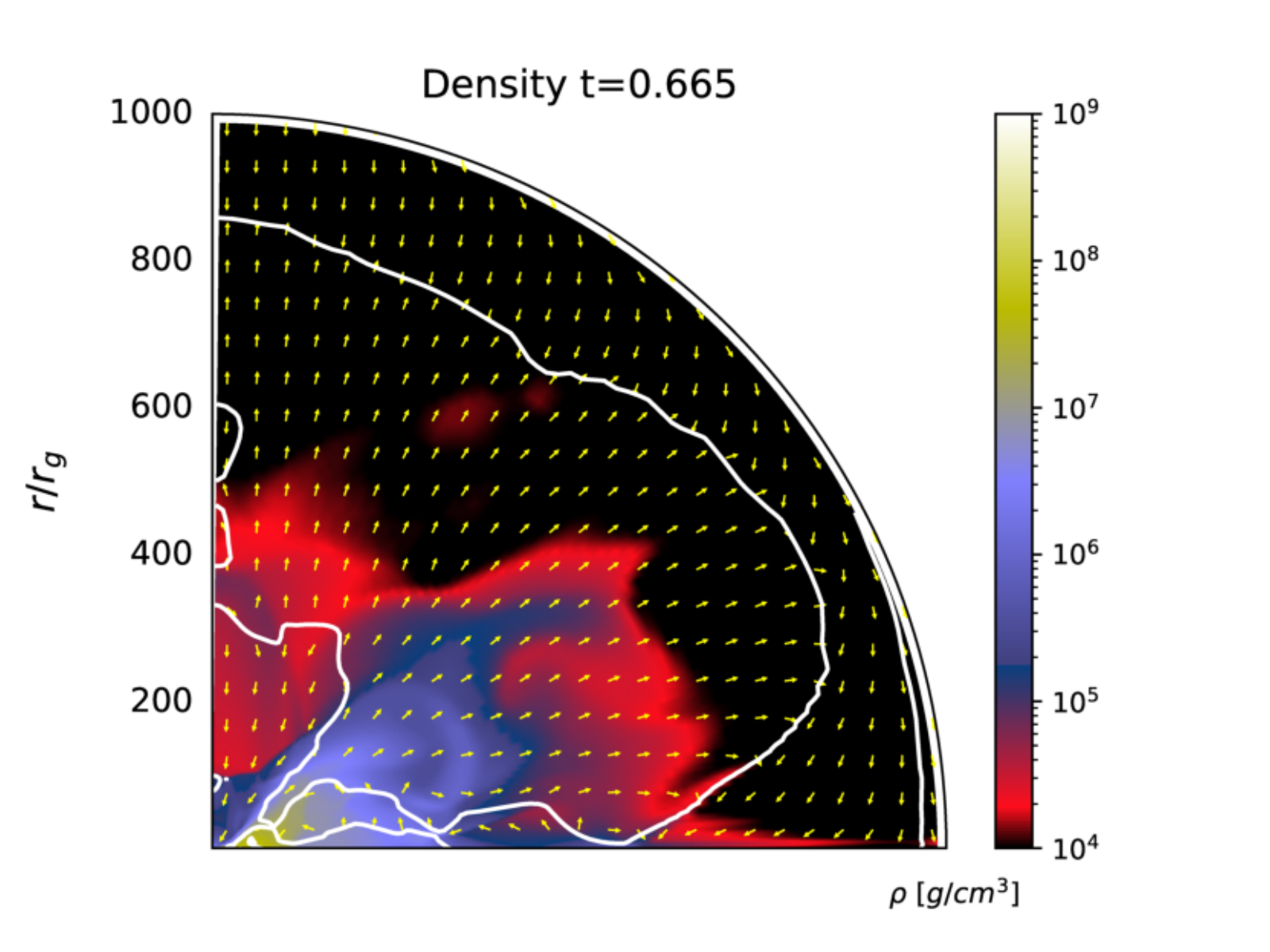}
    \includegraphics[scale=0.31]{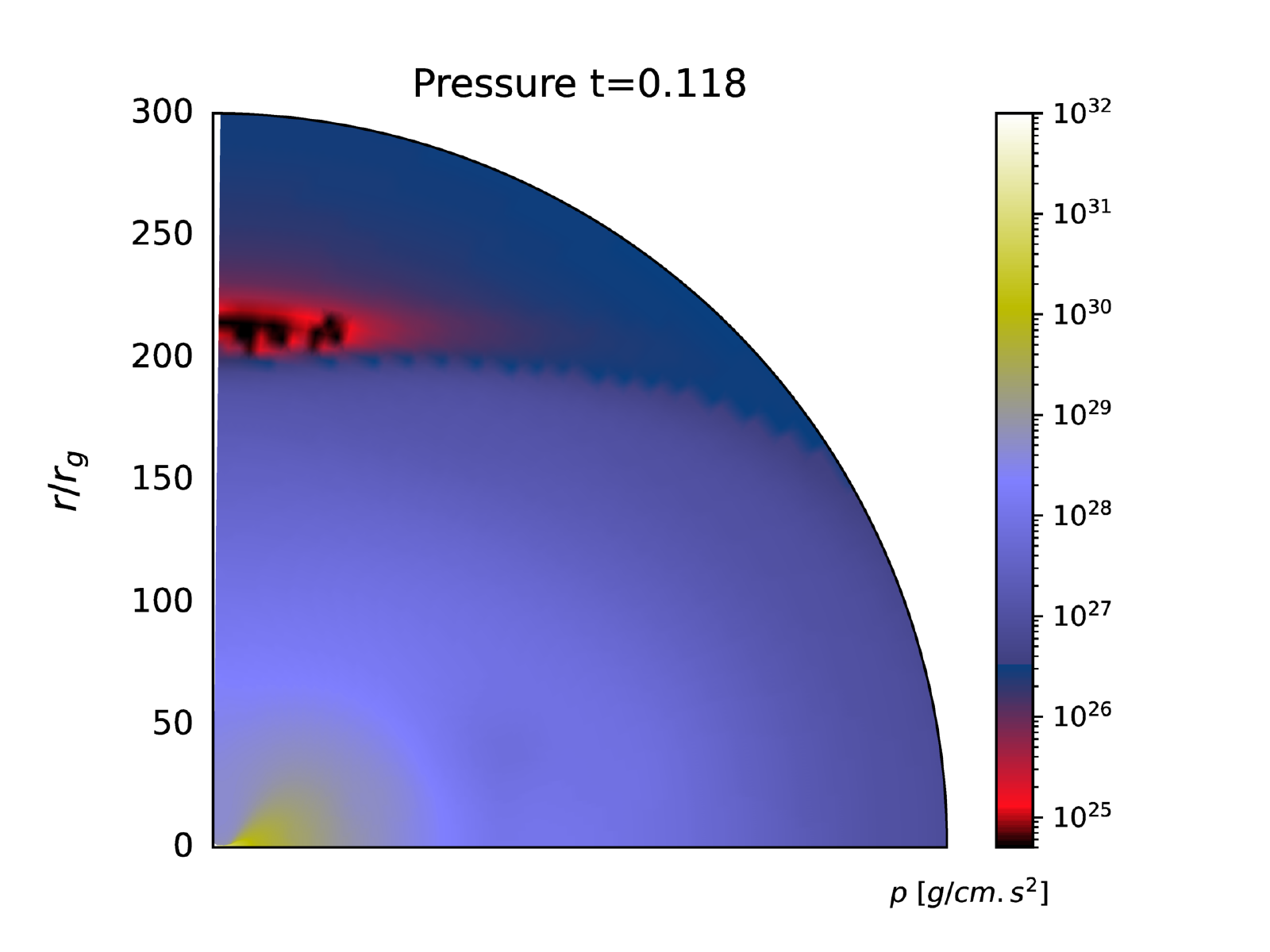}
    \includegraphics[scale=0.31]{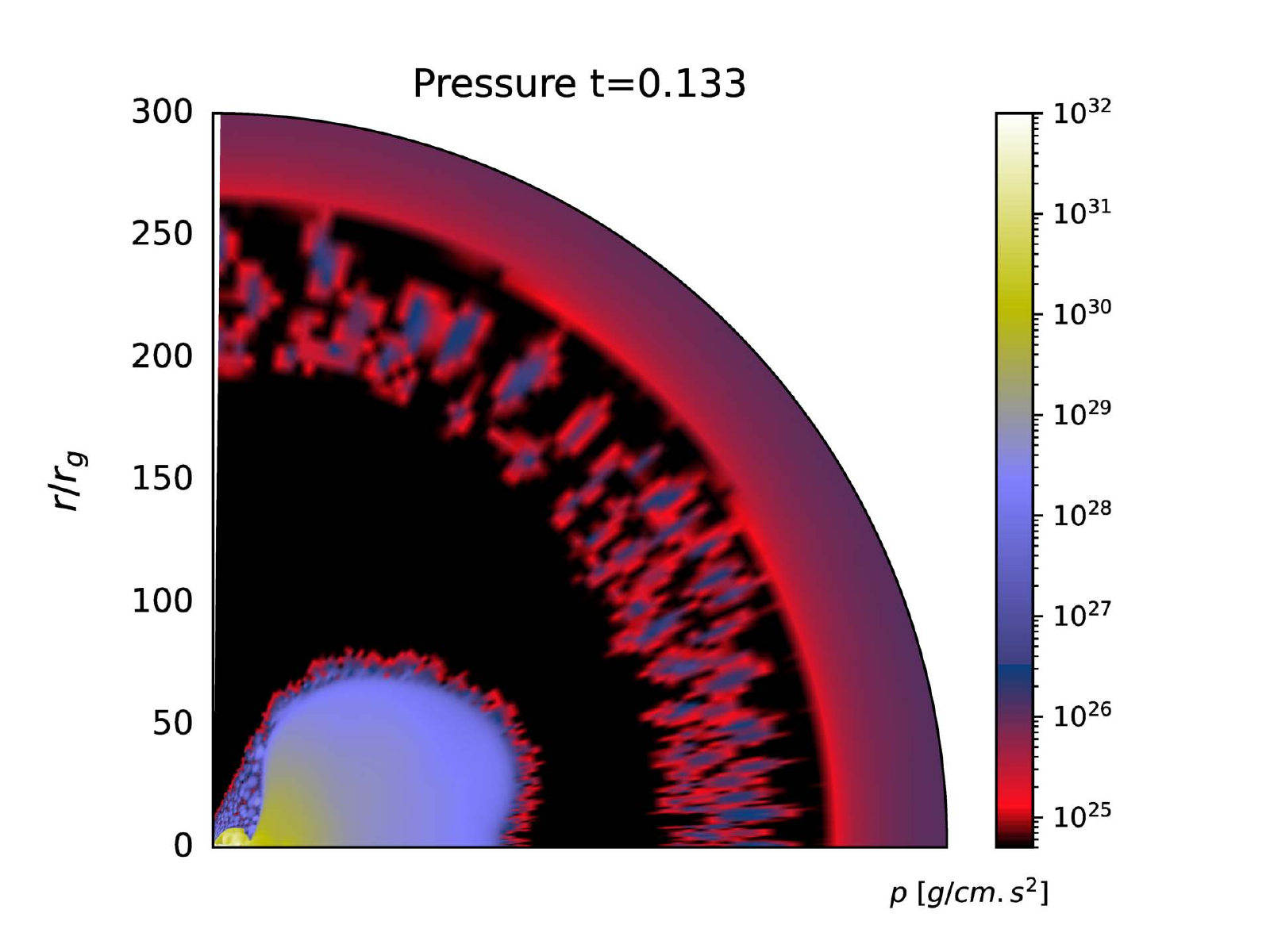}
    \includegraphics[scale=0.31]{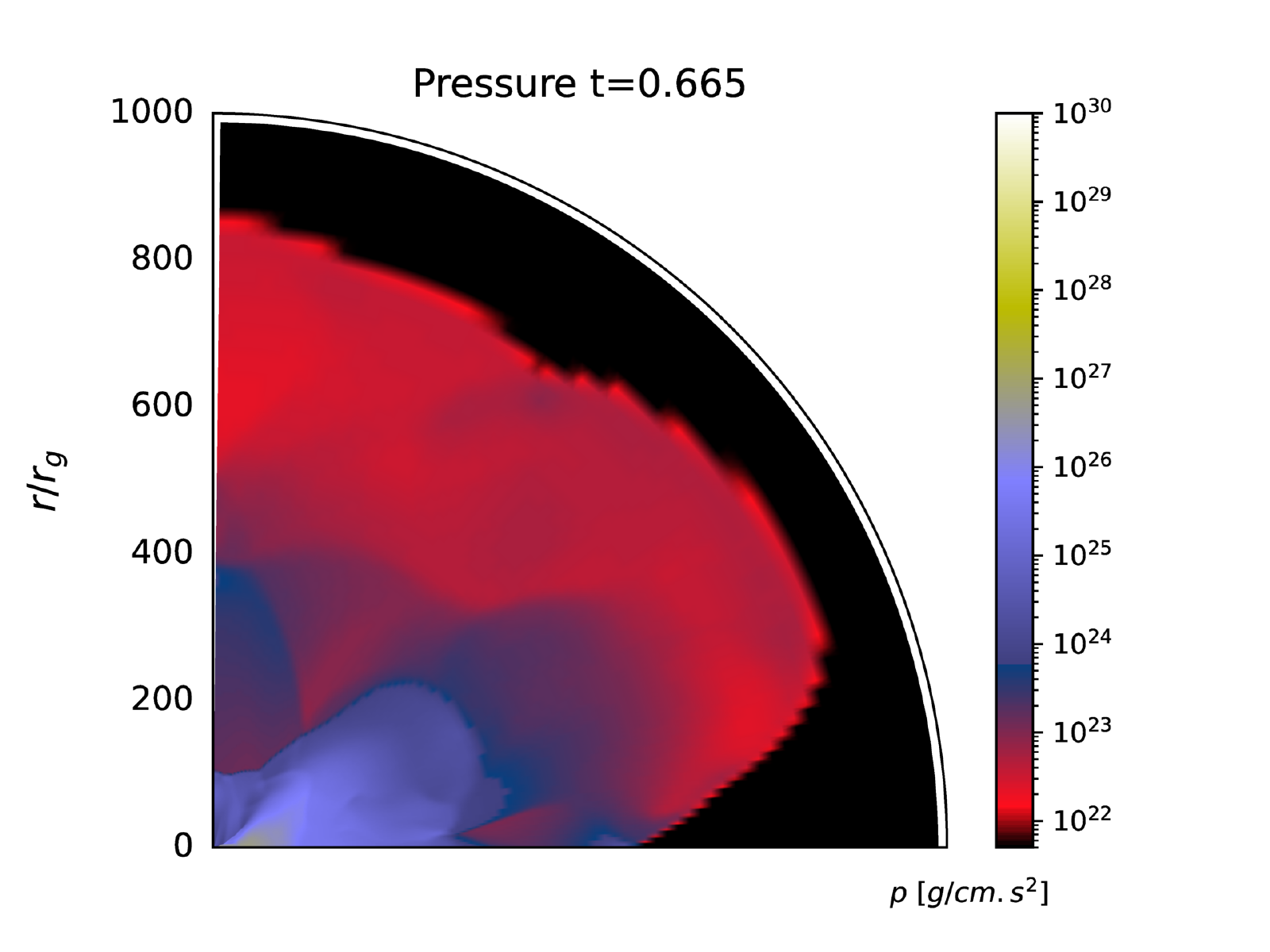} 
    \caption{Snapshots of density and pressure profiles similar to Fig (\ref{RT/SGI_s1.4}), for the model with rotation parameter $S=2$ and initial black hole spin $A_{0}=0.5$. The third snapshot ($t=0.665s$) is zoomed out to a larger area of about $1000 r_{g}$ while the first two time snapshots illustrate a zoomed in to $300 r_{g}$ for a clear representation of both outflow regions at final time steps and the visibility of inhomogeneities.  Model shown is A05-S20-SG-R12, as listed in Table \ref{tab:modele}.} 
    \label{RT/SGI_s2}
 \end{figure*}

In Figure \ref{M_S_A} we have shown time profiles of accretion rate through the black hole horizon, which exhibit a sudden increase at early times of the simulation, followed by an oscillatory pattern in the cases with self-gravity. 
We now show in Figure \ref{RT/SGI_s1.4} the corresponding snapshots of density profiles in addition to velocity vector field. We notice that they reflect formation of some special structures, i. e. an equatorial outflow of matter, which reaches the radii up to about 80 $r_{g}$ and then is stalled in the transonic shock. 
Furthermore, we notice some small inhomogeneities in the density at the chosen time intervals, visible in more detail in Fig. \ref{RT/SGI_s2}. The assumed black hole spin parameter in both models was $A_{0}=0.5$, while the rotation parameter is either $S=1.4$, or $S=2$, for Fig. \ref{M_S_A} or Fig.\ref{RT/SGI_s1.4}, respectively. Both models are including self-gravity impacts. In either Figure, the top three images depict density snapshots of the collapsing stellar core overlaid with normalized velocity vector fields and contour of Mach number $M=1$ (i.e. sonic surface) for the sake of a better representation of the transonic shock and equatorial outflow regions. Furthermore, we provide the pressure maps (the three bottom profiles in both figures) which are corresponding to the same time steps as those of density profiles. In this way, we investigate the possibility of some types of hydrodynamical interfacial instabilities (i.e., Rayleigh–Taylor or Self-gravity Interfacial (SGI) instability). 

In the density profiles during the early time steps, we found an increase of density in the inner regions ($r<100 ~ r_{g}$) which show an accumulation of mass at the equatorial region. Such an accumulation refers to the formation of outflow as shown through backward velocity vectors surrounded by contours of Mach number $M=1$. This appeared in density snapshots at times $t=0.089 ~s$, or $t=0.118 ~s, 0.133 ~s$, for the case with $S=1.4$, or $S=2$, respectively. We interpret the creation of outflow at the equatorial plane as an impact of centrifugal force, which is found for supercritical rotation. Consequently, a transonic shock is located within $\simeq 100 ~ r_{g}$ radius. In the following time steps, the density profiles show that the incoming material from outside the disk is finally channeled into this inner region, leading to a rise in the accretion rate at about $0.1 ~s$ (cf. the early time sharp increase of the accretion rate, as shown in the thick pink and green curves in Figure \ref{M_S_A}). As expected, for the case of higher rotation parameter $S=2$ the outflow region has a larger size, and survives for a longer time ($\simeq 0.133 ~s$). Additionally, another outflow arises at the end of simulation, for the case of higher $S=2$ parameter, which leads to the outflow of matter from the envelope. This is demonstrated in the third density profile  of Figure \ref{RT/SGI_s2} ($t=0.665 ~s$). This causes a more oscillatory behavior of the mass accretion rate during the last time steps of our simulation (cf. thick green curve in Figure \ref{M_S_A}). 
In comparison, for non self-gravitating cases, in \cite{Dominika2021}, we found a considerably longer lasting disk structures at the equator, which were spread out through larger radii. This confirms that there exists an interplay between self-gravity and centrifugal force (rotation), and consequently its suppressive impact on the outflow regions.  

We notice that an inhomogeneous structure appears in both density and pressure profiles (as well as other quantities, such as specific angular momentum and Much number). This behavior is seen at time step $t=0.123 ~s$ for $S=1.4$, and more evident at $t=0.133 ~s$ for $S=2$, respectively. To have a better representation, we depicted a snapshot showing only an upper  hemisphere in the case with $S=2$, for which there is a symmetric structure with respect to the equator (in contrast to the case of $S=1.4$). One can find that the inhomogeneities start growing from the interfaces with significant discontinuities in density. As a result, interfacial instabilities seem likely to be responsible for such a structure. 

In general, instabilities may be divided into two types, global instabilities (such as Jeans) and interfacial instabilities (examples are Kelvin–Helmholtz and Rayleigh–Taylor (RT) instabilities). Among interfacial instabilities RT instability occurs when density and pressure gradients act in opposite directions. This criterion can be identified through the linear growth rate proposed to examine the RT-unstable regions, which reads \citep{kifonidis2003non}:
\begin{equation}
\sigma_{RT} = \sqrt{-\frac{p}{\rho}\frac{\partial ln \rho}{\partial r}\frac{\partial ln p}{\partial r}}.
\label{RT}
\end{equation}
Moreover, 
\cite{hunter1997kelvin, hunter1998stability} introduced another type of interfacial instability driven by self-gravity, called Self-gravity Interfacial instability (SGI). The linear growth rate for the SGI instability is supposed to be \citep{hunter1997kelvin, hunter1998stability}
\begin{equation}
\sigma_{SGI} = \sqrt{\frac{2\pi G (\rho_{2}-\rho_{1})^{2}}{(\rho_{2}+\rho_{1})}}.
\label{SGI}
\end{equation}
The RT and SGI instabilities result in very similar configurations at density snapshots. However, they have their own characteristics which allow us to differentiate between these types of hydrodynamical instabilities. Since self-gravity knows no preferred direction, it is destabilizing across all density interfaces, while an interface is RT-unstable only if the heavy fluid is on top of the light fluid. It has been also confirmed that RT instability is characterized by dense spikes penetrating the tenuous fluid, whereas the SGI develops with tenuous spikes streaming into the denser fluid \citep{hueckstaedt2005parameter}.           

As one can infer from Figure \ref{RT/SGI_s1.4}, SGI instability seems to dominate over RT instability and produces the inhomogeneities. First, density and pressure profiles at $t=0.123s$ confirm the fact that density and pressure gradients over the discontinuity, that appeared at $r\simeq20R_{g}$, act in the same direction. To have a better intuition, some data for the criteria (\ref{RT}) and (\ref{SGI}), are provided in Table \ref{growth_rate}, which quantifies the growth rates of RT and SGI instabilities around the boundary with the emergence of growing unstable configurations (i. e., $r\sim20R_{g}$). It also confirms the positive sign of $\frac{\partial ln \rho}{\partial r}\frac{\partial ln p}{\partial r}$ at these regions. Moreover, a comparison between the growth rate of RT and SGI instabilities represents a considerably faster growing manner for the SGI modes, in the mean that this type of instability seems a plausible candidate that may account for the inhomogeneous structure. Second, it appears that tenuous bubbles penetrate into denser matter which also points out a less dense matter lies on the top of denser fluid, regarding direction of the flow in this region. Therefore, we argue that it is SGI instability which seems to be of growing modes rather than RT. Based on similar evidence, we also believe SGI instability can be a more probable mechanism to cause the inhomogeneous structure in the case with $S=2$, as shown in Figure \ref{RT/SGI_s2}.     

It is worth discussing that such a configuration does not seem to be a non-linear turbulent structure, since we found no mixing among bubbles and spikes so that their size changes remarkably before falling into the black hole. Moreover, the velocity vector fields also confirm no detectable interaction between these small oscillatory patterns, i. e. there seem not to be any chaotic direction in the velocity vectors. Therefore, we believe that this pattern can be treated as the linear growth of SGI modes which is due to SG, without representing any non-linear growth that should occur as a result of interactions between small bubbles and spikes.

\subsubsection{Self-gravity and angular momentum transport}
\label{sec:SG_angular_momentum}

\begin{figure*}[t]
    \centering
    \includegraphics[scale=0.45]{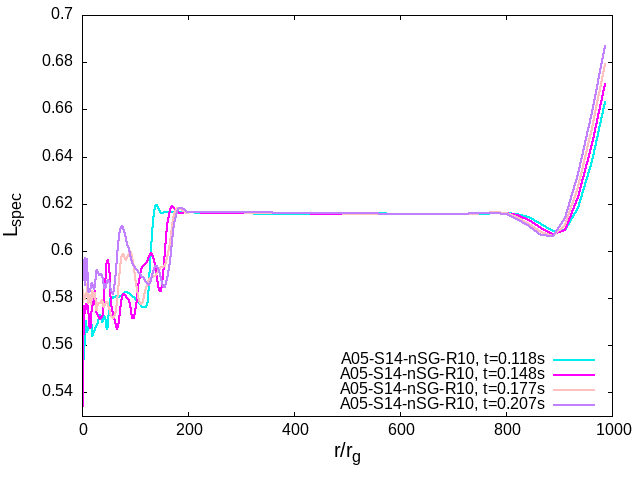}
    \includegraphics[scale=0.45]{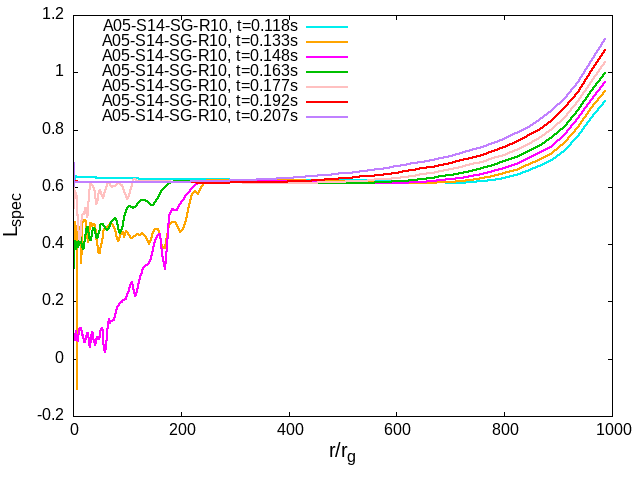}
    \includegraphics[scale=0.45]{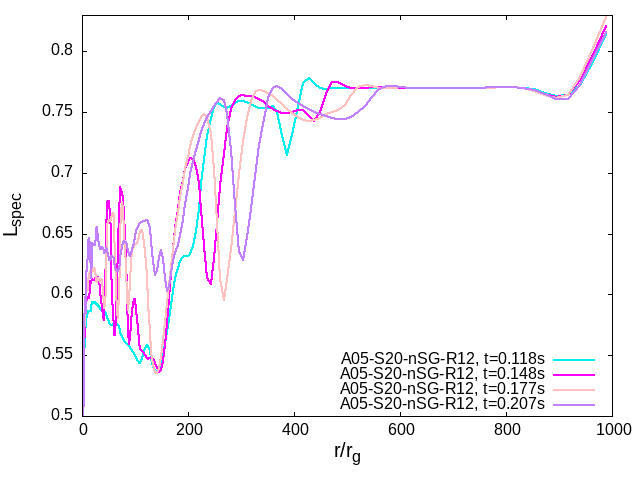}
    \includegraphics[scale=0.45]{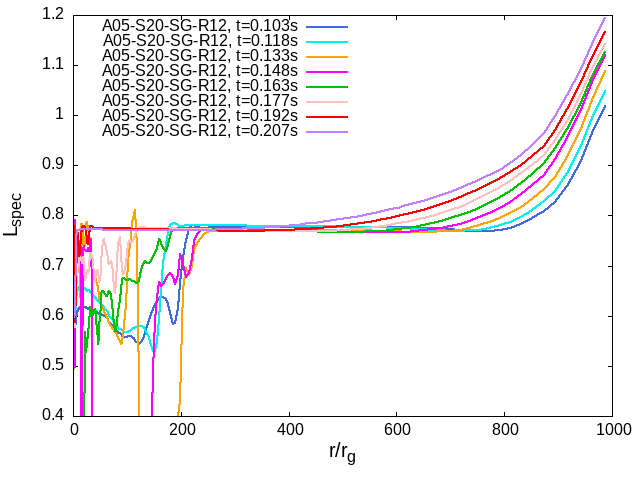}
    \caption{ Specific angular momentum at the equator, taken at several time steps corresponding to the inhomogeneous structure in self-gravitating case (right panels in both rows). The cases without self-gravity are also presented in the left two panels. Two cases of $S=1.4$ and $S=2$ with $A_{0}=0.5$ are shown in the top and bottom panels, respectively. During the emergence of inhomogeneities, self-gravity seems to transfer the angular momentum into the black hole. Models are labeled in all panels with symbols referring to Tab \ref{tab:modele}.}
    \label{lspec}
\end{figure*}

Self-gravity is expected to play a role in the angular momentum transport, e.g. in protoplanetary disks \citep{2011ARA&A..49..195A}. 
In more detail, the transport of angular momentum can be possible via Hydrodynamics (HD) or Magneto-Hydrodynamics (MHD) instabilities to produce a turbulent structure leading to such a transition. Self-gravity can also provide this through gravitational instabilities \citep{Lodato2008}, in addition to SGI instability discussed above. We postulate that self-gravity facilitates the transfer of angular momentum in our collapsing scenario. In Figure \ref{lspec} we show the specific angular momentum of the envelope at the equator for two cases, namely with and without self-gravity. The top panels show models with $S=1.4$, while plots on the bottom present models with $S=2$. Comparing those two cases with non-SG models (with plots located in the left), we show that self-gravity has paved the way for the angular momentum to be transferred outwards as appeared in the larger radii. On the other hand, during the intervals when the density inhomogeneities emerge (i.e., between $t=0.118 ~s -0.148 ~ s$), there seem to be a sudden decrease in the innermost regions, in the profile of the angular momentum. It can be interpreted as a sudden transport of the mass and angular momentum towards the black hole. However, it indicates an upward trend once the inhomogeneities disappear (i.e., from $t=0.163s$ afterwards).

More precisely, we think that self-gravity has two major impacts (SG) on the specific angular momentum. First, SG models accelerate the evolution of envelope considerably, so that the inward mass and outward angular momentum transfer occur within a shorter period of time with respect to non-SG model. This results in a larger increase of the outer region’s specific angular momentum, from a time step to another, in comparison to the non-SG models. Therefore, we may attribute differences in the specific angular momentum at the larger radii, illustrated in Figure \ref{lspec} for the cases with and without self-gravity, to this issue. We believe that the faster evolution in SG models is due to the suppressing impact of SG on the centrifugal force, which also explains the longer lasting outflow of matter around the equator in non-SG models (one may attribute the production of outflow to the centrifugal force, that results in the slower evolution of non-SG models).

\begin{figure*}[t]
    \centering
    \includegraphics[scale=0.37]{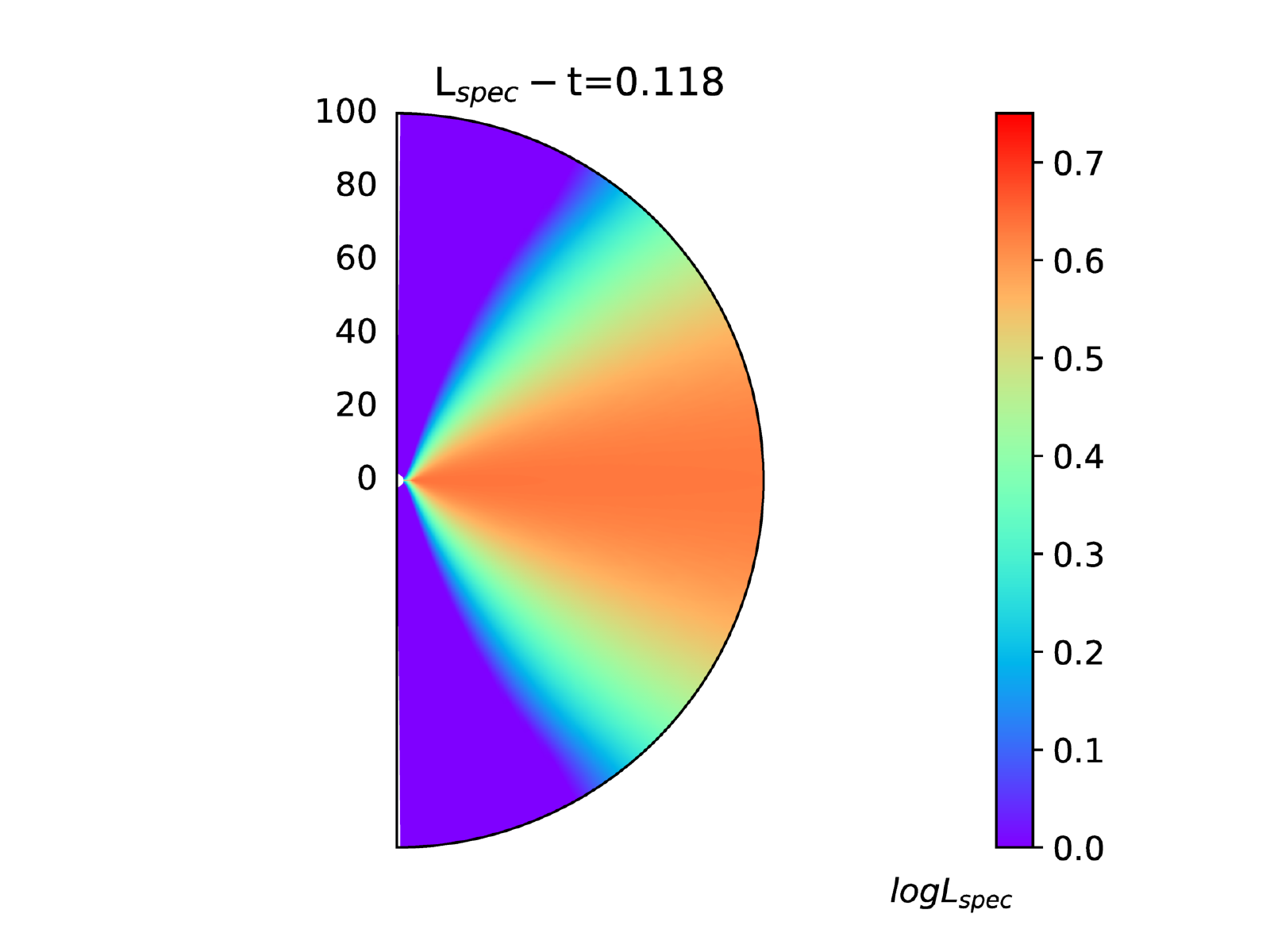}
    \includegraphics[scale=0.37]{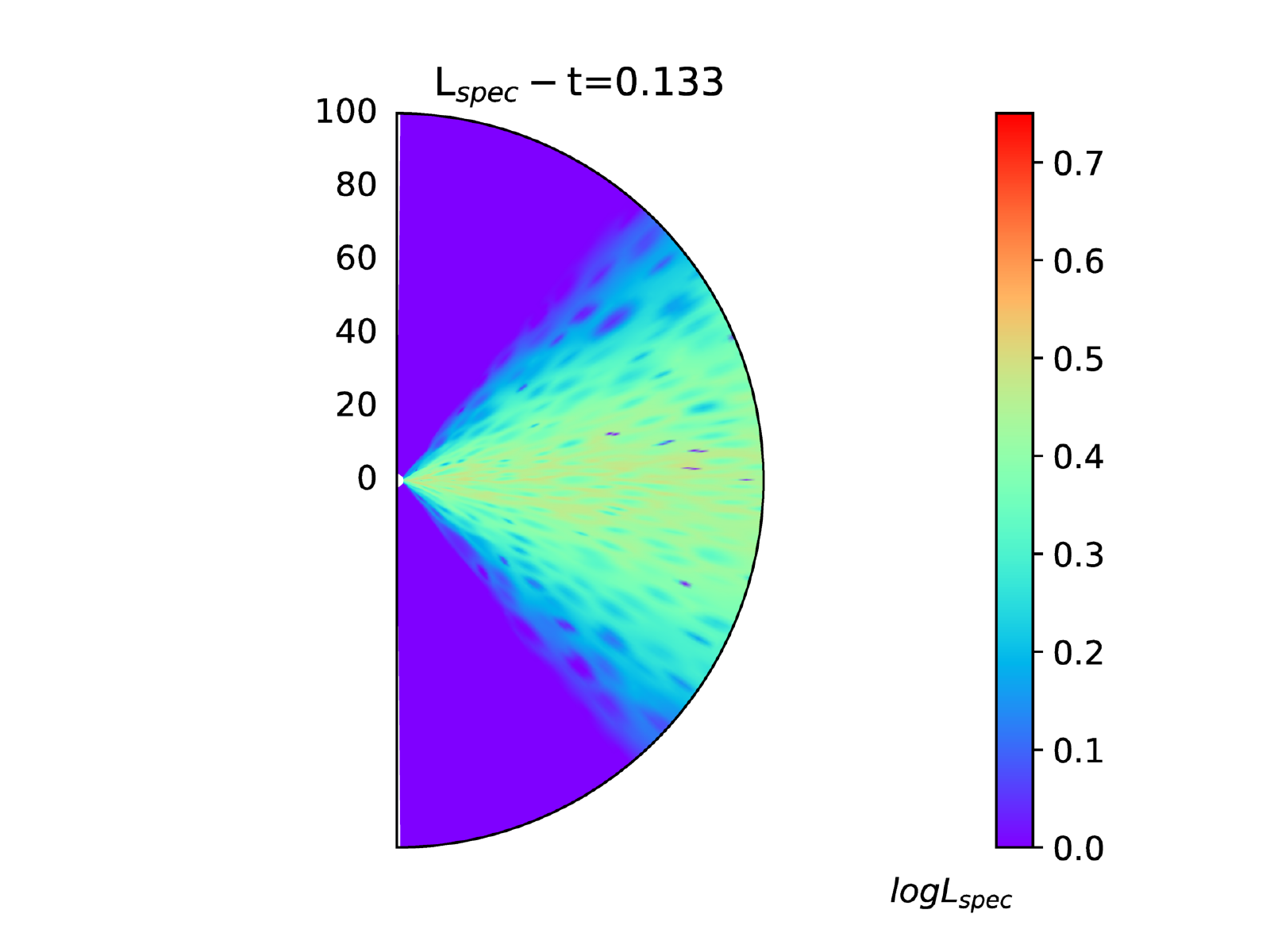}
    \includegraphics[scale=0.37]{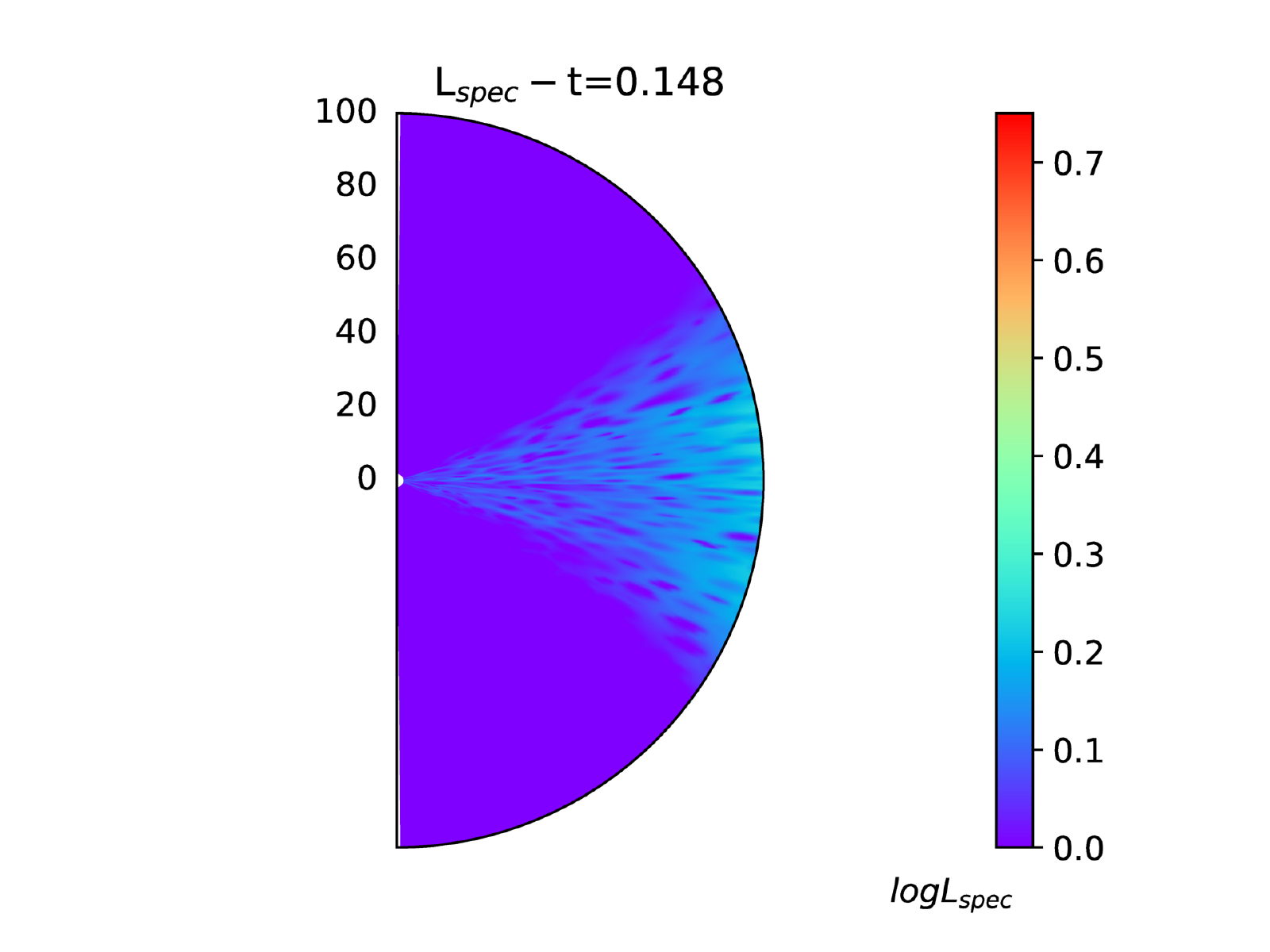}
    \includegraphics[scale=0.37]{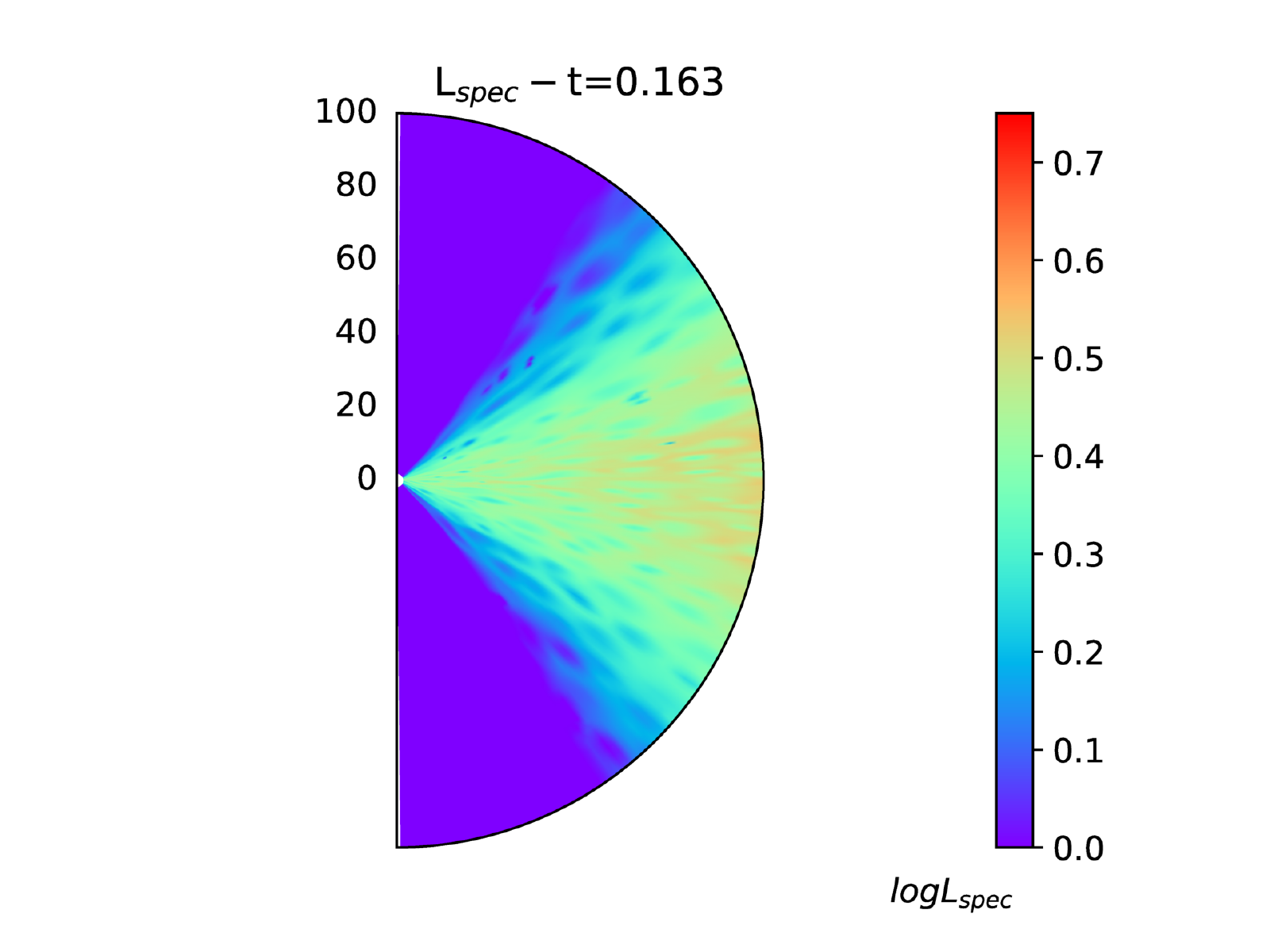}
     \includegraphics[scale=0.37]{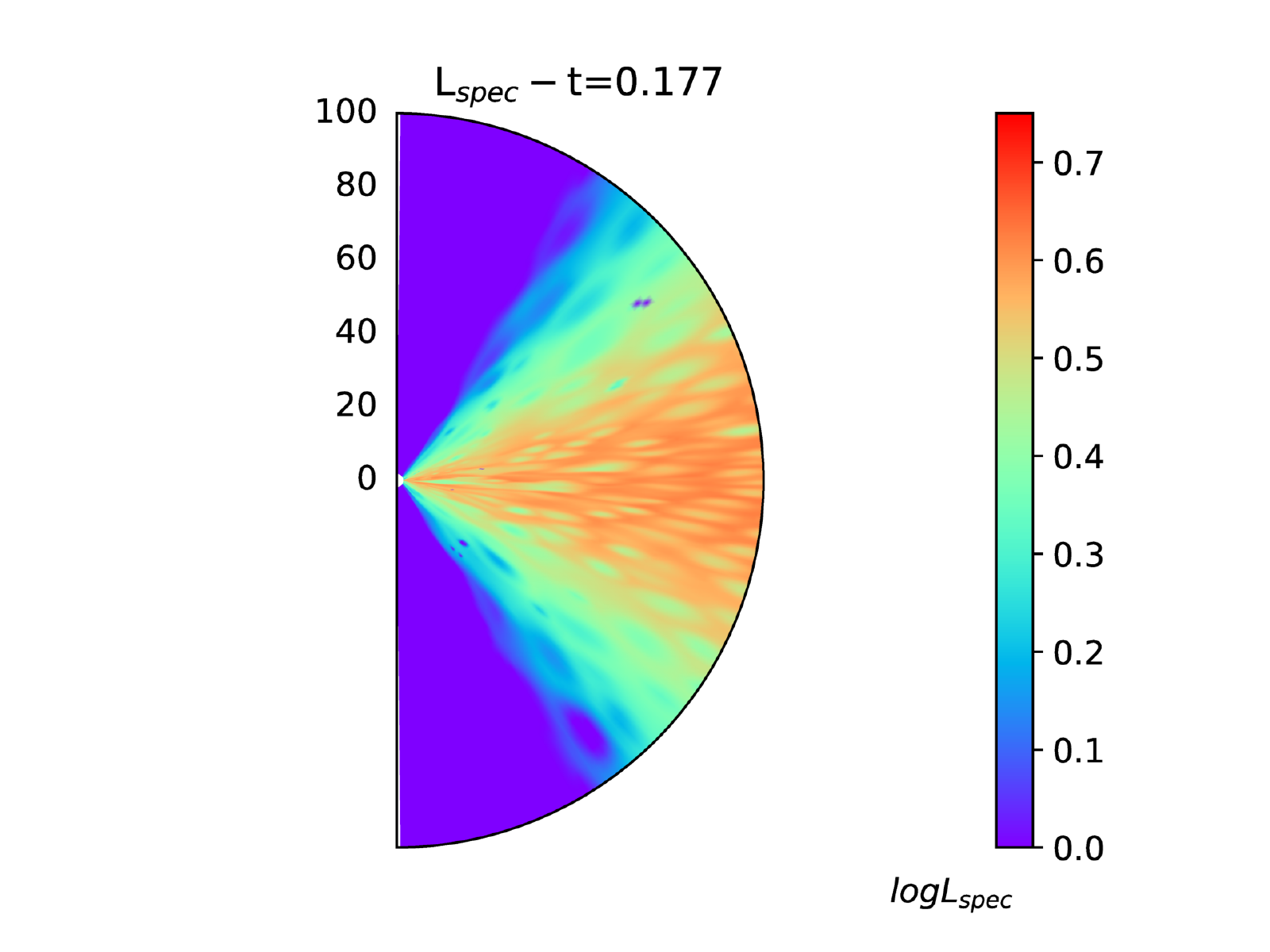}
      \includegraphics[scale=0.37]{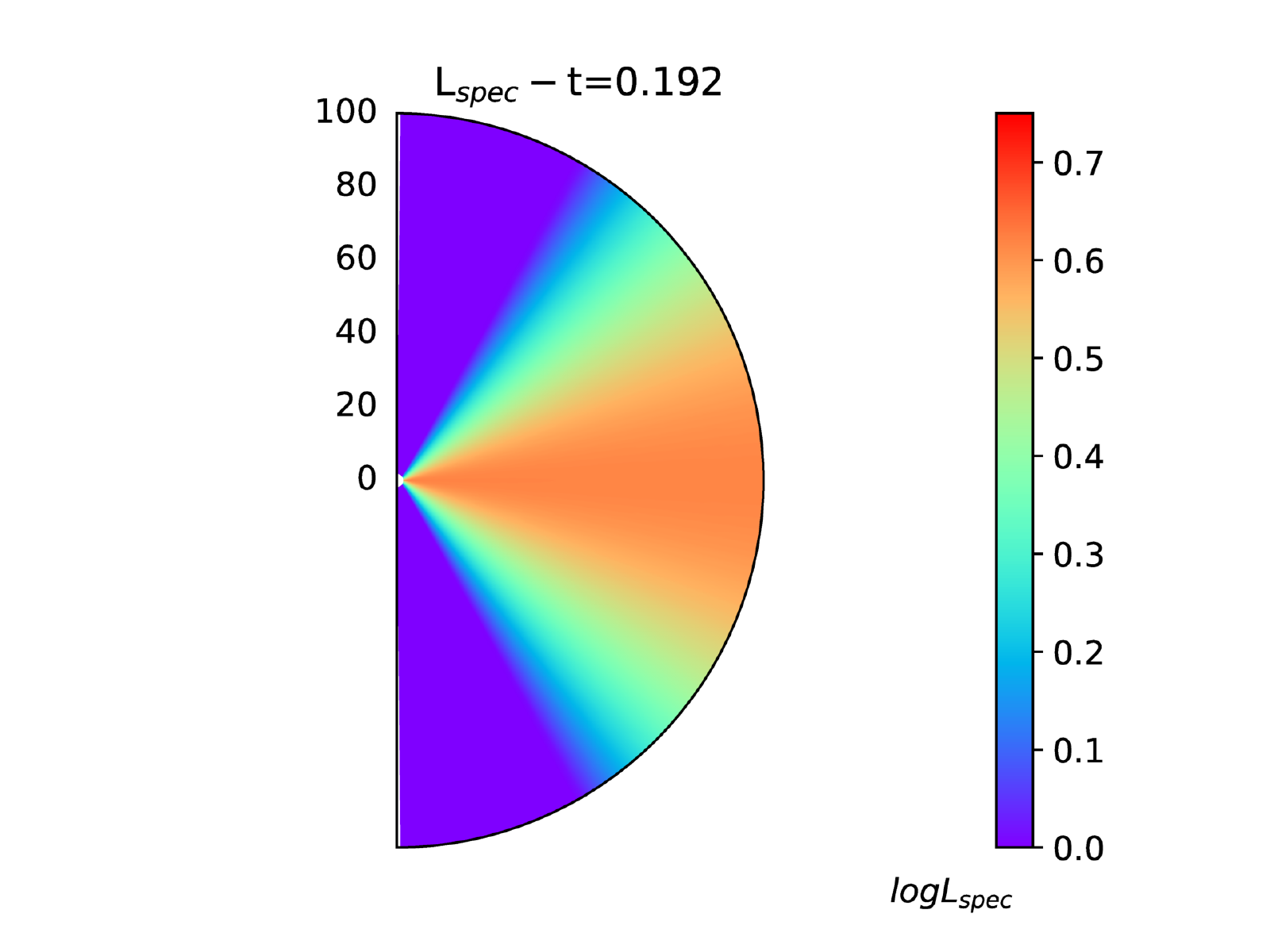}
    \caption{Specific angular momentum snapshots, for the model with self-gravitating collapsing stellar core with $S=1.4$ and $A_{0}=0.5$. In a time interval between the second and the fifth snapshot, an inhomogeneous structure in the inner region can be detected indicating the effects of self-gravity. Model shown is A05-S14-SG-R10, as listed in Tab. \ref{tab:modele}.  }
    \label{lspec_pro_s1.4}
\end{figure*}

\begin{figure*}[t]
    \centering
    \includegraphics[scale=0.37]{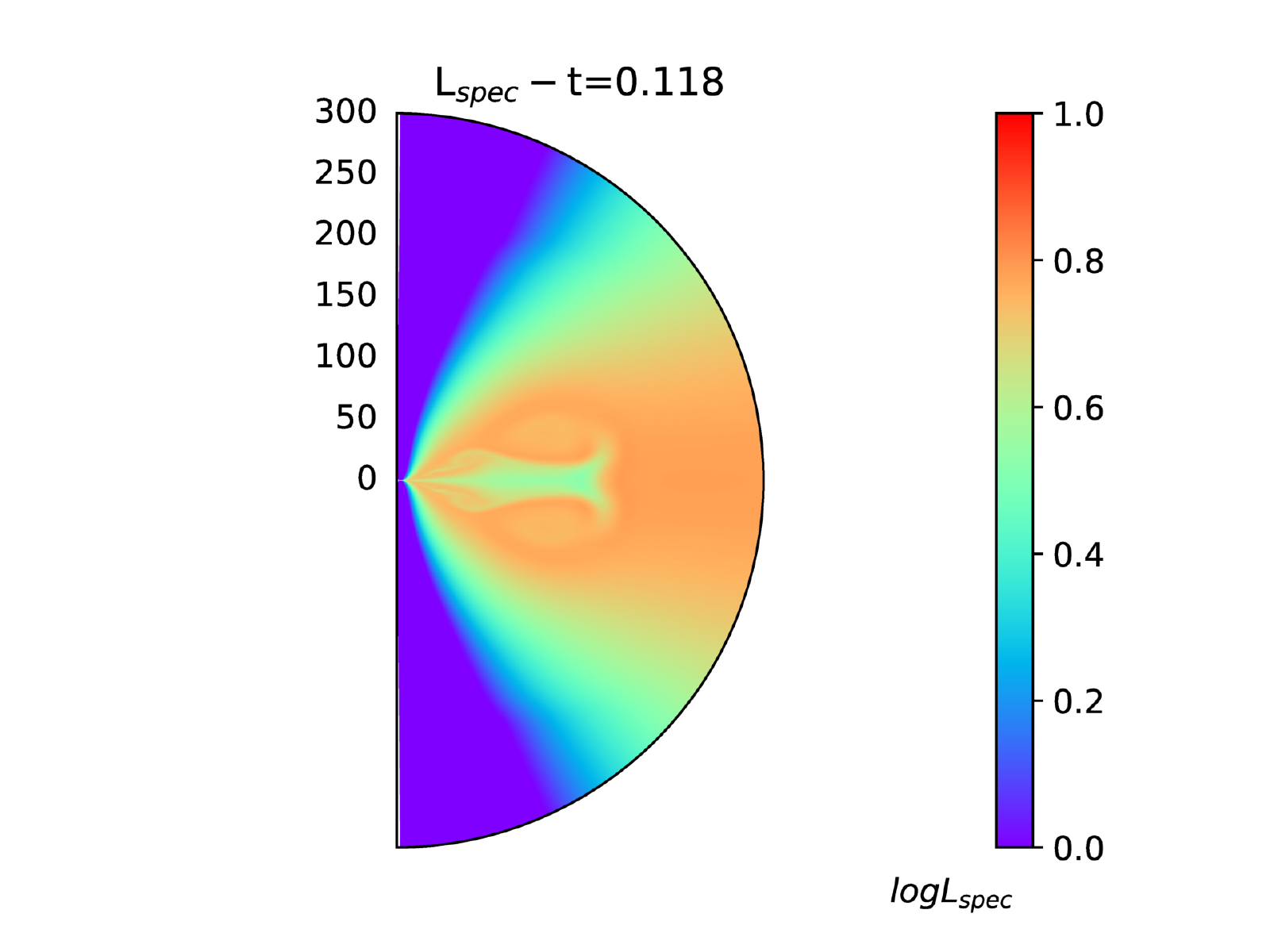}
    \includegraphics[scale=0.37]{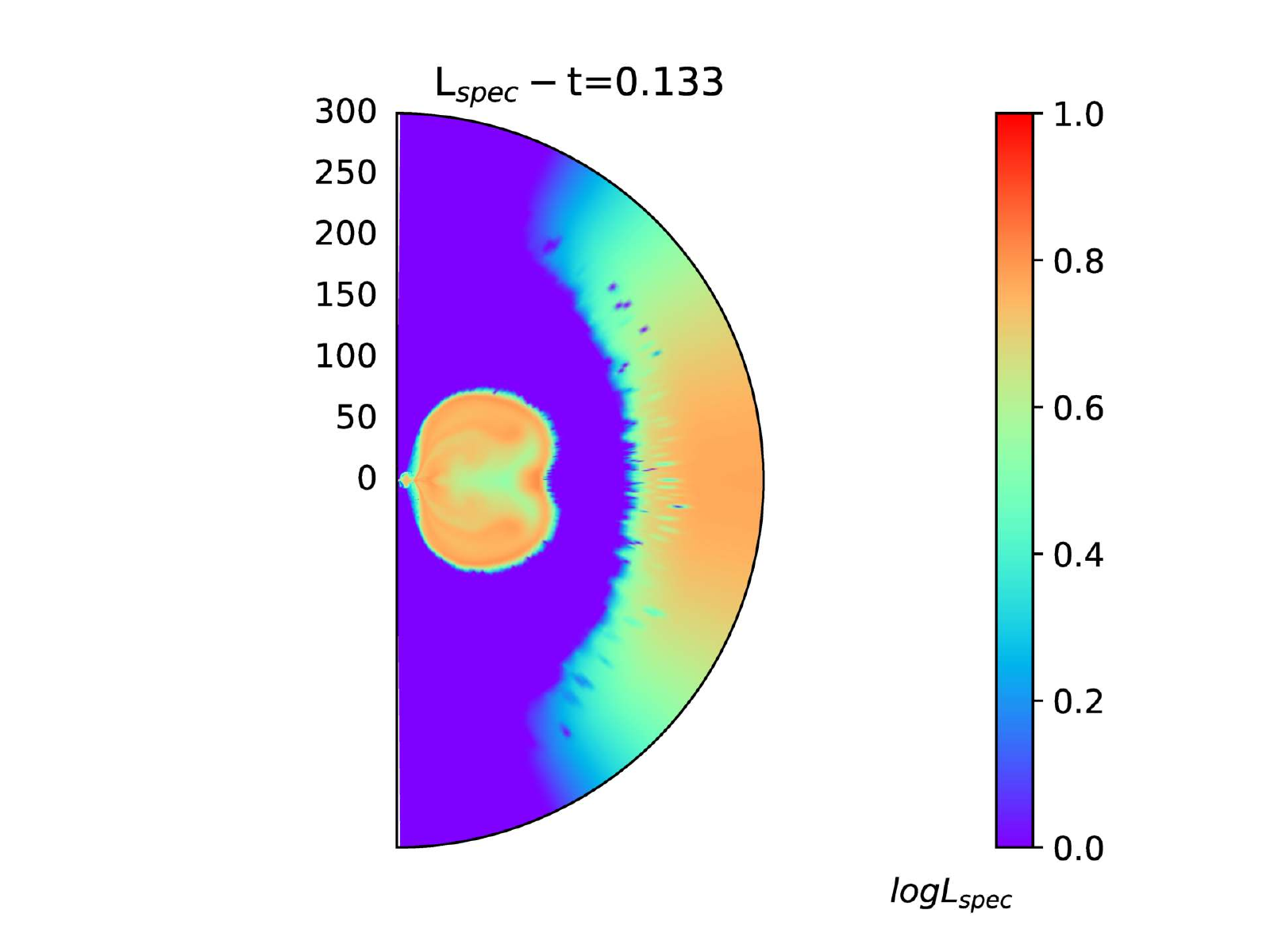}
    \includegraphics[scale=0.37]{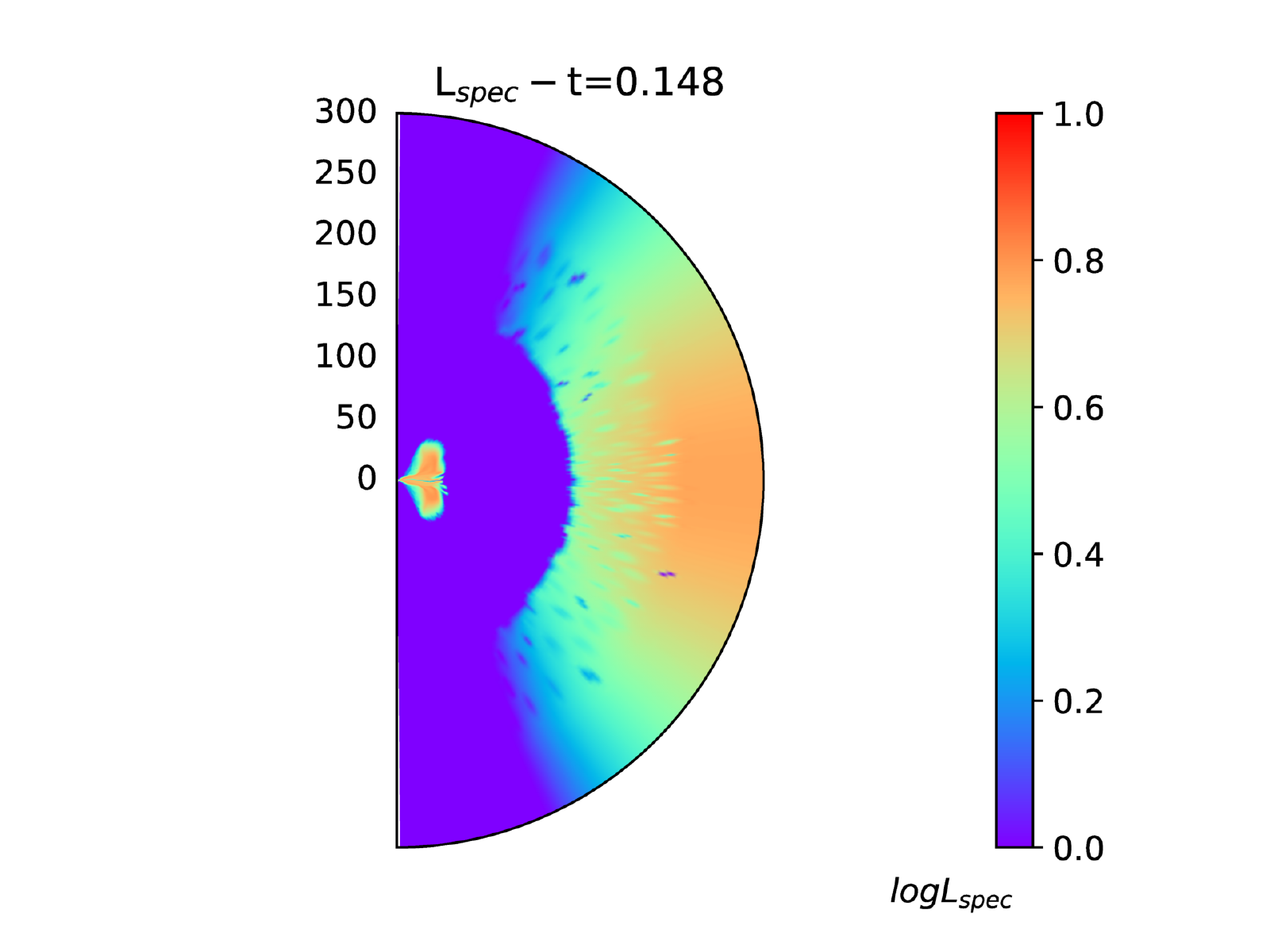}
    \includegraphics[scale=0.37]{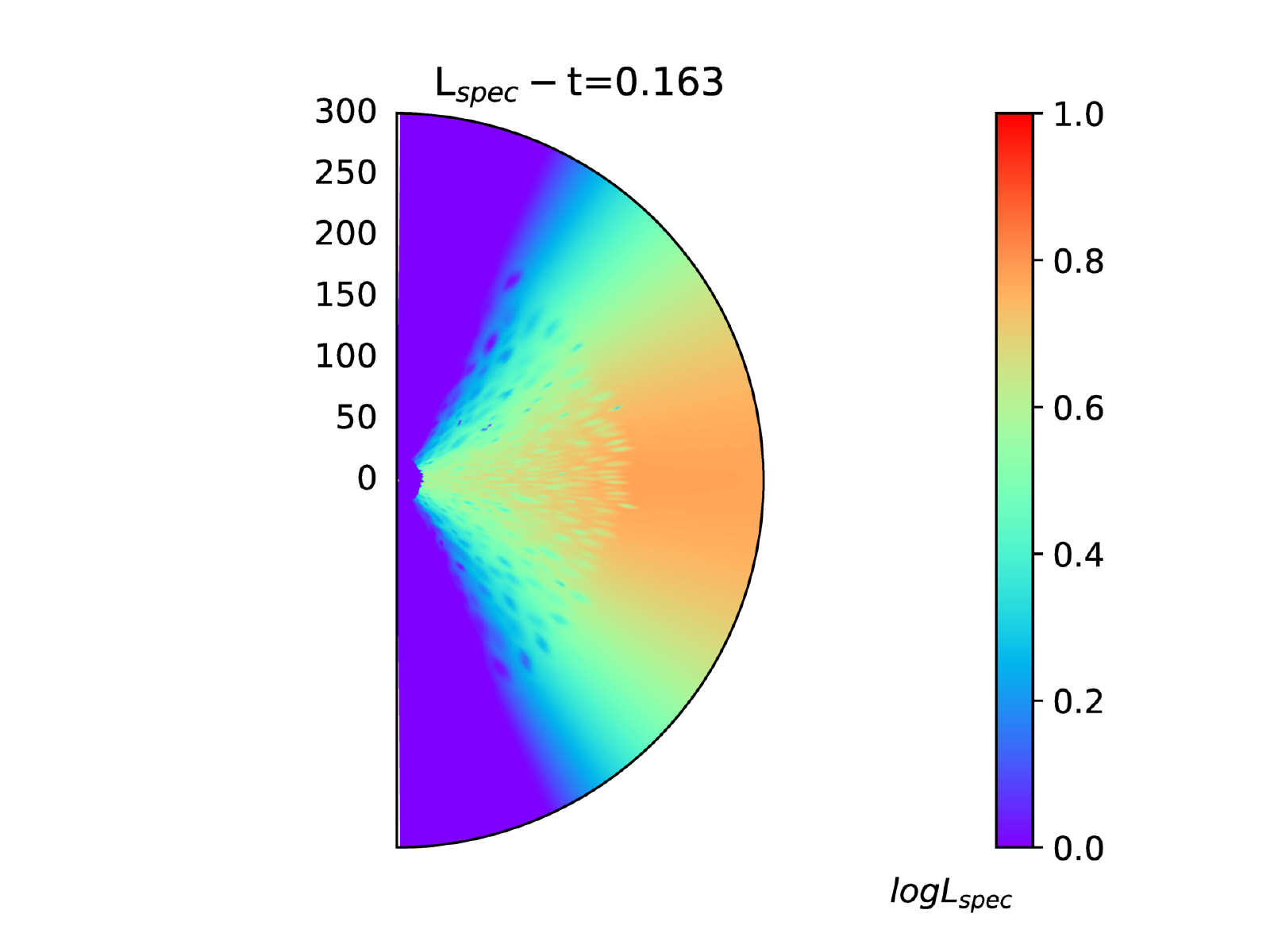}
    \includegraphics[scale=0.37]{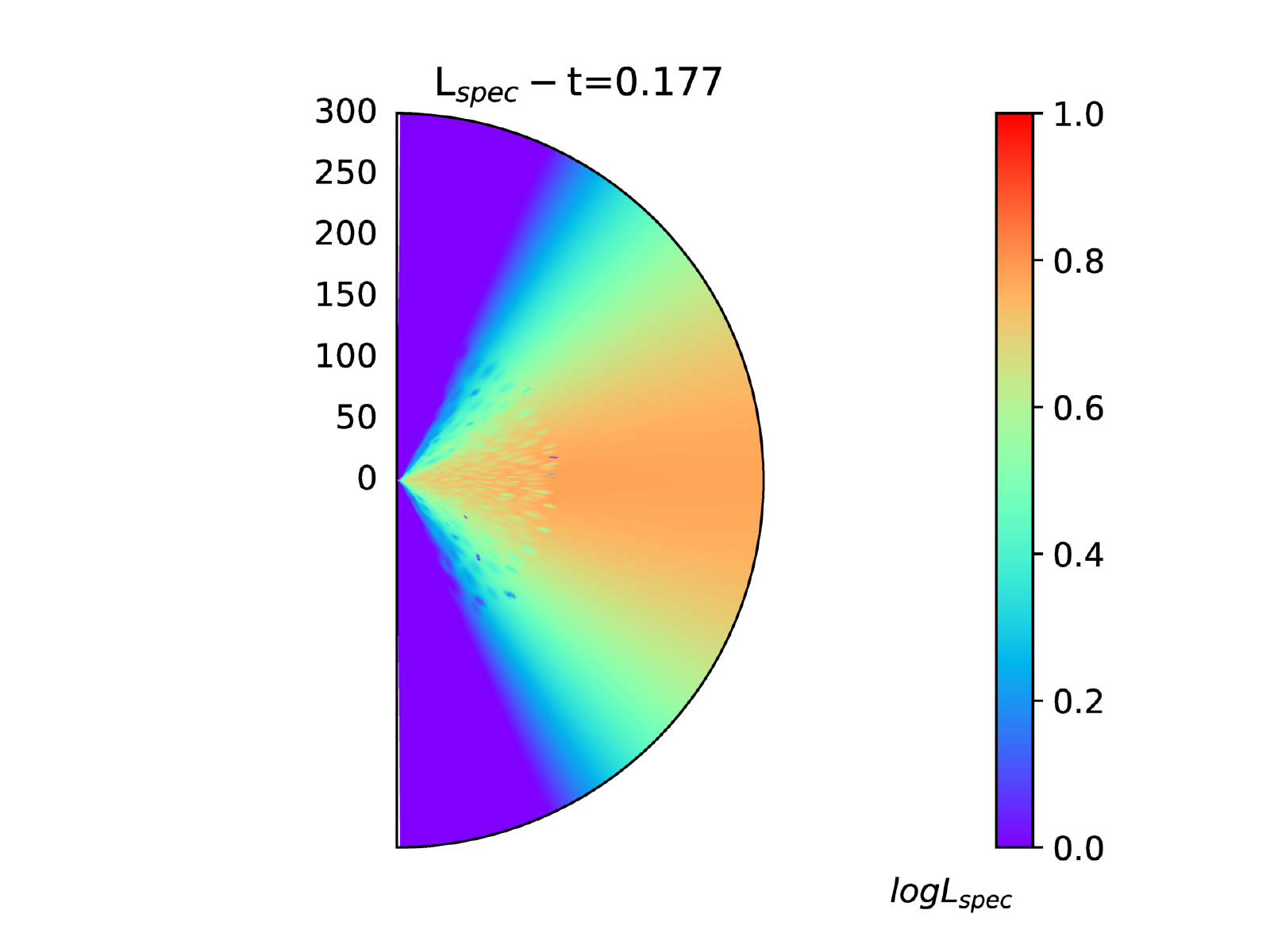}
    \includegraphics[scale=0.37]{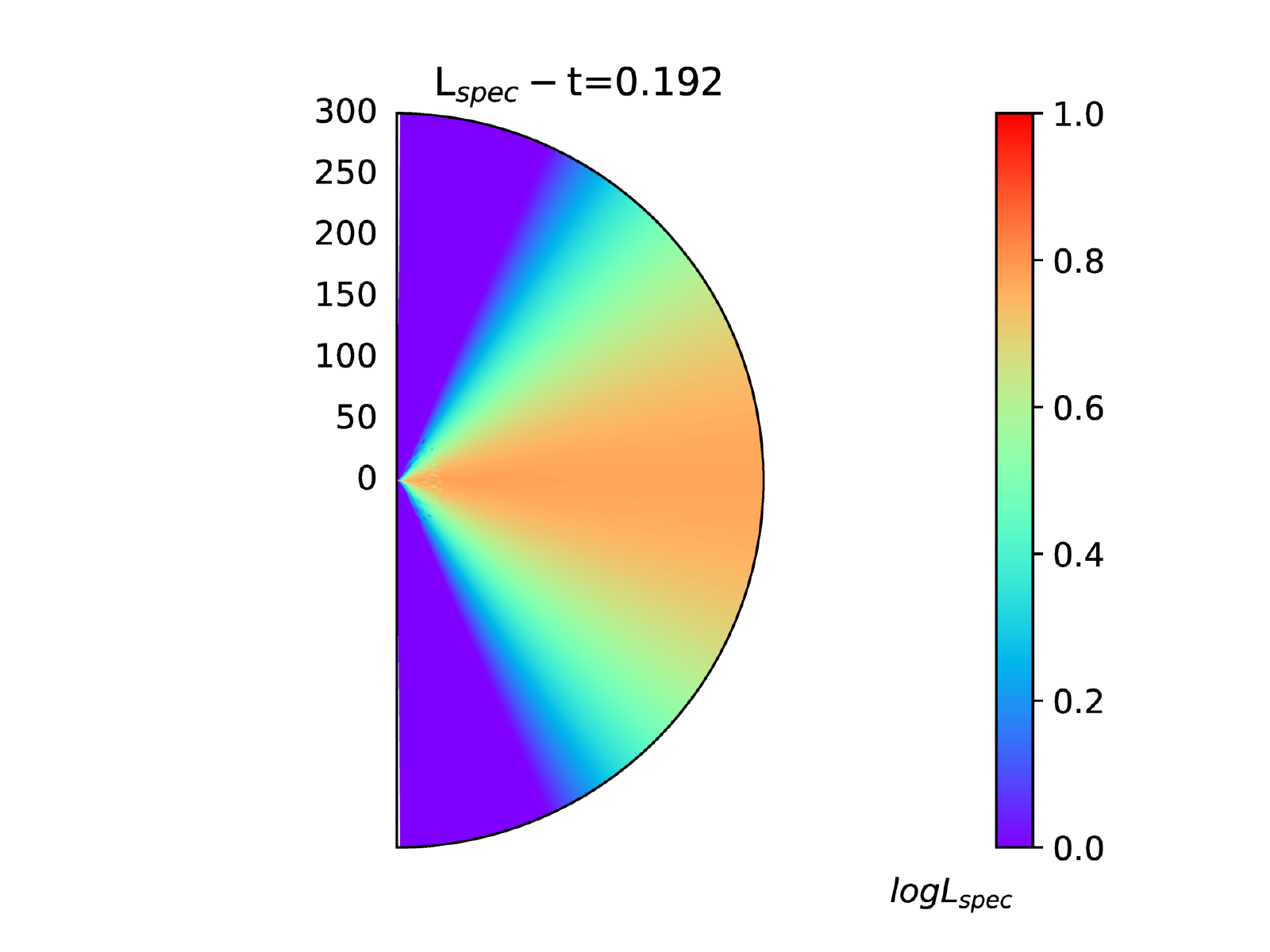}
    \caption{ Specific angular momentum snapshots, for the model with self-gravitating collapsing stellar core with
    $S=2$ and $A_{0}=0.5$. The inhomogeneities demonstrate self-gravity impacts, as seen in the second and third time snapshot. Model shown is A05-S20-SG-R12, as listed in Tab. \ref{tab:modele}.  }
    \label{lspec_pro_s2}
\end{figure*}

Second, the instabilities that occur due to SG effects, i.e. ring-like gravitational instability (which leads to a very small scale ring-like structures, see below),
followed by SGI instability (that causes an inhomogeneous structure of the inner envelope), can affect the inner region's angular momentum transport, and consequently, controls the early-time accretion rate onto the black hole (that can be traced in time domain $0.1-0.2~s$ in Figure \ref{M_S_A}). More precisely, the ring-like gravitational instability sounds to stop the accretion rate from being of a highly increasing manner, through a rise in the angular momentum of the inner regions, producing a very transient ring-like structure. Afterwards, as the inhomogenities grow in the inner region (from $\sim 0.118s$ to $\sim 0.148s$) the bubbles with lower angular momentum (this can be detected from images in Figures \ref{lspec_pro_s1.4} \& \ref{lspec_pro_s2}) increase in numbers at the equatorial plane as well as other zones. We argue that such a mixing of layers with different densities and angular momentum besides the tendency of having the lowest energy configuration cause the angular momentum to be transferred from the spikes (with larger amount of angular momentum) into the bubbles (with lower angular momentum), when they meet at the same radii. This yields an inward transport of angular momentum. Similar discussion about the inward transformation of angular momentum when the fluid elements are mixing, can be found in \cite{balbus2003enhanced} and references therein. In contrast, at time steps during which the inhomogeneous structure start disappearing (from $\sim 0.163s$ up to $\sim 0.192s$) one can find a rise in the specific angular momentum, getting back to a stabilized configuration.

In Figures \ref{lspec_pro_s1.4} and \ref{lspec_pro_s2} we show snapshots of the specific angular momentum distribution for these two models. 
We show here the self-gravitating models. It can be easily traced that the specific angular momentum starts decreasing through an inhomogeneous configuration from $t=0.118 ~s$ until $t=0.148 ~s$ and adopts an increasing trend afterwards, as pointed out earlier. 
In the model with higher rotation of the star's envelope, the inhomogeneous structures in angular momentum distribution seem to be smoothed out more quickly, and by $t=0.177 ~s$ they disappear.

\subsection{Analysis of gravitational stability}
\label{sec:toomre}

In self-gravitating collapsing stellar cores, gravitational instability can provide different structures like axisymetric ring formation as well as nonaxisymetric spiral arms, called I-modes, or fragmentation, that can be referred as J-modes and identified with the Jeans mechanism of instability \citep{Hachisu1987,christo1992}. 
The so-called Toomre parameter, $Q=\frac{\kappa c_{s}}{\pi G \Sigma}$ with $\kappa$ is the epicyclic frequency, $c_{s}$ the local sound speed, and $\Sigma$ the surface density, has been introduced for the axisymetric local instabilities in geometrically thin disks \citep{Toomre1964}.
Later on, \cite{Hachisu1987} proposed a universal criterion for gravitational instability which is valid in both thick and thin systems, as the following:
\begin{equation}
{\tilde{Q}=\frac{\kappa^{2}}{\pi G \rho}} <1, 
\label{eq10}
\end{equation}
where $\kappa^{2}=4\Omega^{2}+rd\Omega^{2}/dr$. They argued that non-axisymetric fragmentation in rapidly rotating systems is generally triggered by the onset of ring formation (as the axisymetric consequence of the gravitational instability).

Considering the importance of gravitational instability in self-gravitating accretion systems, we probe for the possibility of the axisymetric disk formation  and ring fragmentation in our 2D collapsar scenario. It can consequently provide us with an estimate for the non-axisymetric fragmentation. In spite of the fact that our stellar core is set up in 2D, the latter can be considered a possible outcome for a 3D set up (planned for our future work).

\begin{figure}
    \centering
    \includegraphics[scale=0.5]{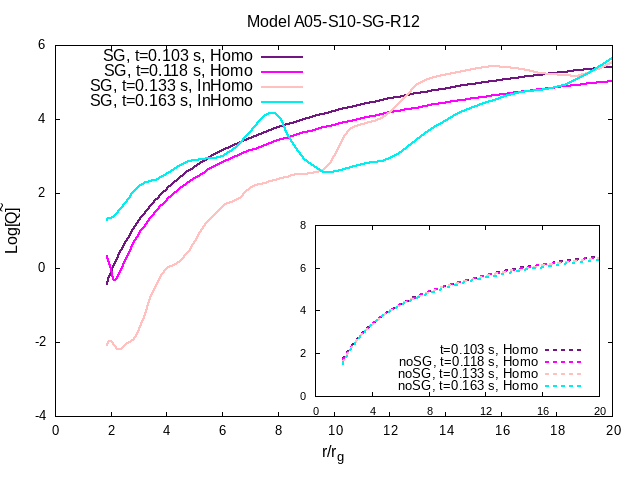}
    \includegraphics[scale=0.5]{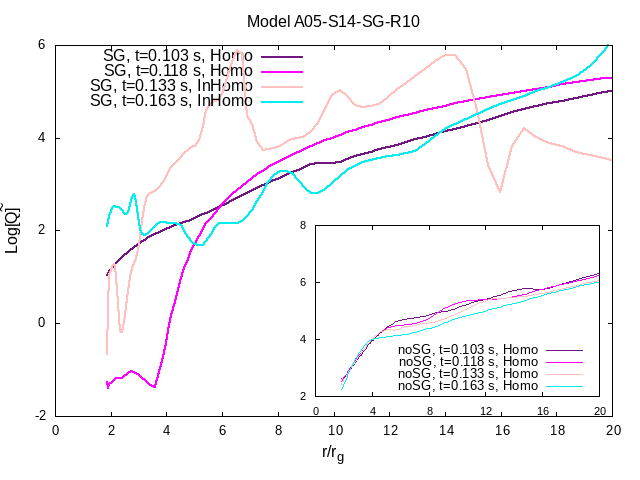}
    \includegraphics[scale=0.5]{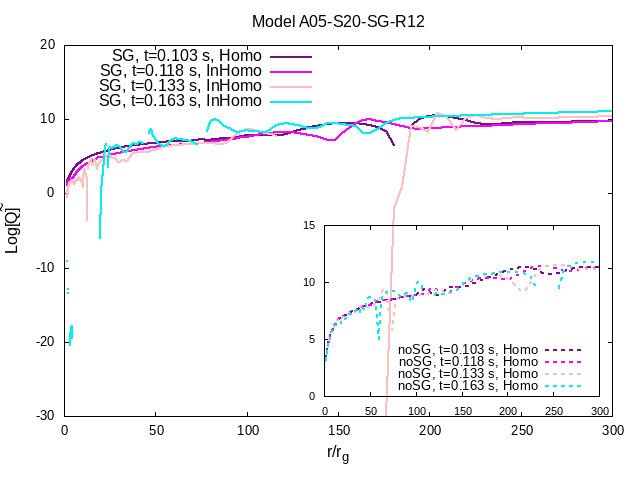}
    \caption{ A demonstration of how the condition for axisymetric modes ($\kappa^{2}/\pi G \rho <1$) is satisfied at the equator in some time steps, considering both cases with (solid lines) and without (dashed inset curves) self-gravity. "Homo" and "InHomo" refer to a homogeneous and inhomogeneous structures, respectively. Notice the logarithmic scale on the vertical axis.  Models are labeled in all panels with symbols referring to Tab \ref{tab:modele}.}
    \label{AxiMode}
\end{figure}

Figure \ref{AxiMode} is a demonstration of the axisymetric mode's condition, given by Eq. \eqref{eq10}, to observe its behavior on the equatorial plane for several time steps. We are comparing here both cases, with and without self-gravity. To consider any possible connection with the emergence of density and angular momentum inhomogeneities, we also mark the profiles with "Homo" and "InHomo" labels, which stand for the homogeneous and inhomogeneous structures, respectively.

Additionally, in cases without self-gravity, we did not find any possibility for axisymmetic modes to form, as shown by the smaller inset plots in Fig. \ref{AxiMode}.

\begin{figure}
    \centering
    \includegraphics[scale=0.5]{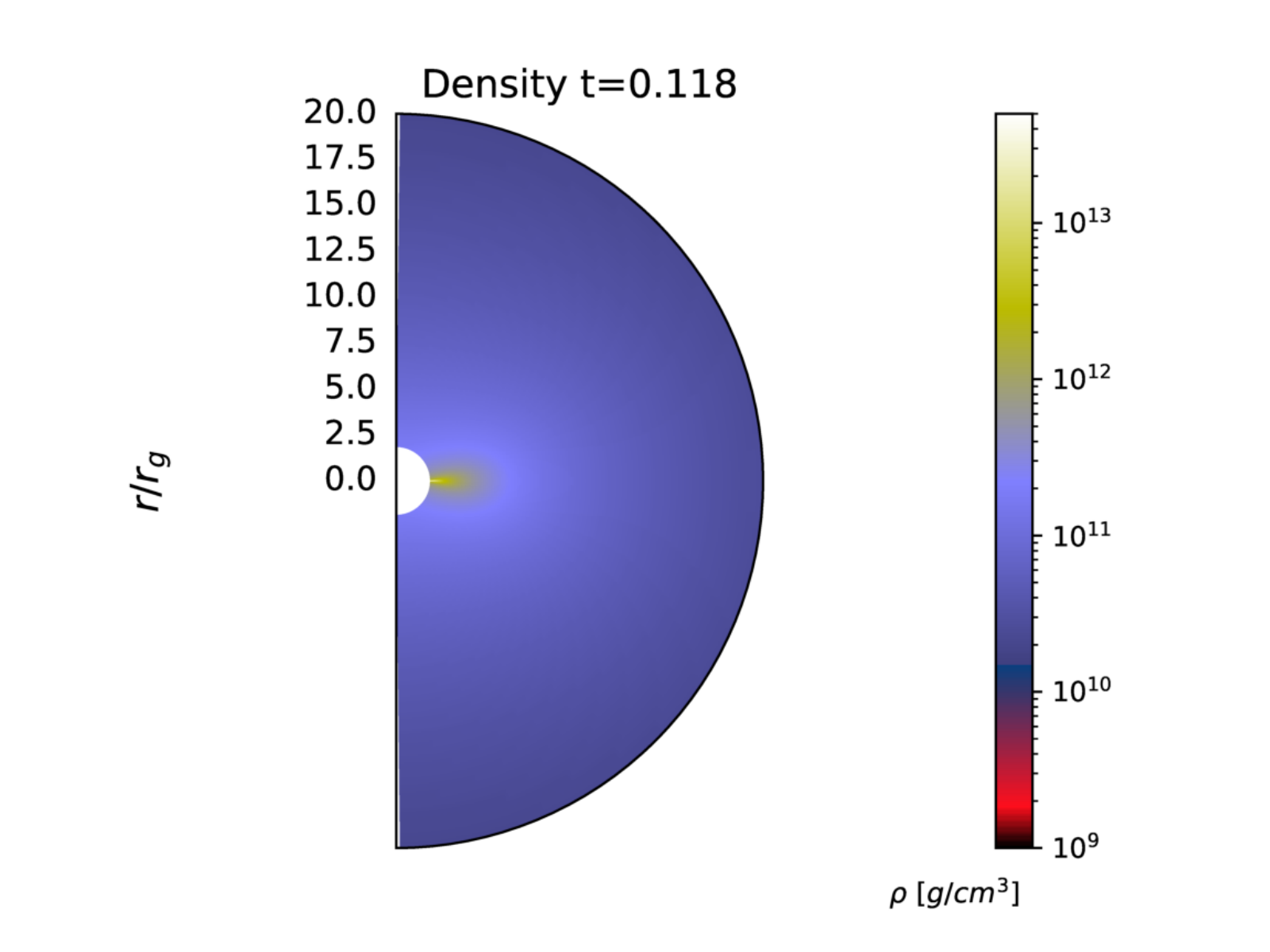}
    \includegraphics[scale=0.5]{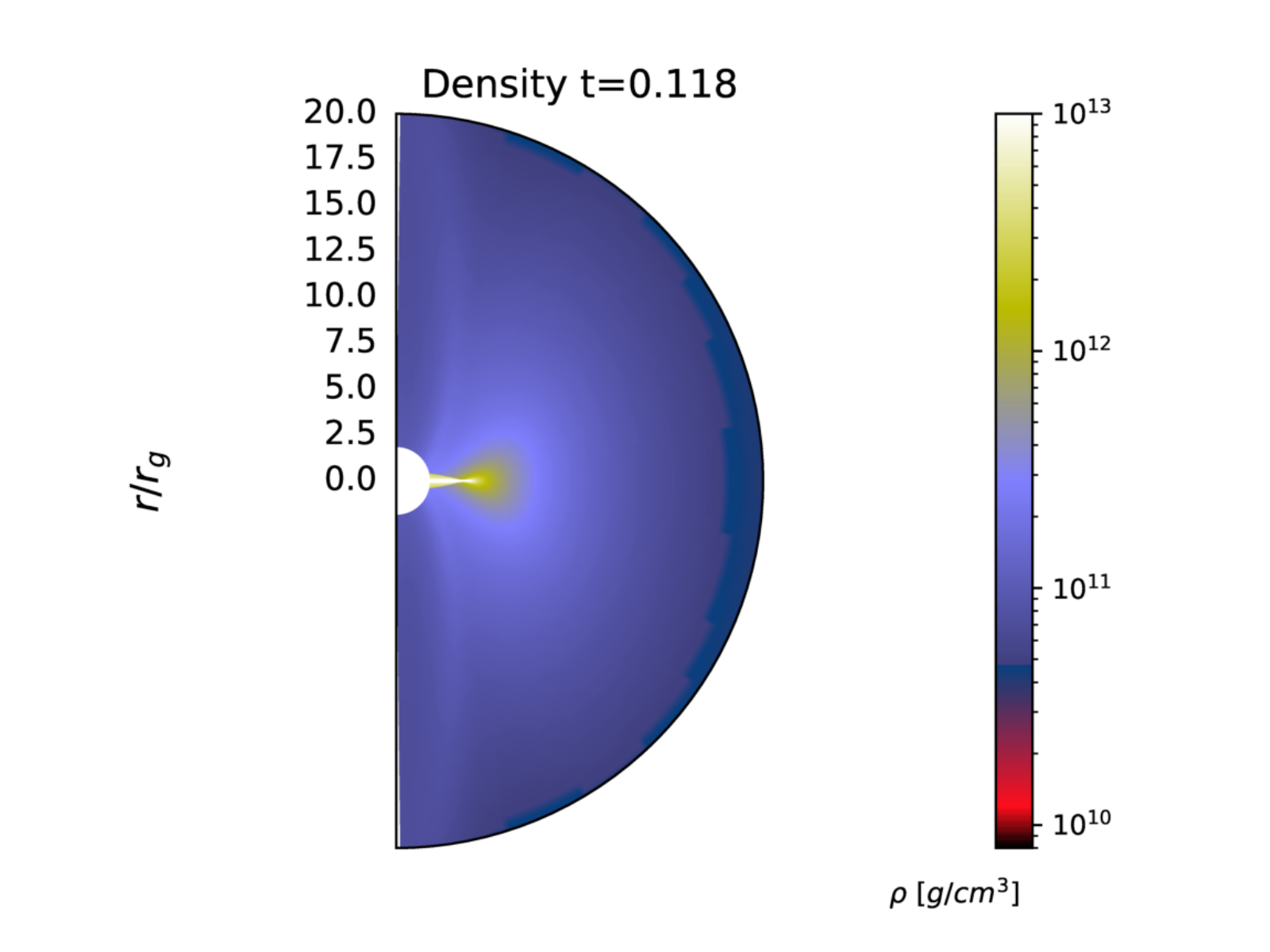}
    \includegraphics[scale=0.5]{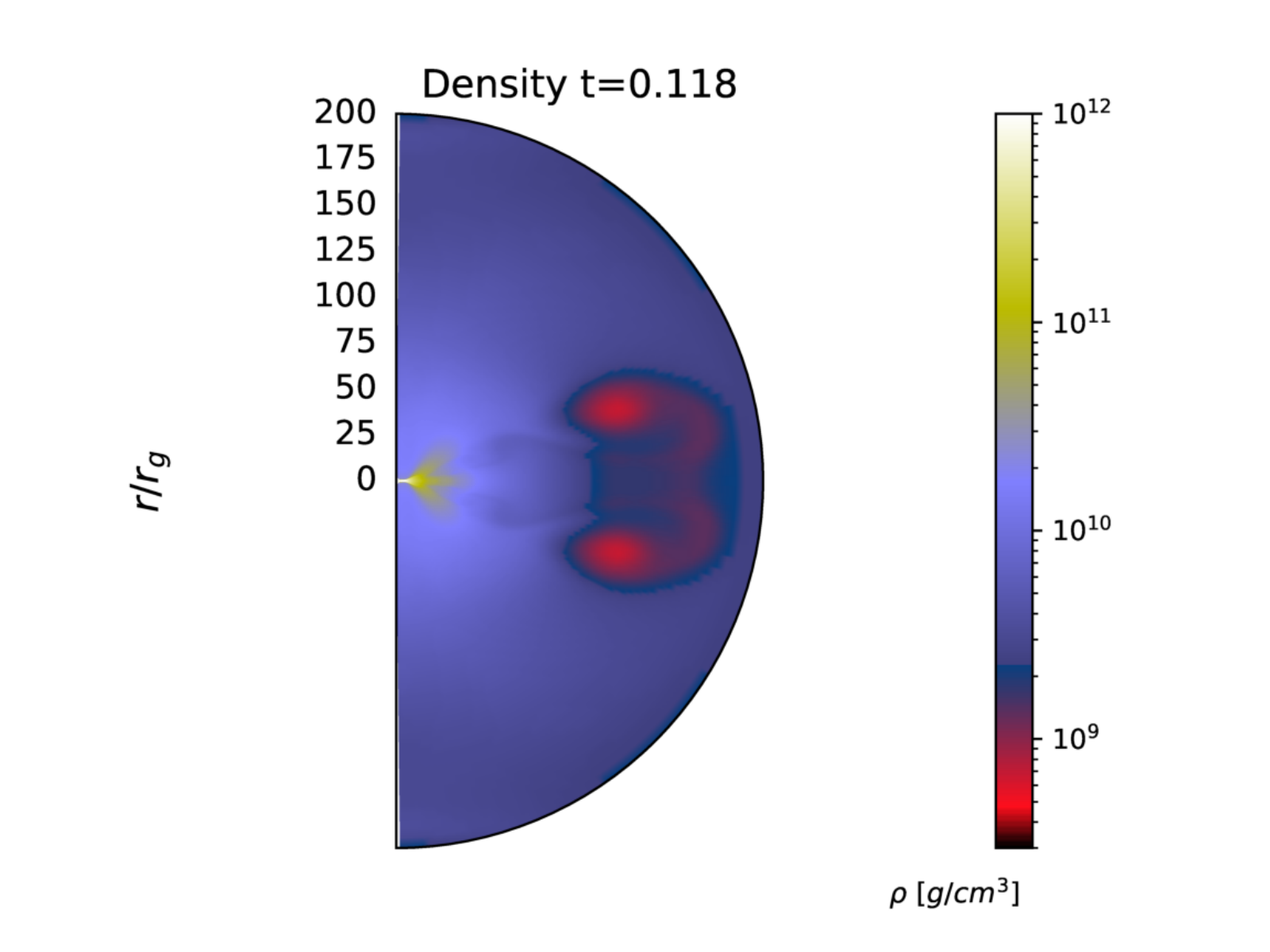}
    \caption{ Density profiles at the time $t=0.118 ~s$ for three cases of $S=1,~1.4$ and $2$, related to the images at the top, middle and bottom, respectively. These snapshots illustrate density structure of the envelope when the ring-like gravitational instability gets possible. The case with $S=2$, last profile, shows no axisymmetric growing mode, however. Models shown from top to bottom, are A05-S10-SG-R12, A05-S14-SG-R10 and A05-S20-SG-R12, as listed in Tab. \ref{tab:modele.} 
    }
    \label{AxiMode_prof}
\end{figure}

One can find that the inner regions are unstable towards ring formation, just before the inhomogeneities start arising. For higher rotation parameter $S$, the unstable region moves outward and the collapsing star is prone to becoming unstable at earlier times.  However, we did not detect any unstable region for the case of $S=2$ as can be found from the plot at the bottom in Figure \ref{AxiMode}. The gaps in this case appear to be related to the emergence of inhomogeneties at the equator in the corresponding time steps. Considering the increase of the spin parameter $A_{0}$, for a given $S$ parameter, a smaller axisymetric instability region would appear. Based on \cite{Hachisu1987}, we argue that as soon as the condition for axisymetric ring formation is met, the non-axisymetric fragmentation can be possible. To investigate it, we would need however a 3D set up of the collapsing stellar core, which is beyond the scope of the present work. To provide an intuition of what happens through the unstable regions, while we consider the axisymetric modes, we present density profiles at the time $t=0.118 ~s$ in Figure \ref{AxiMode_prof}. It seems to be an unstable time snapshot for these three cases of rotation parameter, $S=1$, 1.4, and 2, from top to bottom, respectively.

\subsection{Simulations of magnetized collapsing stars}
\label{sect:magnetized}

\begin{figure*}[ht]
      \centering
   \includegraphics[scale=0.28]{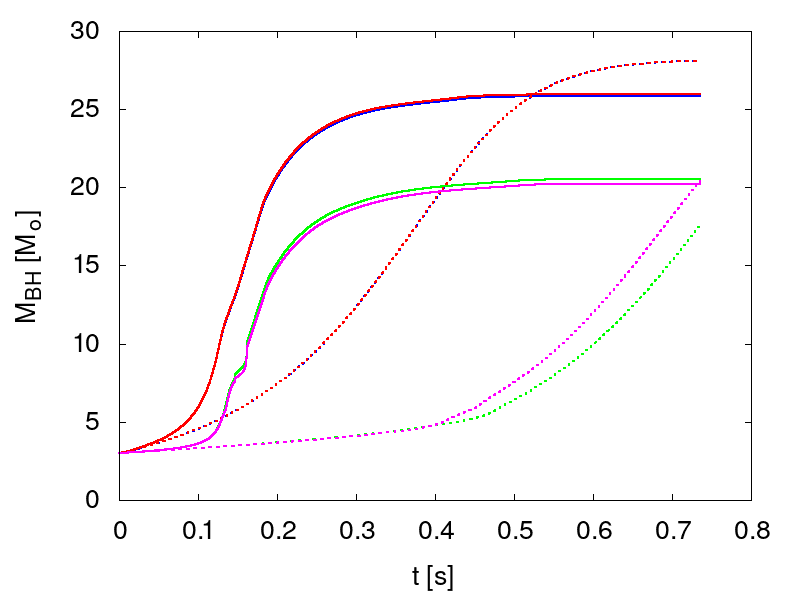}
   \includegraphics[scale=0.28]{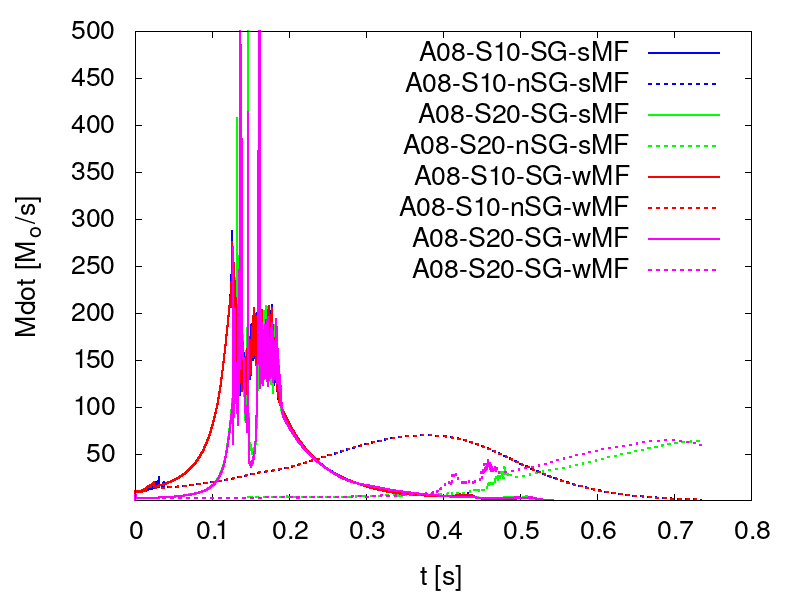}
   \includegraphics[scale=0.28]{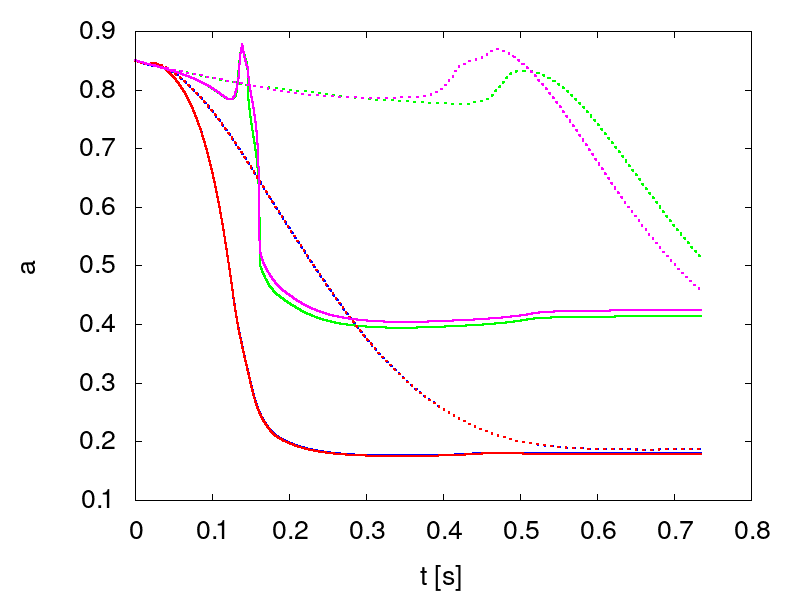}
     \caption{Time evolution of the black hole mass (left), accretion rate on the black hole horizon (middle) and black hole spin (right) for the rotation normalized with specific angular momentum at ISCO $S=1.0$ and $S=2.0$. ( Models are labeled in the middle panel with symbols referring to Tab. \ref{tab:modele}). Initial spin of the black hole was $A_{0}=0.85$, and the initially vertical magnetic field given by Eq. \ref{eq:vertical} was normalized with $\beta=10$ or $\beta=100$ at the ISCO radius.
             }
             
    \label{fig:models_mag_a085}
\end{figure*}

We  now investigate the role of the magnetic field and its importance from the point of view of gravitational stability of the collapsing core. 
Here we present the general trends of the system evolution, starting from the simplest case of a weak uniform magnetic field, given by Eq. \ref{eq:vertical}.
In Figure \ref{fig:models_mag_a085} we plot the time evolution 
 of the black hole mass, spin, and mass accretion rate onto the black hole, for the two values of specific angular momentum in the collapsing star, $S=1.0$ and $S=2.0$, and we compare the system evolution with and without self-gravity term. The initial gas-to-magnetic pressure ratio id $\beta=10$, and the initial black hole spin is $a=0.85$ in all models.
 In general, the larger specific angular momentum results in smaller final black hole mass and larger final black hole spin
(although net spin-down is found in all cases). The self-gravitating  stellar core evolves much faster, and the final value of black hole spin is reached already after the initial $\sim 10000$ $t_{g}$. Also, in case of self-gravitatingcollapsing cores, the instantaneous accretion rate presents much larger amplitudes of oscillations, during the period of steep black hole mass increase.

We have also examined the influence of the self-gravity on the global density profile and the shape of the magnetic lines. Self gravity modifies 2D profiles of density or other quantities  noticeably on two stages of the evolution. For all of the combinations of $A_{0}$, $S$ and gas-to-magnetic pressure values the self-gravity shows its effect for the first time  around $t=0.133 ~s$ by creating inhomogenities.  On this stage the presence of the magnetic field does not change significantly the self-gravity influence. Their structure and lasting times are similar in the simulations with and without magnetic field. They are  visible only in the small scale around $\sim100 ~r_{g}$. Around $t\sim0.148 ~s$ we see drop of the angular momentum which is not reflected in the density profile, and around $t\sim0.163\textrm{s}$ higher values appear again. Inhomogenity  disappear around $t\sim0.192 \textrm{s.}$. Magnetic field does not have an significant influence on the evolution at that time. Inhomogenities have the same structures and morphology as seen in Fig. \ref{lspec_pro_s1.4} for a non-magnetised case (model A05-S14-SG-R10).

\begin{figure*}[ht]
      \centering
     \includegraphics[width=\textwidth]{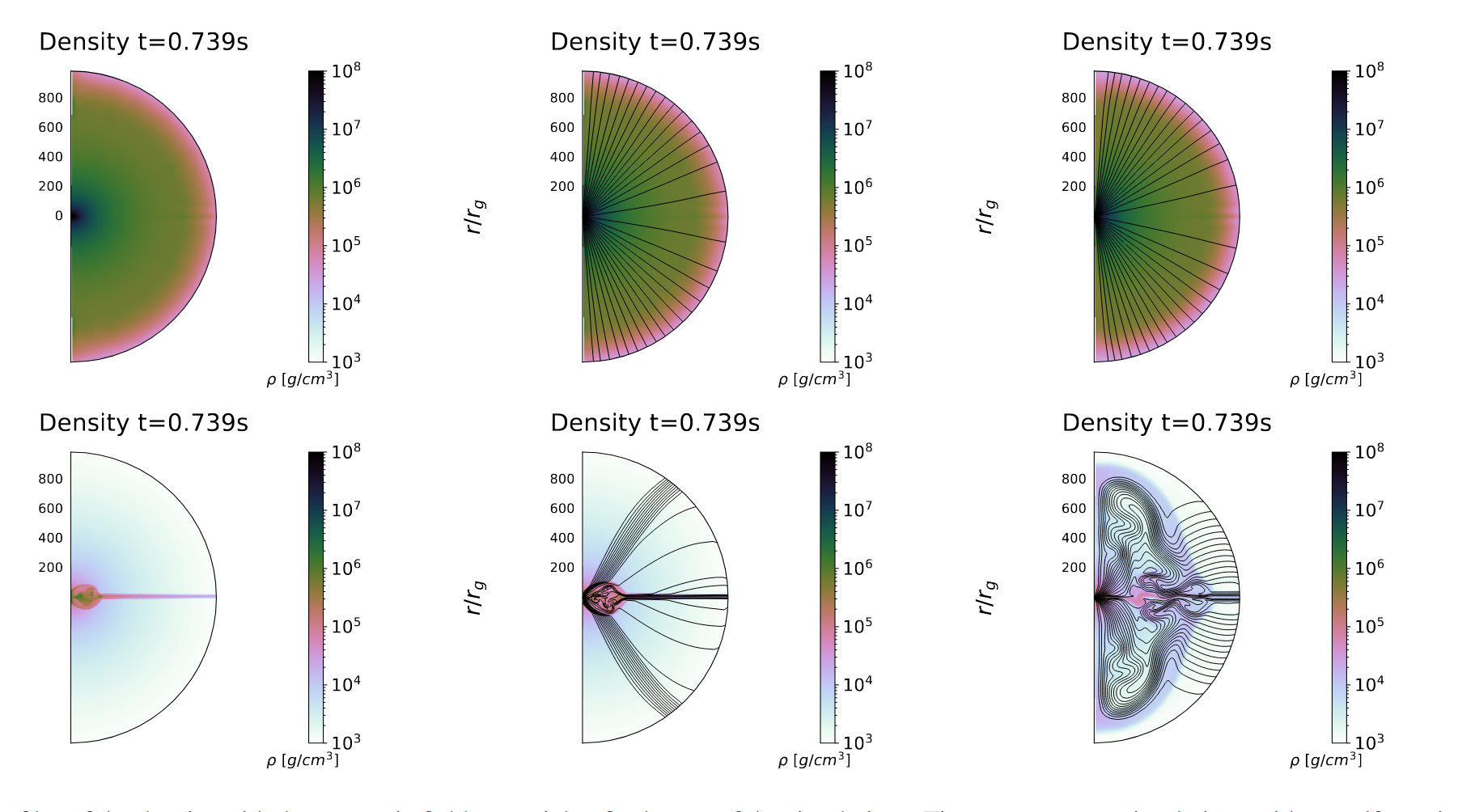}
\caption{Profiles of the density with the contours of the magnetic field vector potential at final stage of the simulations. First row presents simulations with no self-gravity and: no magnetic field (left panel), $\beta=100$ (middle panel) and $\beta=10$ (right panel). Second row presents simulations with self-gravity and: no magnetic field (left panel), $\beta=100$ (middle panel) and $\beta=10$ (right panel).
Simulation assumed vertical magnetic field configuration. Parameters: $A_{0}=0.3, S= 1.0$. Models shown are A03-S10-nSG-R12, A03-S10-nSG-wMF, and A03-S10-nSG-sMF (top row), and A03-S10-SG-R12, A03-S10-SG-wMF, and A03-S10-SG-sMF (bottom row)} as listed in Table \ref{tab:modele}.
    \label{fig:prof}
\end{figure*}

 Self-gravity makes its presence known for the second time at the end of the simulations. This time scale of the effect depends on the magnetic filed presence and strength. For the simulations without self-gravity in the final stage of the simulation more or less spherically symmetrical structure of the density is preserved, and magnetic potential lines are radial. Situation change for the self gravity profiles. With the absence of the magnetic field we observe thin disk-like structure which forms at the end of the simulations. For the magnetic field characterized by $\beta=100$ similar structure is formed and its shape is followed by the magnetic potential lines. It is however slightly bigger then in non-magnetized case. More magnetized envelope with $\beta=10$ results in structure which is visible at scales of up to $r\sim 800 ~r_{g}$. Moreover, simulations with self-gravity leave the evolved stellar core much less dense. We present exemplary profiles illustrating those stages of the simulations for $A_{0}=0.3$ and $S=1.0$ in the Fig. \ref{fig:prof}.  
We show here models A03-S10-nSG-R12, A03-S10-nSG-wMF and A03-S10-nSG-sMF in top ow, while bottom row shows models 
A03-S10-SG-R12, A03-S10-SG-wMF and A03-S10-SG-sMF.

\begin{figure*}[ht]
      \centering
   \includegraphics[scale=0.42]{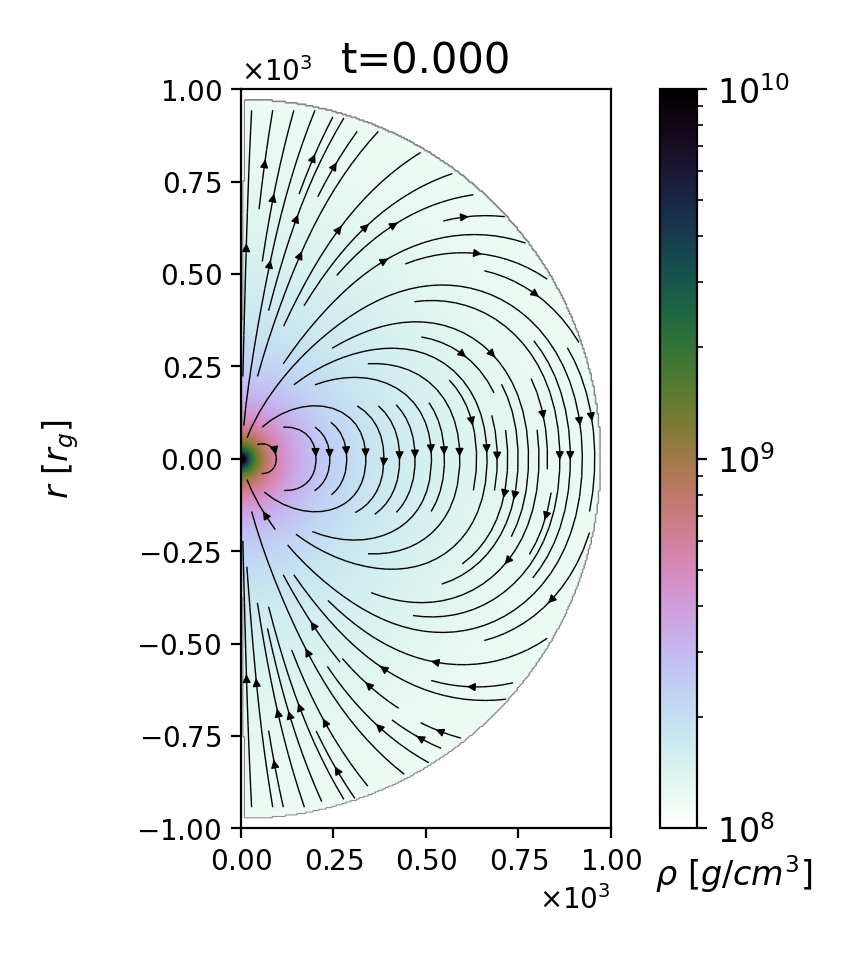}
   \includegraphics[scale=0.42]{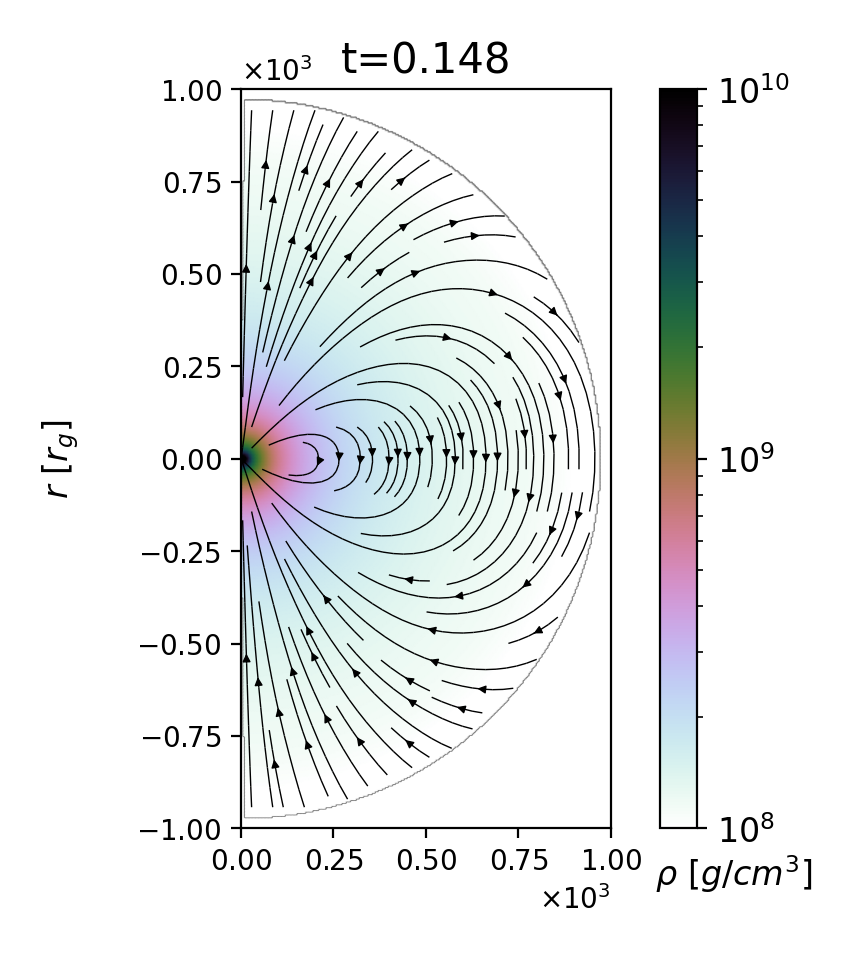}
   \includegraphics[scale=0.42]{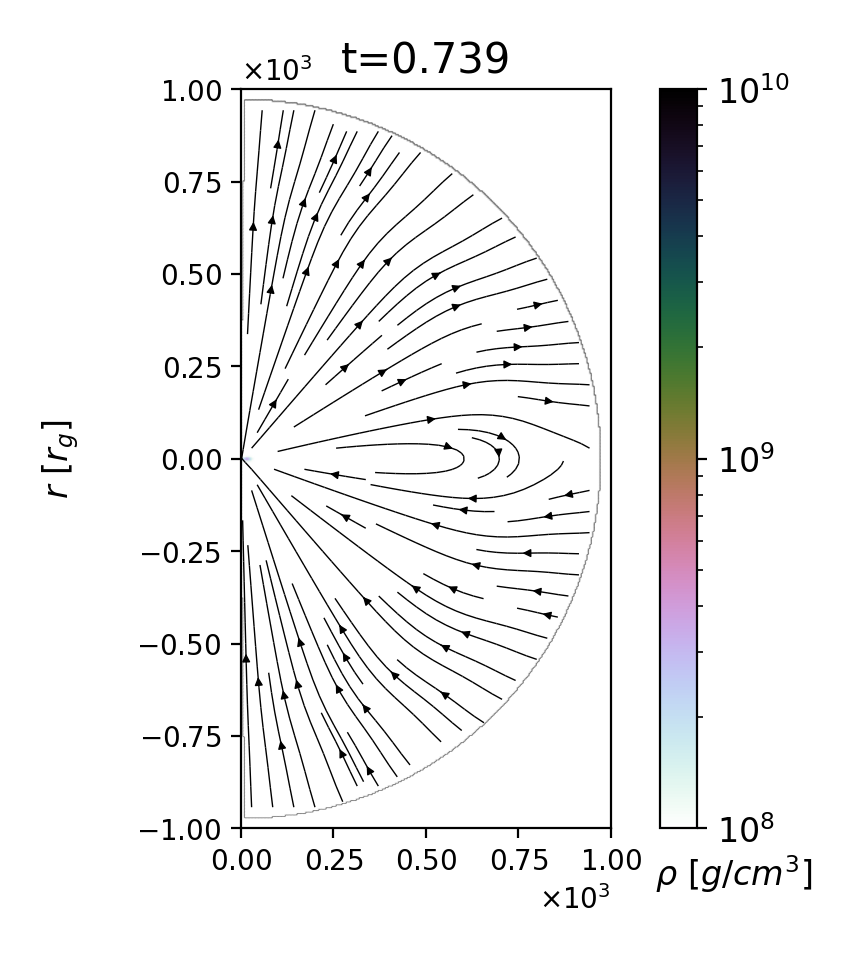}
   \includegraphics[scale=0.42]{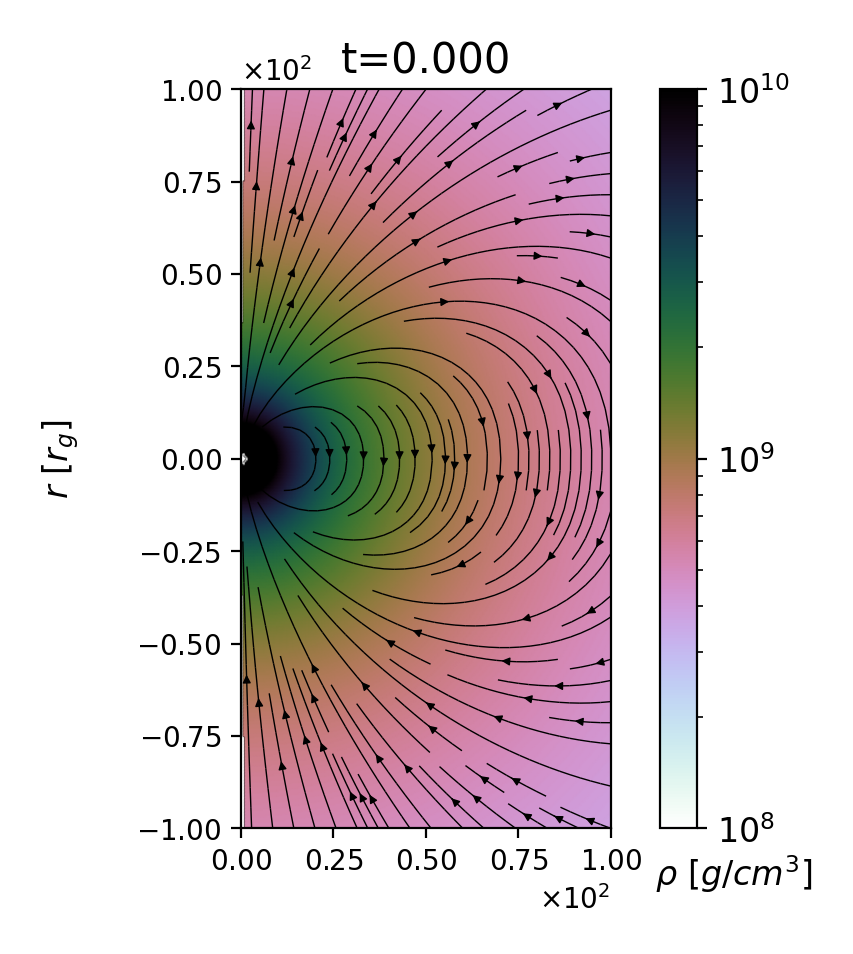}
   \includegraphics[scale=0.42]{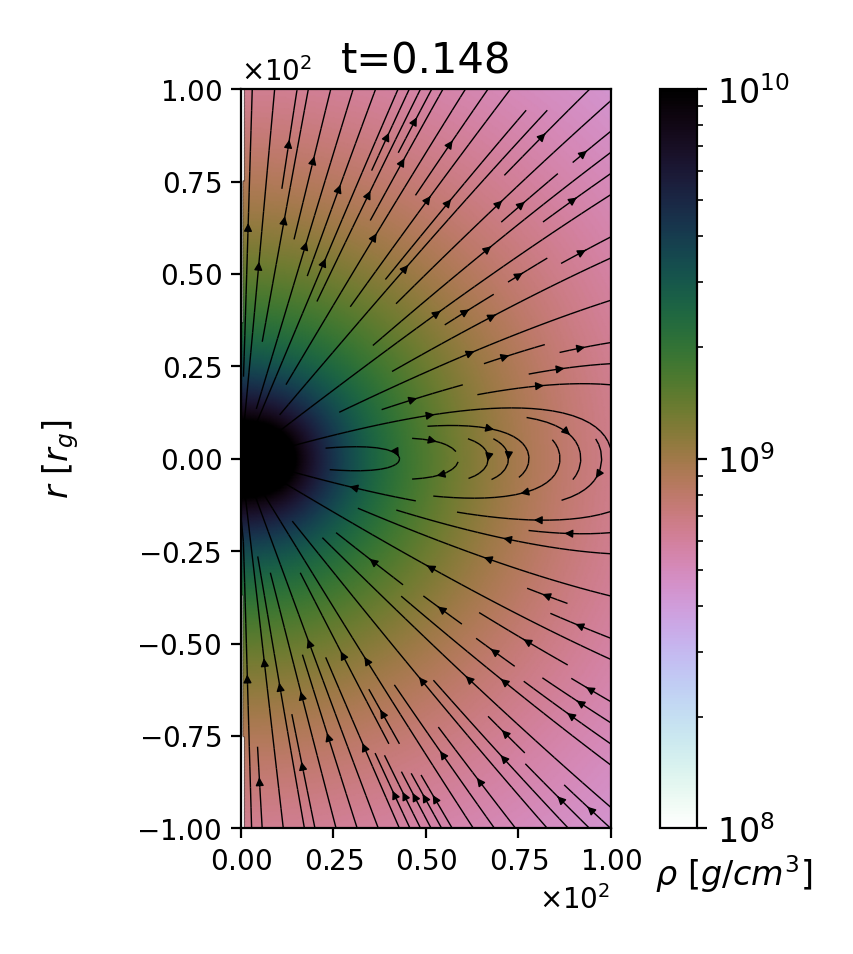}
   \includegraphics[scale=0.42]{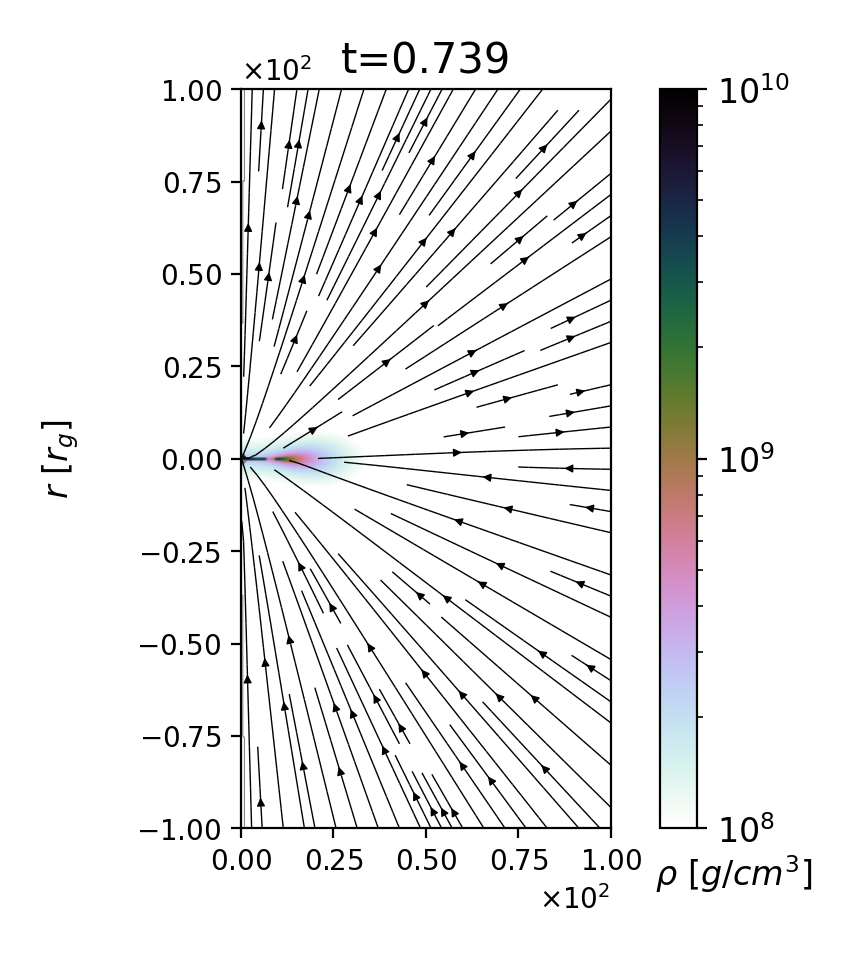}
             \caption{Distribution of density and magnetic field  vector potential contours at t=0 (left), and the short-time evolved snapshots taken at time $t=0.148 ~s$ (middle), and at the end of simulation time $t=0.739 ~s$ (right).
             The magnetic field of dipole configuration is adopted and normalized with $\beta=50$. Top row is scaled to the outer radius of the domain, at 1000 $r_{g}$, and bottom row is zoomed in to 100 $r_{g}$. Simulation was done in an evolving Kerr metric, but without self-gravity effect. Parameters: $A_{0}=0.85$, $S=1.0$. Model shown is D08-S10-nSG-b50, as listed in Tab. \ref{tab:modele}.
             }
    \label{fig:profiles_dipole}
\end{figure*}

\begin{figure*}[ht]
      \centering
   \includegraphics[scale=0.42]{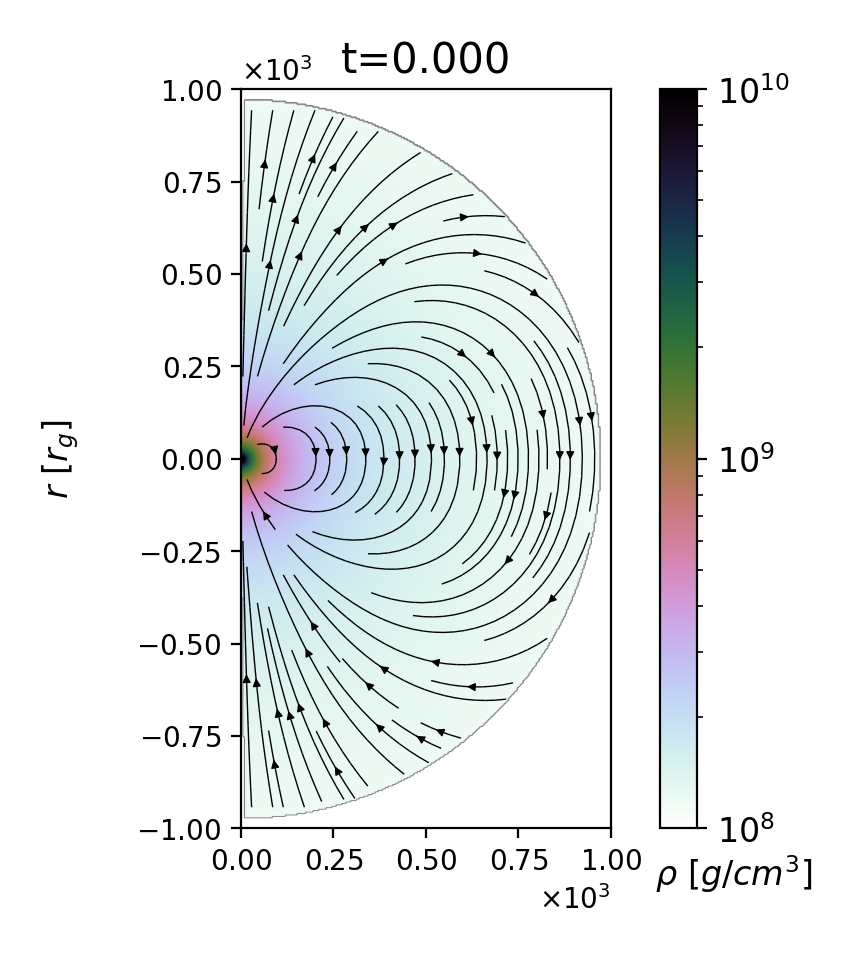}
   \includegraphics[scale=0.42]{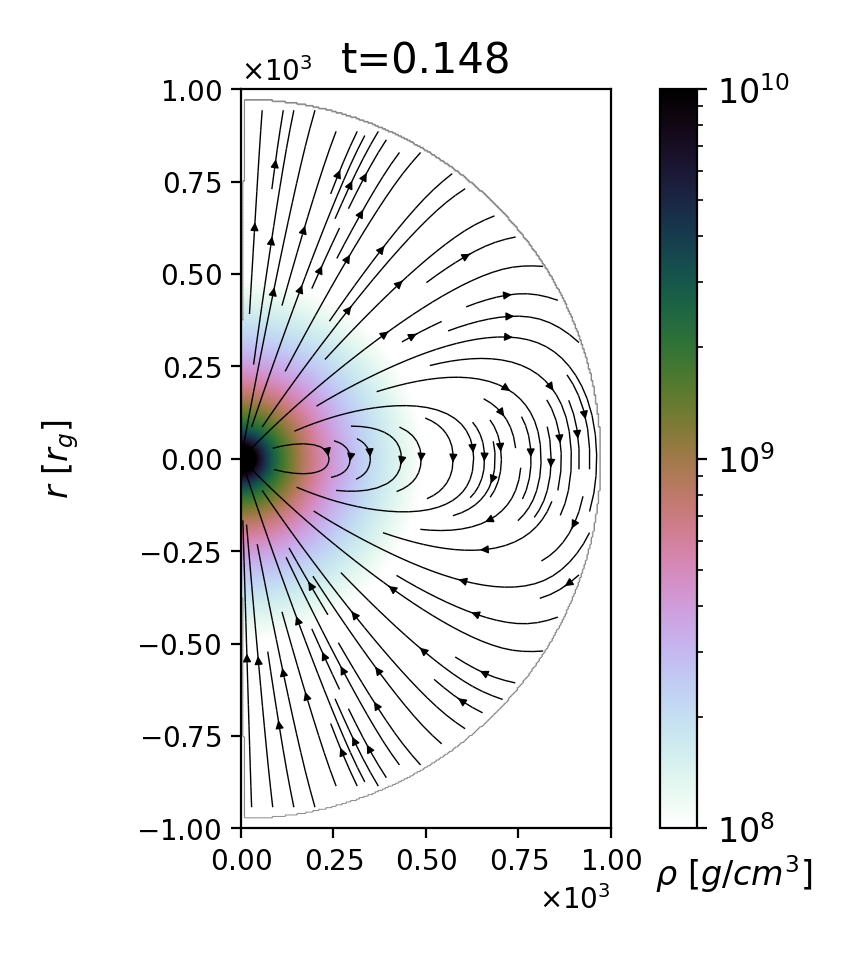}
   \includegraphics[scale=0.42]{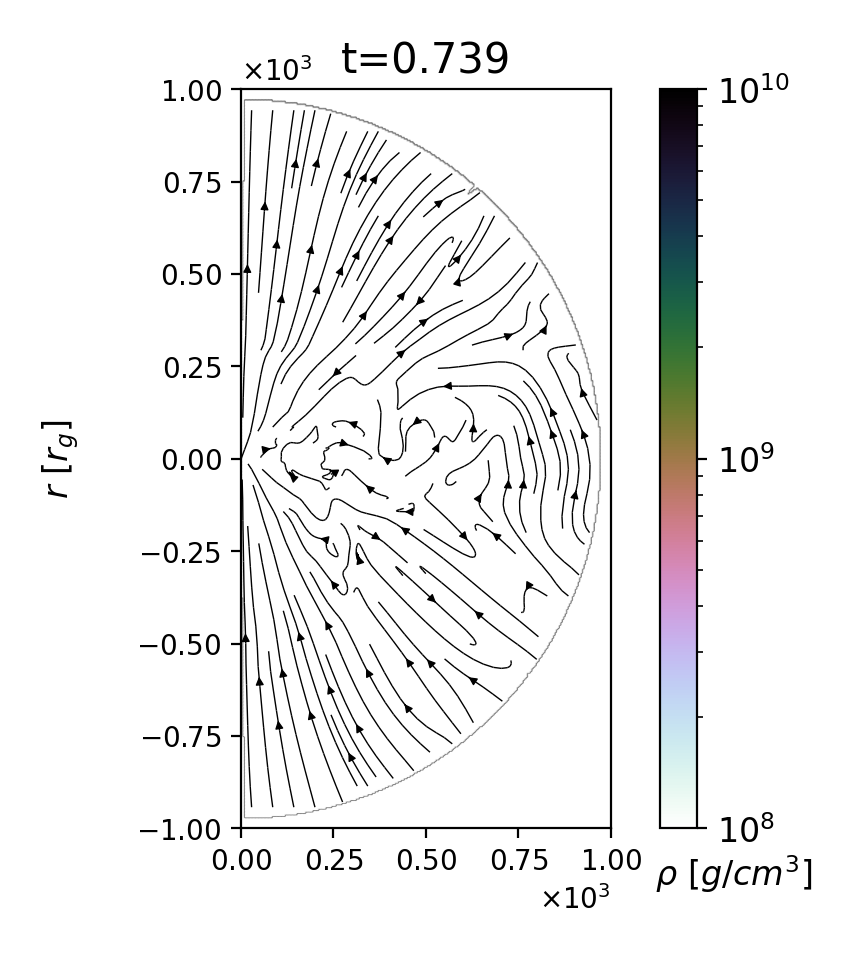}
   \includegraphics[scale=0.42]{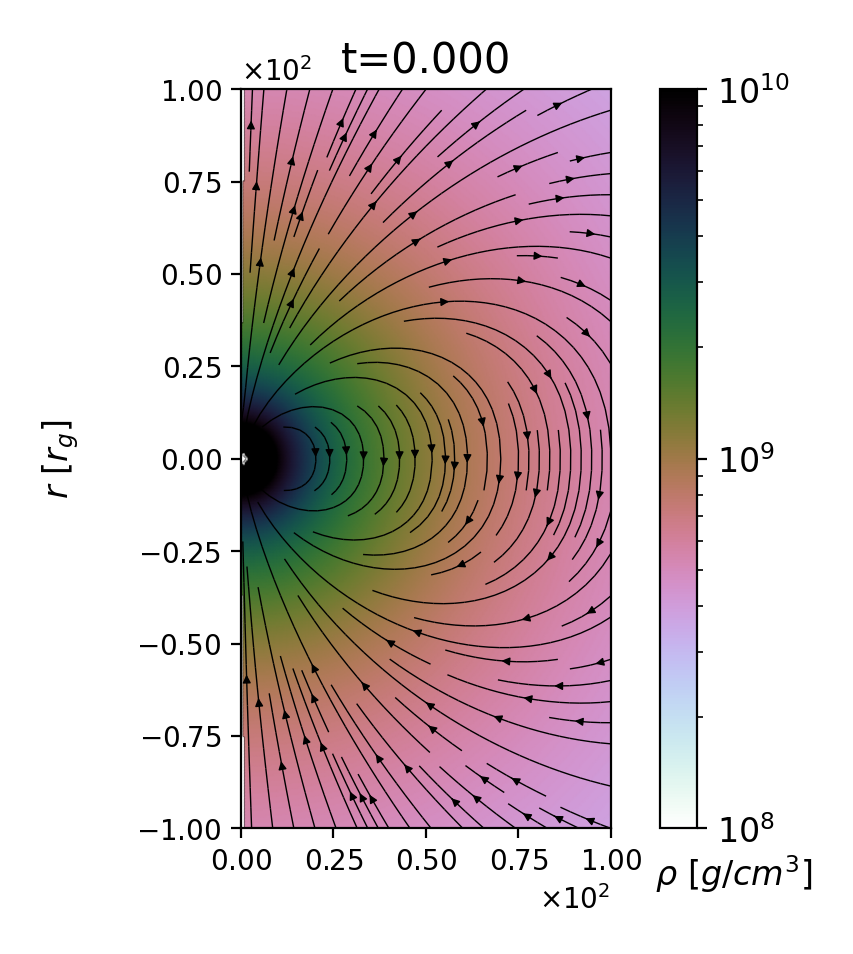}
   \includegraphics[scale=0.42]{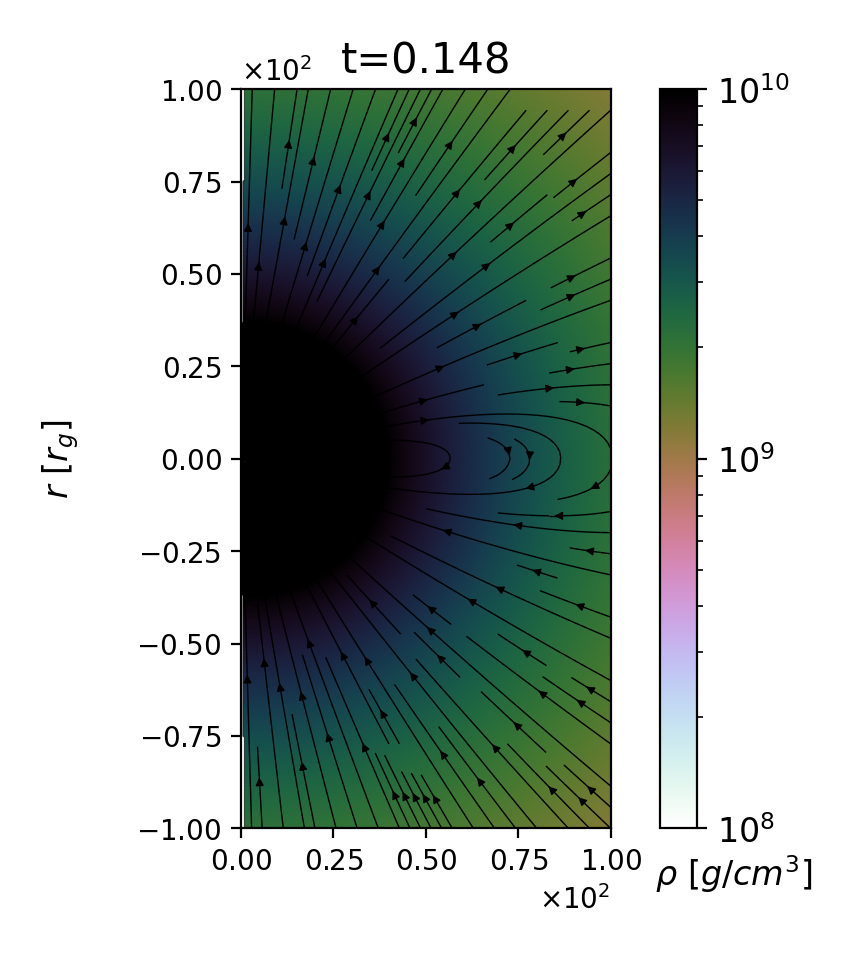}
   \includegraphics[scale=0.42]{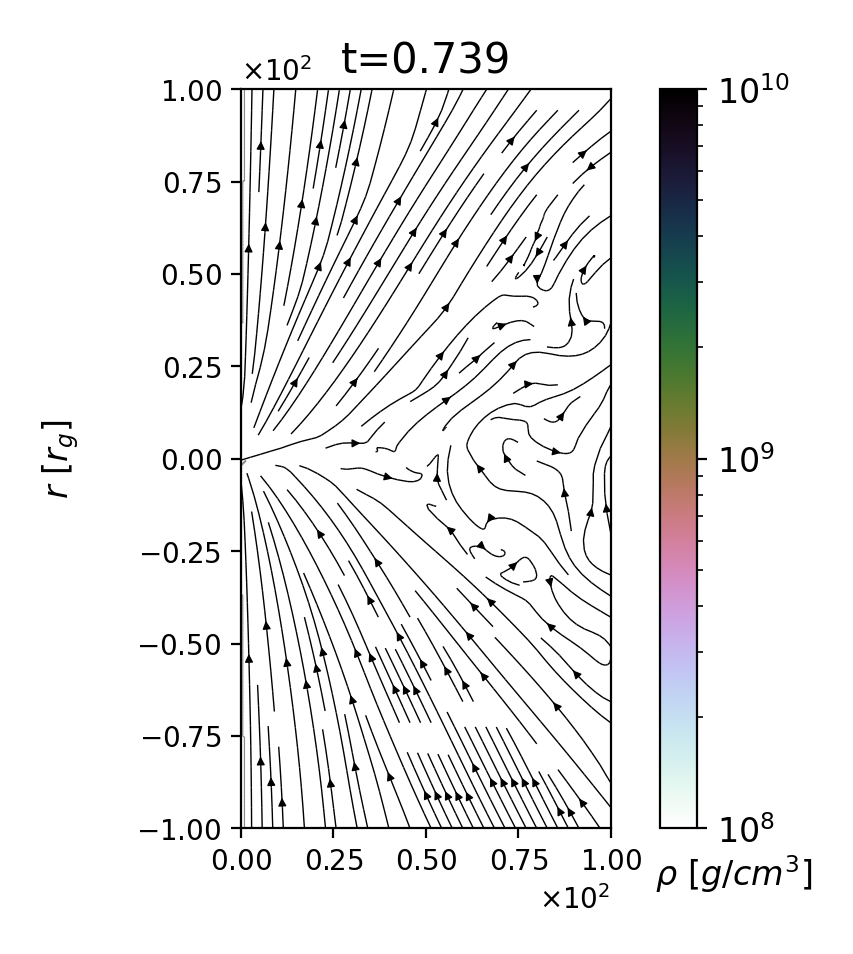}
             \caption{Distribution of density and magnetic field  vector potential contours at t=0 (left), and the short-time evolved snapshots taken at time $t=0.148 ~s$ (middle), and at the end of simulation time $t=0.739 ~s$ (right).
               The magnetic field of dipole configuration is adopted and normalized with $\beta=50$. Top row is scaled to the outer radius of the domain, at 1000 $r_{g}$, and bottom row is zoomed in to 100
               $r_{g}$. Simulation was done in an evolving Kerr metric, plus with self-gravity effect.
             Parameters are the same as in Fig. \ref{fig:profiles_dipole}. Model shown is D08-S10-SG-b50, as listed in Tab. \ref{tab:modele}. 
             }
    \label{fig:profiles_dipole_sg}
\end{figure*}

In addition to the vertical magnetic field configuration, we computed the evolution of collapsing star embedded in the dipole magnetic field, given by Eq. \ref{eq:dipole}.
In Figures \ref{fig:profiles_dipole} and \ref{fig:profiles_dipole_sg} we present the initial and evolved states of magnetized models with a dipole configuration. This configuration seems more natural for a stellar structure, at large scale in the envelope. It was not considered in our previous study \citep{Dominika2021}.
In the presented models the initial black hole spin was assumed equal to $A_{0}=0.85$, and specific angular momentum was normalized with $S=1.0$ or $S=2.0$. In Figure \ref{fig:profiles_dipole} we show the snapshots from model with neglected the self-gravity effects, while in the Figure \ref{fig:profiles_dipole_sg} the self-gravity effect is included.
The dipole field is a prospective configuration of the field, in the context of stellar collapse models \citep{Burrows2022}.
We notice, that in comparison to the uniform field configuration, the general evolution of the system is similar. There appear small quantitative differences as for the 
final black hole mass and spin, as well as the average accretion rate values. 
In non-self-gravitating models, the maximum accretion rates (we probed only the case of initial spin $A_{0}=0.85$ and envelope rotations $S=1.0$ or $S=2.0$) seem to be slightly larger for dipole field than for vertical one. The maximum black hole spin value does not change, however the final spin of the black hole in general can be a bit larger (by $\sim 0.02$) while the final black hole mass gets slightly smaller, for $S=1.0$. In case of $S=2.0$, the trend reverses, and final black hole mass is larger, while final spin is smaller (even by $\sim 0.06$). Detailed values are given in Table \ref{tab:modele}.
Noticeably, for dipole field normalized to a maximum gas-to-magnetic field ratio at horizon, the results wit dipole field do not depend on this normalisation.

\begin{figure*}[ht]
      \centering
   \includegraphics[scale=0.42]{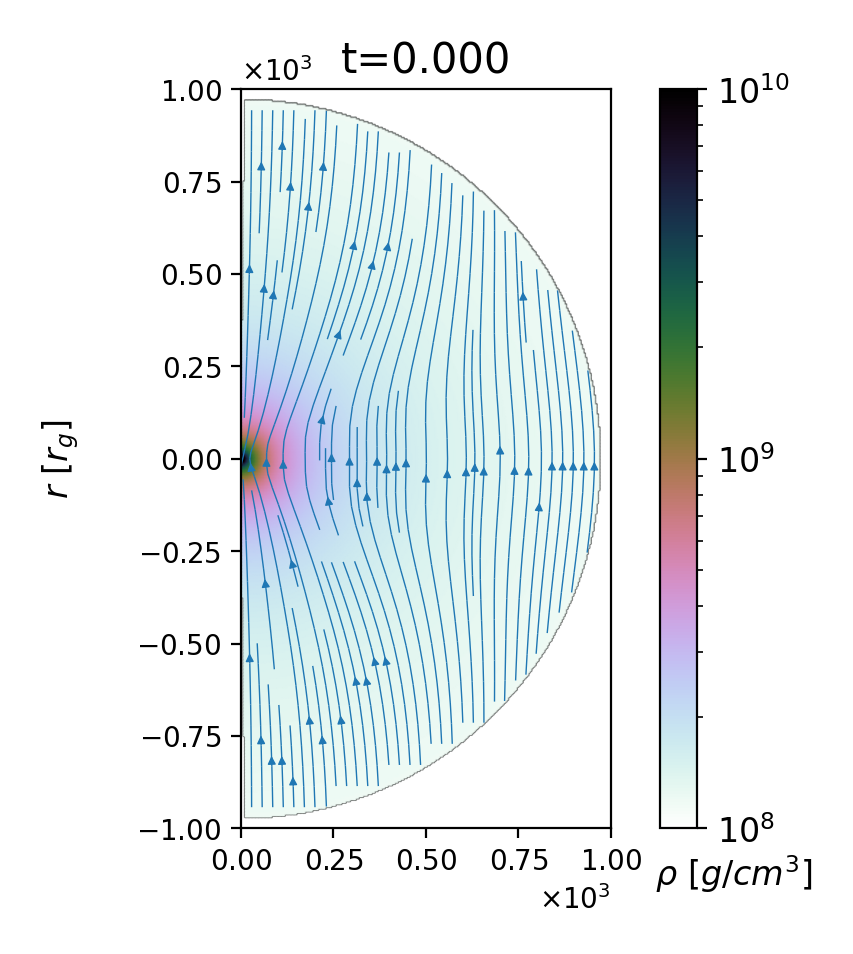}
   \includegraphics[scale=0.42]{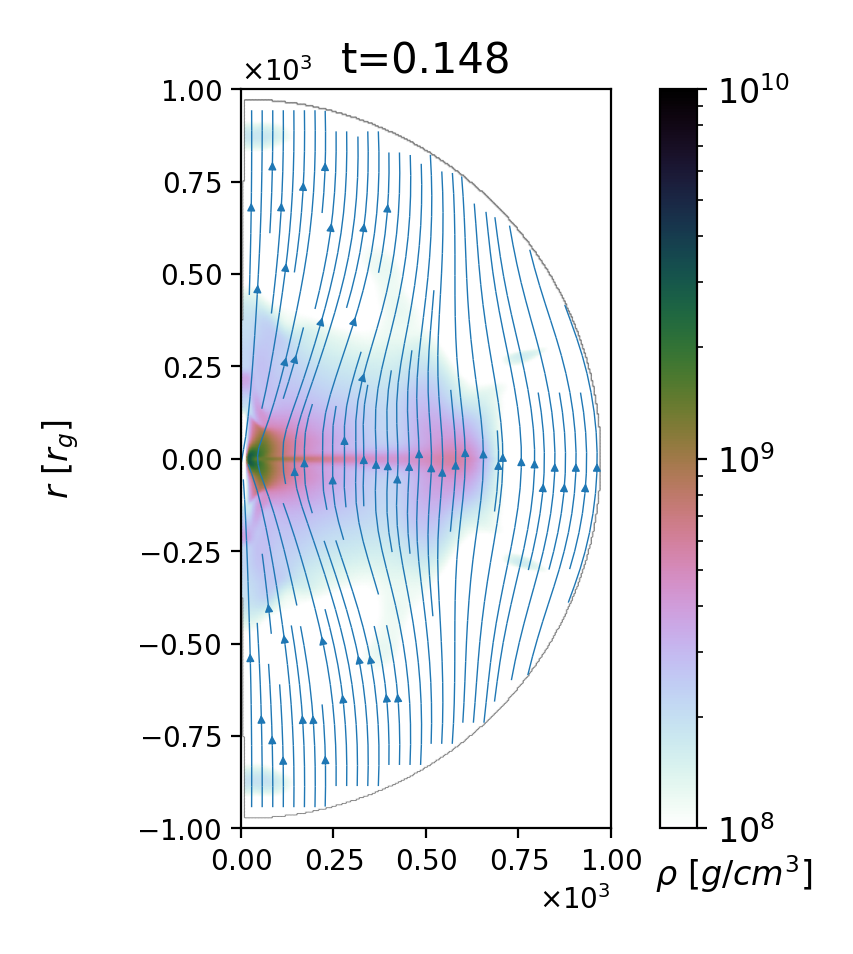}
   \includegraphics[scale=0.42]{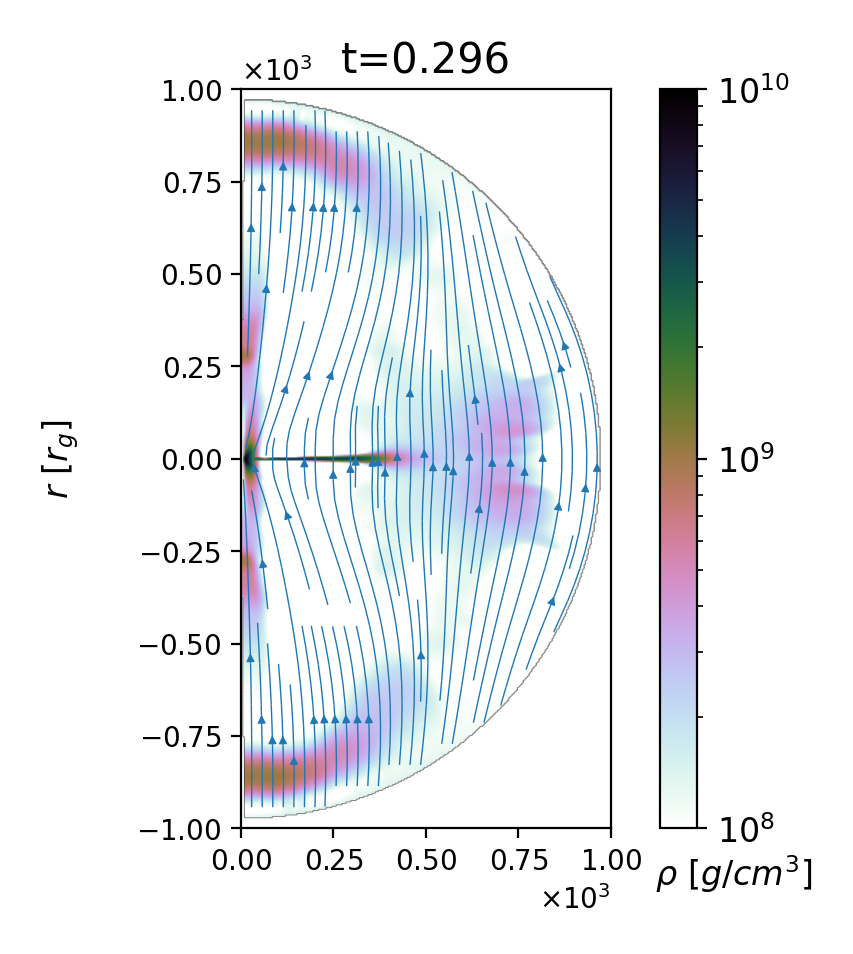}
   \includegraphics[scale=0.42]{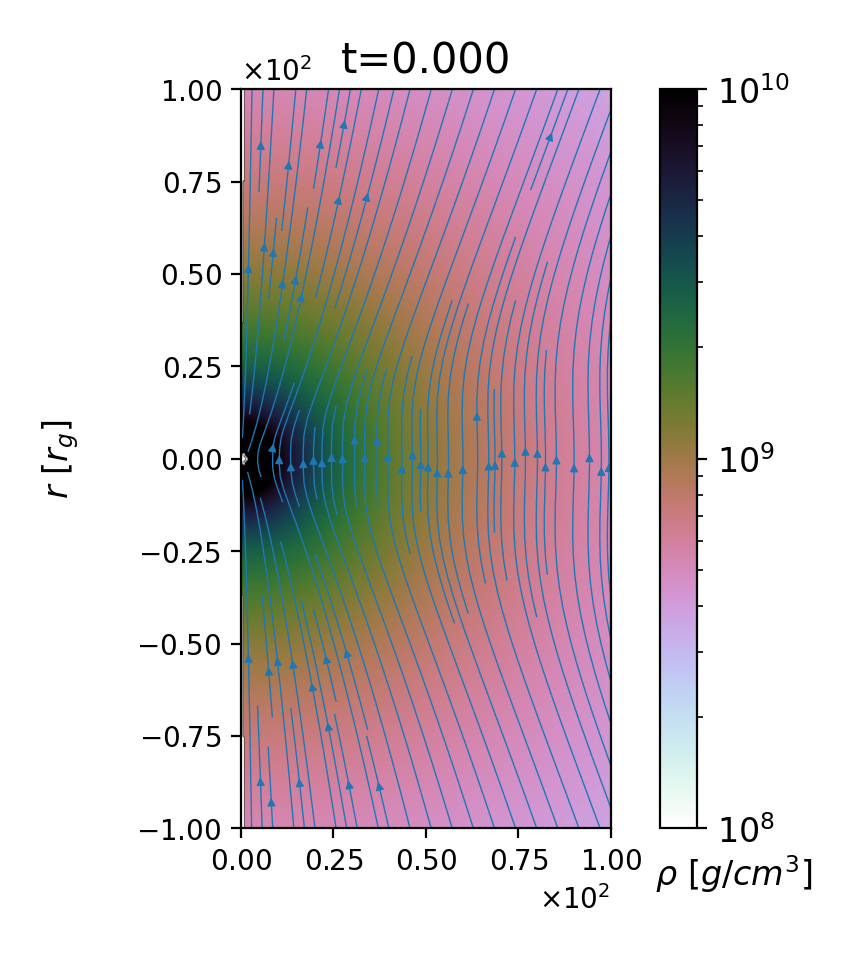}
   \includegraphics[scale=0.42]{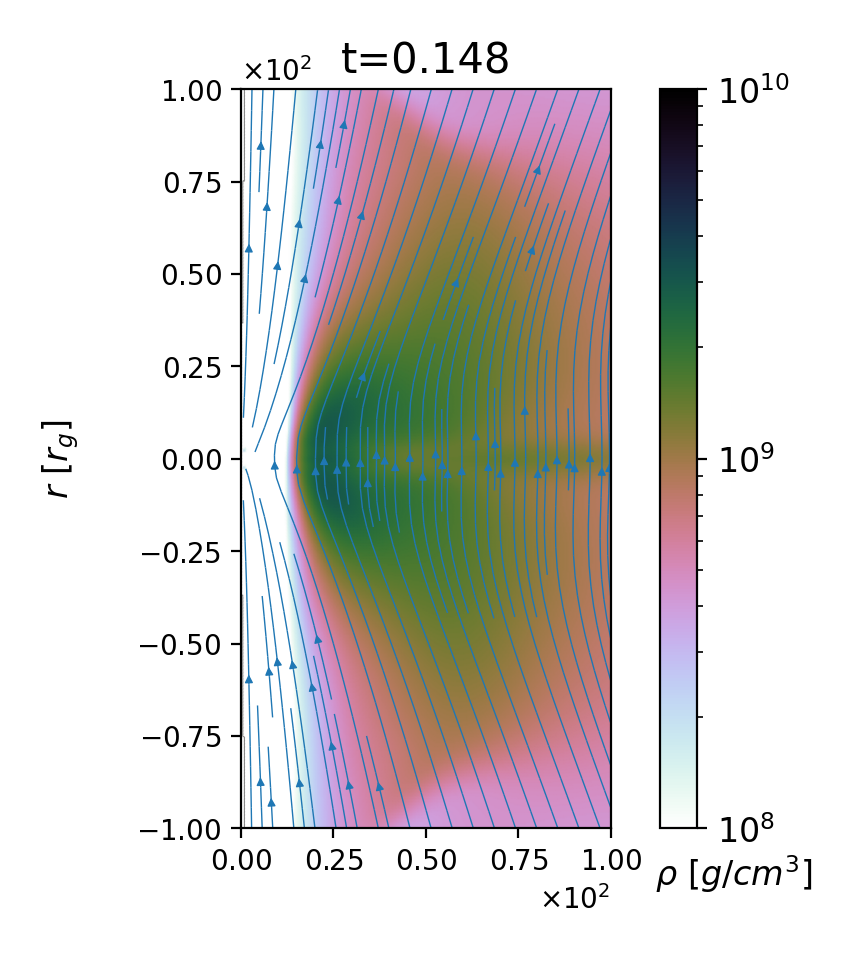}
   \includegraphics[scale=0.42]{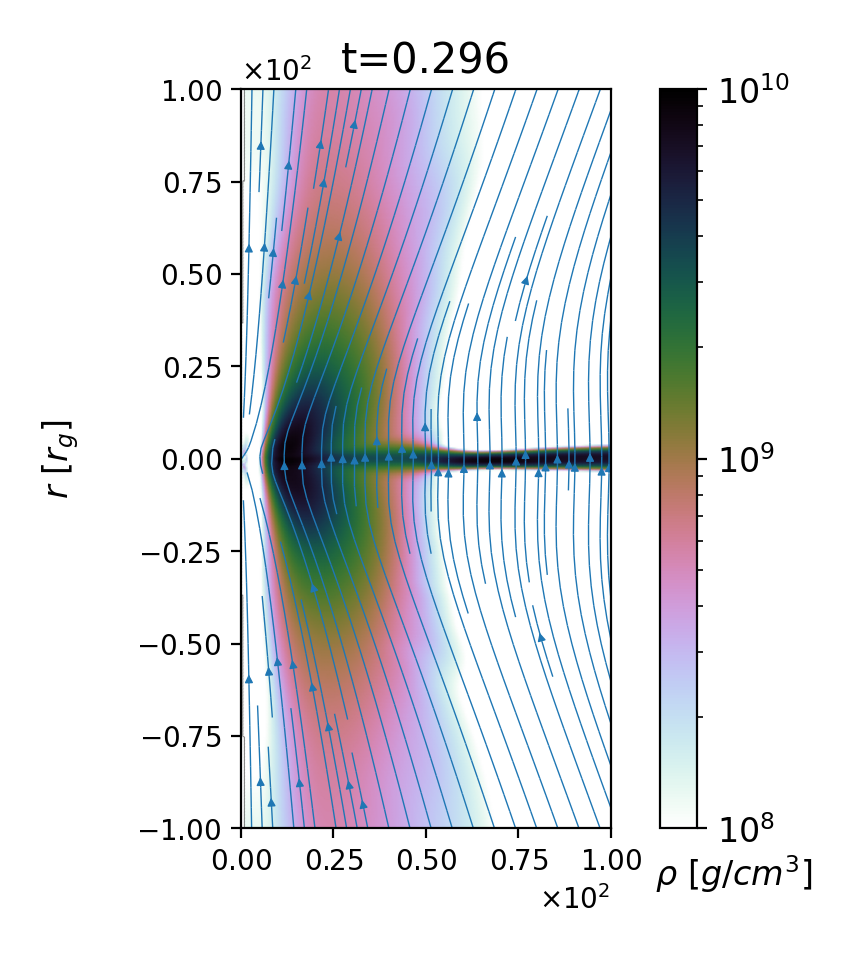}
             \caption{Distribution of density and magnetic field  vector potential contours at t=0 (left), and the short-time evolved snapshots taken at time $t=0.148 ~s$ (middle), and at the end of simulation at time $t=0.296 ~s$ (right).
             The magnetic field of dipole configuration is adopted and normalized with $\beta=50$. Top row is scaled to the outer radius of the domain, at 1000 $r_{g}$, and bottom row is zoomed in to 100 $r_{g}$. Simulation was done in an evolving Kerr metric, but without self-gravity effect. Parameters: $A_{0}=0.85$, $S=1.0$. Model shown is W08-S10-nSG-b50, as listed in Tab. \ref{tab:modele}
             }
    \label{fig:profiles_wald}
\end{figure*}

\begin{figure*}[ht]
      \centering
   \includegraphics[scale=0.42]{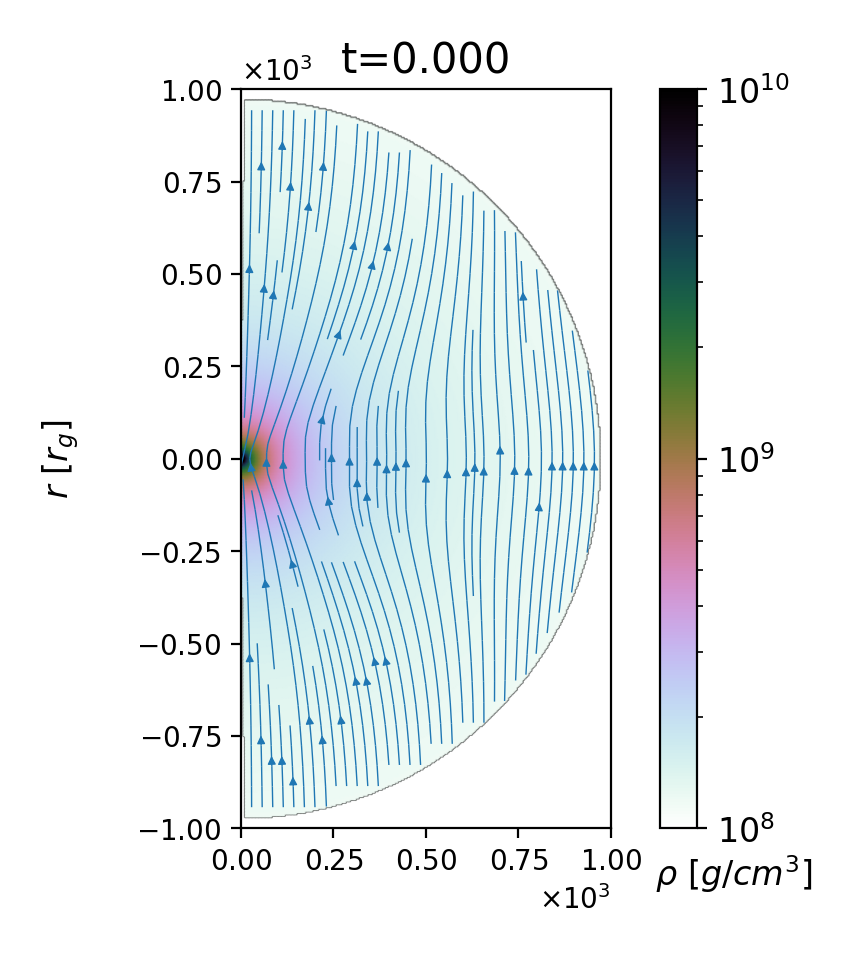}
   \includegraphics[scale=0.42]{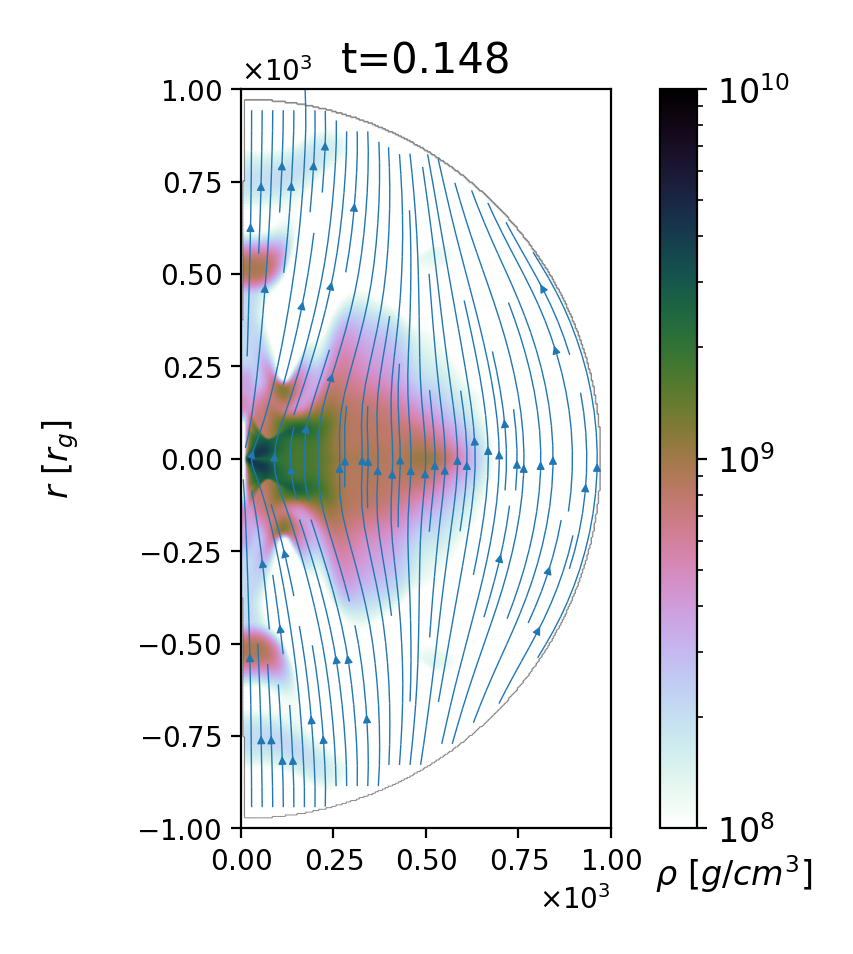}
   \includegraphics[scale=0.42]{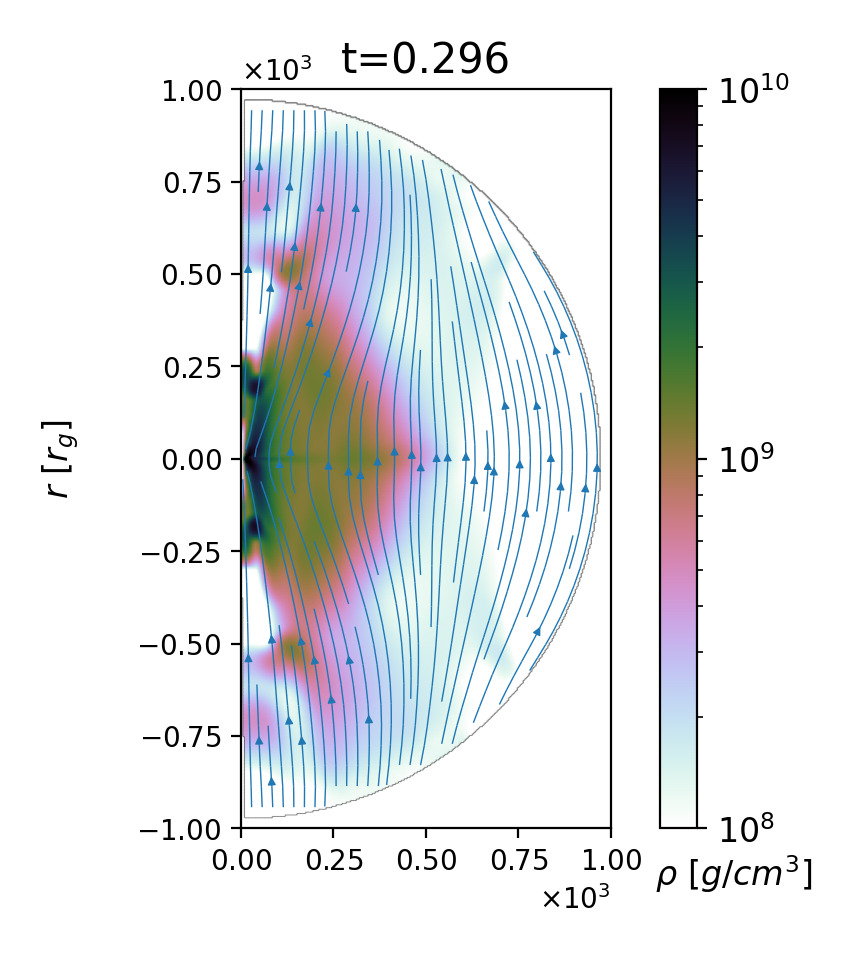}
   \includegraphics[scale=0.42]{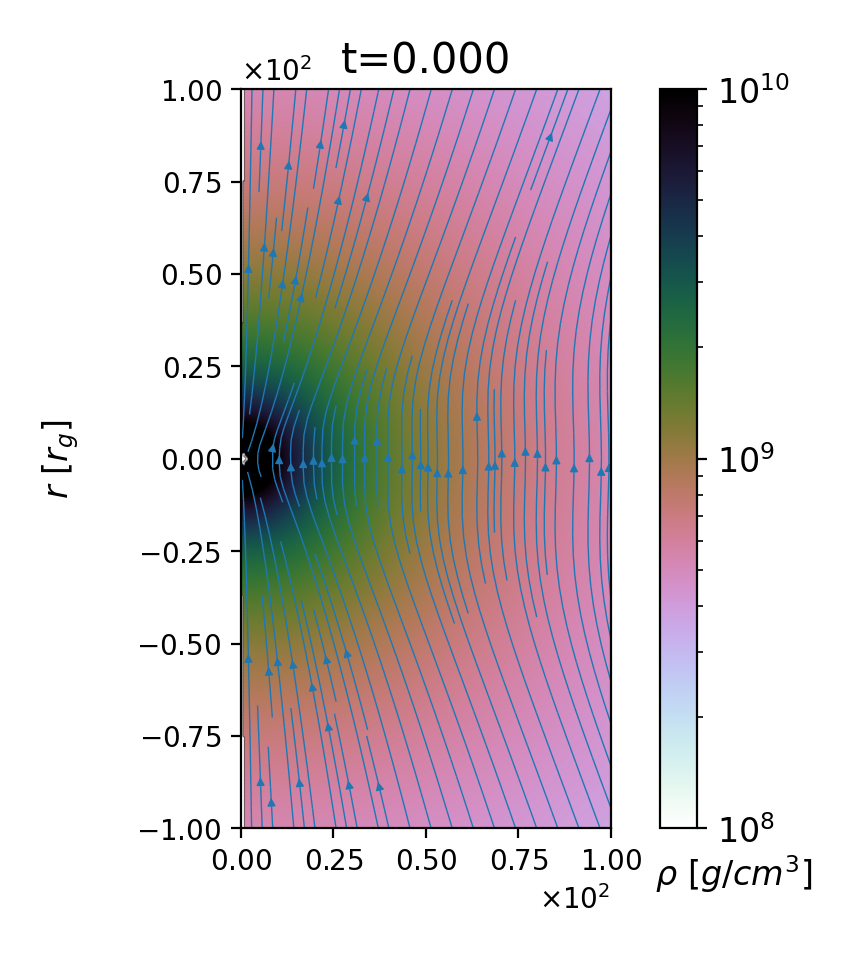}
   \includegraphics[scale=0.42]{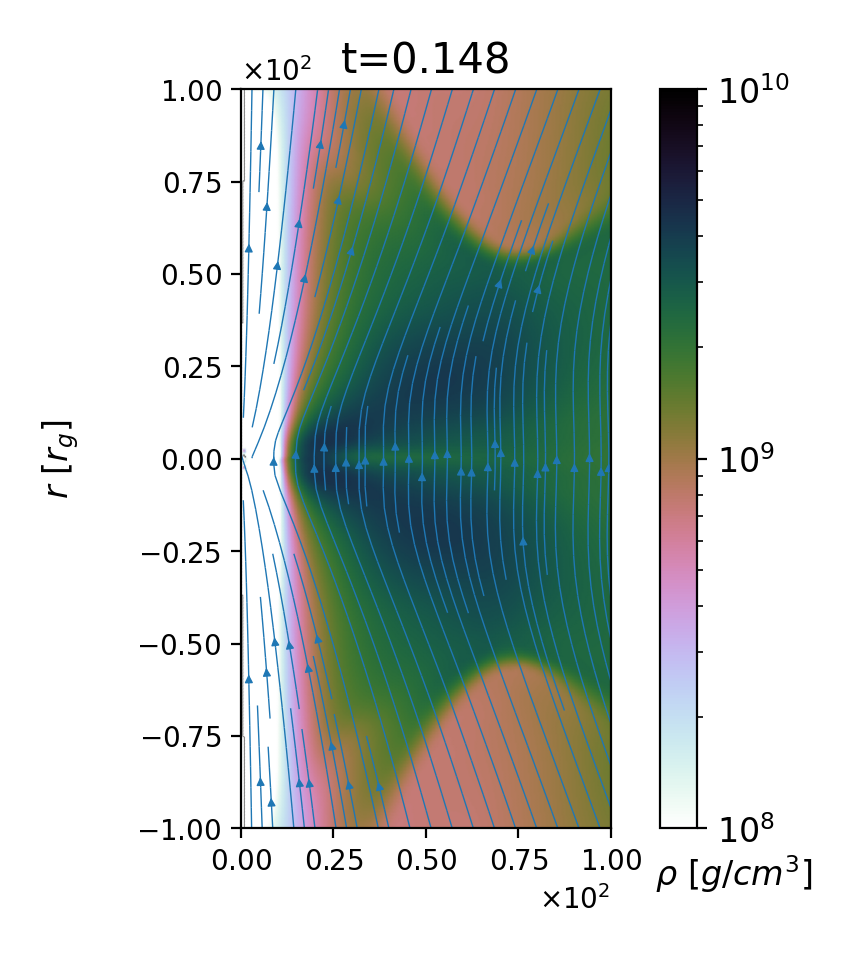}
   \includegraphics[scale=0.42]{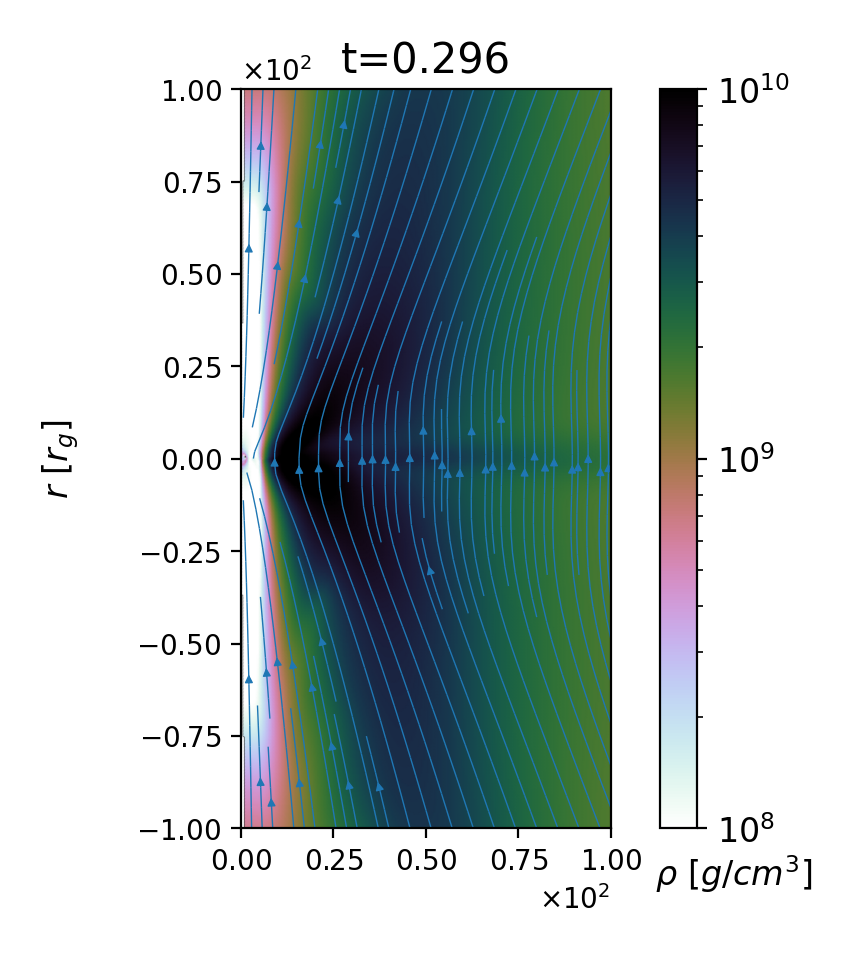}
             \caption{Distribution of density and magnetic field  vector potential contours at t=0 (left), and the short-time evolved snapshots taken at time $t=0.148 ~s$ (middle), and at the end of simulation at time $t=0.296 ~s$ (right).
             The magnetic field of dipole configuration is adopted and normalized with $\beta=50$. Top row is scaled to the outer radius of the domain, at 1000 $r_{g}$, and bottom row is zoomed in to 100 $r_{g}$. Simulation was done in an evolving Kerr metric, plus with self-gravity effect. Parameters: $A_{0}=0.85$, $S=1.0$. Model shown is W08-S10-SG-b50, as listed in Tab. \ref{tab:modele}
             }
    \label{fig:profiles_wald_sg}
\end{figure*}

In Figures \ref{fig:profiles_wald} and \ref{fig:profiles_wald_sg} we present the initial and evolved states of magnetized models with a Wald configuration as given by Eq. \ref{eq:Wald}. We notice that in this case magnetic field acts as a barrier and prevents material from accreting onto black hole. A repulsive effect of the black hole magnetosphere is seen in both simulations, with and without self-gravity. Hence, black hole mass does not change during the simulation (cf. Table \ref{tab:modele}).

In case of the dipole magnetic field, shown in Fig. \ref{fig:profiles_dipole}, the accreting torus is formed at the equatorial region close to the black hole. Its size is rather small. 
The magnetic flux brought to the black hole horizon is not large enough, to be able to power the successful jet. We checked that the dimensionless magnetic flux, i.e. scaled to the mass flux on the black hole horizon:
\begin{equation}
    \phi_{BH} =  {\Phi_{BH} \over {\sqrt{\dot M r_{g}^{2} c}}} = 
    { {\int{B_{r} dA} \over  {\sqrt{\dot M r_{g}^{2} c}}}}
    \label{eq:phiBh}
\end{equation}

is at most about $\phi_{BH}\sim5$ in all models, and very quickly drops to zero during the evolution.
On the other hand, larger $\phi_{BH}>15$, is presumably needed to form a magnetically arrested state and help launching the relativistic jets from the collapsar's central and power a gamma ray burst
\citep{JaniukJames2022}.

It is beyond the scope of the present paper to investigate in more detail the evolution of a magnetically arrested state of the accretion flow, especially in the case of a self-gravitating collapsar. We plan to study this scenario in a separate work, and verify whether such configuration can possibly give rise to an long-lasting jet launched from the black hole horizon. 
For now, we only verified, that for self-gravitating models embedded in dipole magnetic field, a slightly larger $\phi_{BH}$ (albeit still smaller than 'canonical' value of 15) is reached at the beginning of the simulation, if only we normalize the models  with maximum magnetisation, $\sigma=b^{2}/\rho$ (instead of the maximum gas-to-magnetic pressure ratio).
We list those models also in the Table \ref{tab:modele}. We conclude, that still a purely dipole field is rather unable to support launching relativistic jets from self-gravitating collapsars, unless the fields at the core of the star is amplified and reconfigured. Such conclusions should be however verified by fully 3D simulations, similar to those presented in \cite{Gottlieb2022}. 
 In their work, the dipole magnetic field prescription has been modified with a factor depending on the radius, to disentangle the magnetic field of the stellar core from the dipole-like field of the envelope(cf. Eq. 10 in their article).

We propose that the magnetic field preserved by the collapsing stellar core which forms a black hole, might be reconfigured during collapse and for a Wald magnetosphere. Hence, a repulsing effect of such field will act on matter, if the black hole is spinning sufficiently fast \citep{Karas2020}.

\begin{figure*}[ht]
      \centering
   \includegraphics[scale=0.33]{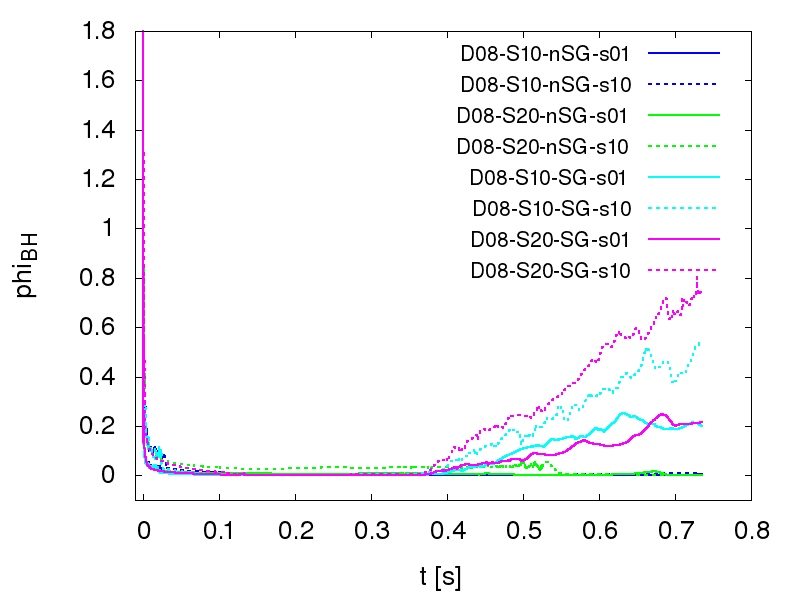}
   \includegraphics[scale=0.33]{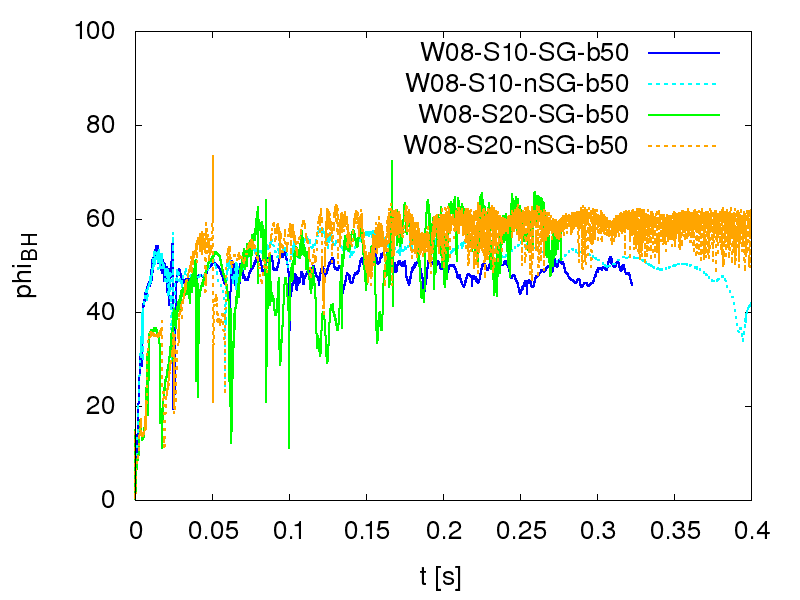}
     \caption{Time evolution of the magnetic flux at the black hole horizon, normalized to the mass flux (see Eq. \ref{eq:phiBh}). Models are normalized with specific angular momentum at ISCO $S=1.0$ and $S=2.0$ (as labeled in the panels). Initial spin of the black hole was $A_{0}=0.85$. The initial magnetic field was dipole given by Eq. \ref{eq:dipole}, and  was normalized with maximum magnetisation of $\sigma=1$ or $\sigma=0.1$ (left panel), or the vertical field given by Eq. \ref{eq:Wald}, and normalized with gas-to-magnetic pressure ratio at ISCO equal to $\beta=50$. Models are labeled in the both panels with symbols referring to Tab \ref{tab:modele}.
             }
    \label{fig:models_phiBH}
\end{figure*}

In Figure \ref{fig:models_phiBH} we show the time dependence of the magnetic flux of the horizon, for the above mentioned configurations with two values of magnetisation, $\sigma= 1$, and 0.1. For comparison, we also checked the magnetic flux level in case of our third magnetic field configuration, the Wald solution given by Eq. \ref{eq:Wald} and supposed to describe accurately the magnetosphere of a fastly spinning black hole. 

In the numerical simulation, the Wald magnetic field confines a large-scale toroidal structure in the equatorial plane, which is present for most of the simulation time. A temporary jet-like structure forms in the polar regions, in the early time of the simulation. A low density funnel is formed along the black hole rotation axis, and reaches distance about $\sim 250 r_{g}$.  
In this simulation, the magnetically arrested state developed, and the dimensionless magnetic flux of $\phi_{BH}\sim 50$ was reached at the horizon region. This high values were obtained in both self-gravitating and non-SG models, regardless of the specific angular momentum content in the envelope.
Still, the magnetic field did not prevent matter from collapsing through the polar and intermediate-latitude regions, hence dense material was later present below and above the torus. This material ultimately halted the jet that was trying to emerge from the black hole.
Again, possibly 3D simulation results might lead to a different outcomes, and will be studied in a future work.
Specifically, the case of highly magnetized collapsar, where initial magnetisation is normalized to $\sigma=0.1$,  dense material should not fall onto the center from the poles, but be expelled outwards. Then, the jet will be more likely to break out, is a form of a persistent, or transient structure, than for less magnetized material.
To study this effect in detail, 3-dimensional simulation with finer resolution, and possible adaptive mesh refinement, will be needed.

\subsection{Global trends across the black hole spin scale, angular momentum content, and the magnetic field strength}

In the three previous subsections, we discussed the dependencies of the collapsing stellar core properties separately for each black hole spin.
We also used several values of the angular momentum content in the pre-collapse star, and several prescriptions of the initial magnetic field configuration and strength.

\begin{figure*}[ht]
      \centering
   \includegraphics[scale=0.33]{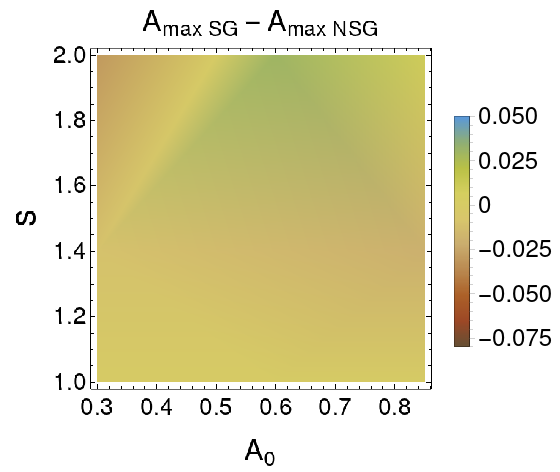}
   \includegraphics[scale=0.33]{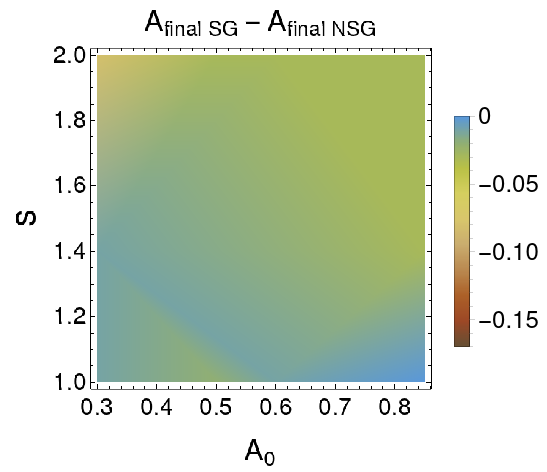}
   \includegraphics[scale=0.33]{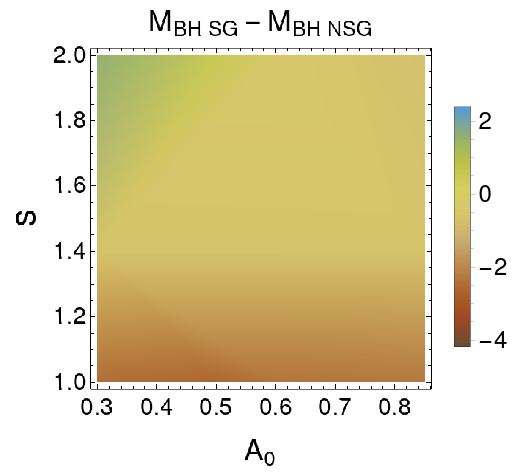}
   \includegraphics[scale=0.33]{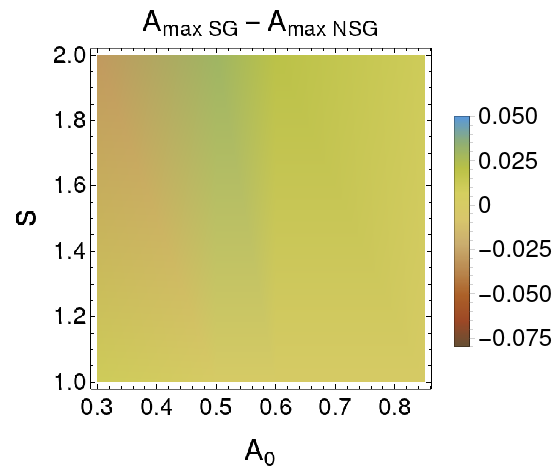}
   \includegraphics[scale=0.33]{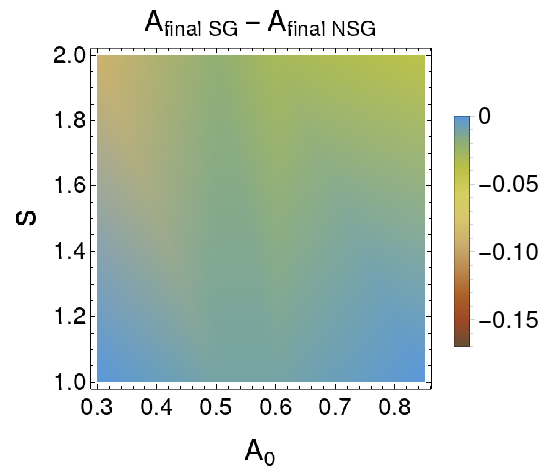}
   \includegraphics[scale=0.33]{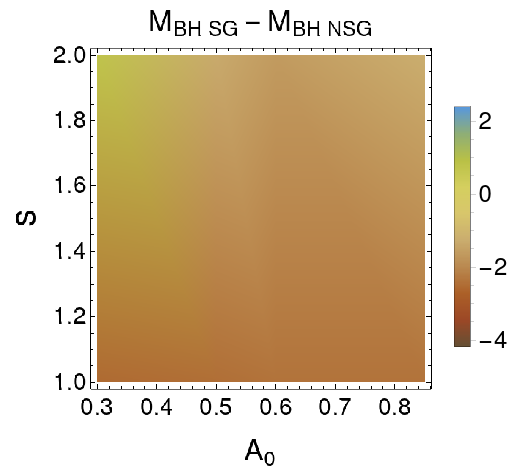}
   \includegraphics[scale=0.33]{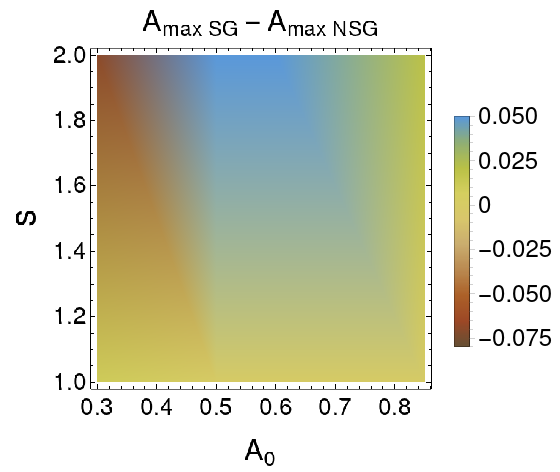}
   \includegraphics[scale=0.33]{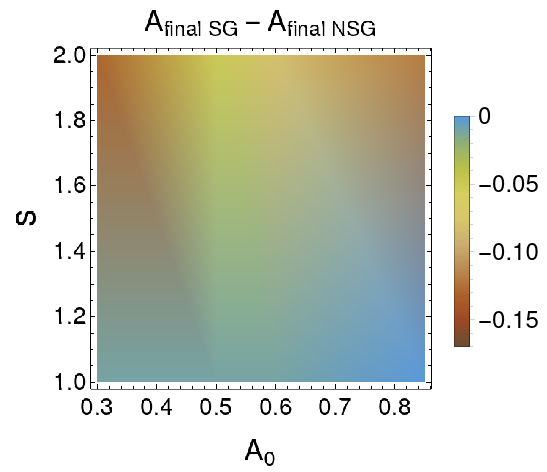}
   \includegraphics[scale=0.33]{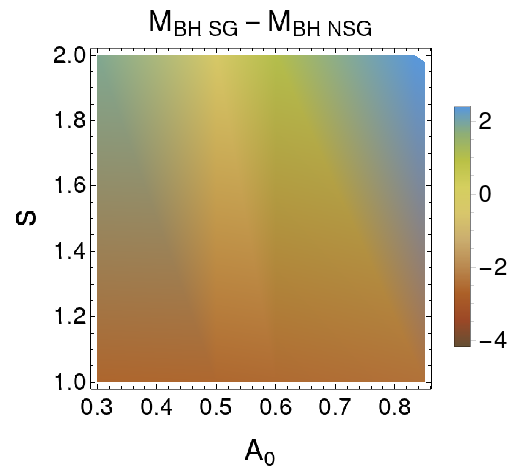}
             \caption{The density plots showing the difference of $A_{final}$ (left panels), $A_{max}$ (middle panels) and $M_{BH}$ (right panels) between Self-Gravitating models and non Self-Gravitating models for simulations without magnetic field (upper row), with magnetic field normalized to $\beta=100$ (middle row) and for $\beta=10$ (lower row).  
             }
    \label{fig:dp}
\end{figure*}

Here we propose a synthetic, quantitative way to examine the influence of the self-gravity on the system. We check the relative differences between global resulting quantities: final black hole spin, $A_{final}$, maximal black hole spin,  $A_{max}$, and the maximal black hole mass, $M_{BH}$, calculated for the simulations with and without self-gravity. 
We show three cases with different input of the magnetic field, as presented in Fig. \ref{fig:dp} (top row is non-magnetized case, middle and bottom rows differ with respect to $\beta_{0}$ parameter that normalized vertical field). 
We show the color maps of those three global quantities, in the parameter space defined by the model parameters: $A_0$ and $S$. 

We notice that the final spin of black hole is always smaller in simulations with self-gravity. The magnetic field input enhances this difference, especially for models with higher rotation parameter, and regardless of the initial black hole spin.
In other words, the relative difference between final spins is negative and spans larger parameter space, if the magnetic field is present in the collapsing stellar core. The reason for that is the action of magnetic barrier, that pushes outward the in-falling gas, hence temporary decrease of the mass accretion rate. The effect on the black hole angular momentum is rather negligible, but the dimensionless spin $A$ is reduced.

The behaviour of the final black hole mass and the maximal spin is more complicated and depends on the combination of the $A_0$ and $S$ parameters. $A_{max}$ is higher for self-gravitating models with $A_0 \sim 0.6$ and high values of $S$, 
for magnetized models.
But when the initial spin is very low or very high, the maximum spin  $A_{max}$ is higher for models without self-gravity.
In weakly magnetized, and non-magnetized models, the relative differences become roughly zero, as shown in the top row of Fig. \ref{fig:dp}.

Similarly, the maximal black hole mass in most cases is lower for the simulations with self-gravity, however, again we can see different behaviour for the simulations with high $S$ and stronger magnetic field.
The relative difference between the resulting quantities is enhanced by the magnetic field. 

To sum up, as shown in Figure {\ref{fig:dp}}, one can interpret the self-gravity as a factor that lowers the final mass and spin transferred into the black hole, while it may increase the maximum spin parameter of the black hole with respect to non self-gravitating case. On the other hand, we found that a sharp decrease of the accretion rate at time $t\gtrsim 0.2s$ which appears in self-gravitating case (see the second plots of both rows in Figure \ref{M_S_A}), coincides with an increase in the specific angular momentum of the inner radii. It suggests that the accretion of matter into the black hole may take a longer time than what we considered for the whole simulation, so that the final mass of the black hole is obviously less than in non self-gravitating case. We believe that it may further cause a decrease in the final spin of the black hole, as well. However, considering the higher maximum spin parameter in the self-gravitating case, we came into conclusion that at the earlier time steps ($\lesssim 0.15s$), the self-force of the envelope seems to pave the way for the angular momentum to be transferred into the black hole via making more amount of mass fall inside the horizon as a result of the higher accretion rate. When it comes to magnetic field effects, it appears to influence these parameters in either cases with and without self-gravity. However, for the self-gravitating case this impact is rather more complicated than the non self-gravitating one. On the one hand, self-force acts against magnetic filed since it causes the matter to get denser which is against the role of magnetic field in small scale. On the other hand, one may expect the magnetic field decreases the accretion of matter towards the black hole due to the large scale effect of magnetic torque on the envelope. This type of influence agrees with the previously mentioned increase in the specific angular momentum of the inner radii when self-gravity is taken into account, at the time steps $\gtrsim 0.2s$. In general, we found that self-gravity impact dominates over magnetic field so that it influences non self-gravitating case more than self-gravitating one. This may result in the increase of the relative difference between black hole features, regarding two cases with and without self-gravity. 

\clearpage
\onecolumn

\begin{center}
  \begin{longtable}{|c|c|c|c|c|c|c|c|c|c|c|c|}
        \caption{List of models calculated for initial black hole of 3 $M_{\odot}$, and initial cloud mass of 25 $M_{\odot}$. All models account for the Kerr metric update due to changing mass and spin of the black hole. Some models (SG) have also the option of self-gravity force acting on the matter due to mass and angular momentum of gas enclosed inside the sphere.
        Most simulations were run until 50000 $t_{g}$, but models marked with asterisk were run until 30000 $t_{g}$. Models names given in column 1 are used in the text and figures labels}
    \label{tab:modele} \\
             \hline
Model&\textbf{$a_{0}$} &\textbf{S}  &\textbf{$R_{in}$} &Self  & Magnetic  &$\sigma_{0}$ &$\beta_{0}$  &$< \dot M > $  &$M_{BH}^{final}$ &
$a^{max}$ & $a^{final}$\\     
 Name & & &  $(R_{g})$ &  Gravity &   Field &  &   &   $(M_{\odot}s^{-1})$  & $(M_{\odot})$ & &  \\  
     \hline
\endfirsthead
\multicolumn{11}{c}%
            {{\bfseries \tablename\ \thetable{} -- continued from previous page}} \\
            \hline
Model&\textbf{$a_{0}$} &\textbf{S}  &\textbf{$R_{in}$} &Self  & Magnetic  &$\sigma_{0}$ &$\beta_{0}$  &$< \dot M > $  &$M_{BH}^{final}$ &
$a^{max}$ & $a^{final}$\\     
 Name & & & $(R_{g})$ &  Gravity &   Field &  &   &   $(M_{\odot}s^{-1})$  & $(M_{\odot})$ & &  \\  
\hline            
      \endhead

\hline \multicolumn{11}{|r|}{{Continued on next page}} \\ \hline
\endfoot

\hline \hline
\endlastfoot
A03-S10-nSG-R10&0.3   &  1.0   & 1.0  & -  & no & - &  - & 33.51 & 28.11  & 0.61 & 0.21 \\ 
A03-S10-nSG-R12& 0.3   &  1.0   & 1.2  & -  & no & - &  - & 33.51 & 28.11  & 0.62 & 0.22 \\ 
A03-S14-nSG-R12&0.3   &  1.4   & 1.2  & -  & no & - &  - & 31.12 &  26.30 & 0.76 & 0.30 \\ 
A03-S20-nSG-R10&0.3   &  2.0   & 1.0  & -  & no & - &  - & 20.97 & 18.75  & 0.84 & 0.56 \\ 
A03-S20-nSG-R12&0.3   &  2.0   & 1.2  & -  & no & - &  - & 20.48 &  18.39 & 0.85 & 0.57 \\ 
A03-S10-SG-R10&0.3   &  1.0   & 1.0  & yes & no & - &  - & 31.21 & 25.58  & 0.62 & 0.20 \\ 
A03-S10-SG-R12&0.3   &  1.0   & 1.2  & yes & no & - &  - & 31.31 & 25.62  & 0.62 & 0.21 \\ 
A03-S10-SG-R17&0.3   &  1.0   & 1.7  & yes & no & - &  - & 31.17 &  25.53 & 0.62 & 0.21 \\ 
A03-S14-SG-R12&0.3   &  1.4   & 1.2  & yes  & no & - &  - & 25.71$^{*}$ &  25.79 & 0.75  & 0.28 \\ 
A03-S20-SG-R10&0.3   &  2.0   & 1.0  & yes & no & - &  - & 24.23 & 20.07  & 0.80 & 0.48 \\ 
A03-S20-SG-R12&0.3   &  2.0   & 1.2  & yes & no & - &  - & 24.01 & 19.97  & 0.82 & 0.49 \\ 
A03-S20-SG-R127&0.3   &  2.0   & 1.7  & yes & no & - &  - & 24.43 & 20.17  & 0.77 & 0.48 \\ 
\hline
A05-S10-nSG-R12&0.5   &  1.0   & 1.2  & -  & no & - &  - & 33.51 & 28.11  & 0.66 & 0.20 \\ 
A05-S10-nSG-R17&0.5   &  1.0   & 1.7  & -  & no & - &  - & 33.51 &  28.11 & 0.67  & 0.21 \\ 
A05-S14-nSG-R10 &0.5   &  1.4   & 1.02  & -  & no & - &  - & 31.75 &  26.90 & 0.78 & 0.28 \\ 
A05-S20-nSG-R12 & 0.5   &  2.0   & 1.2  & -  & no & - &  - & 23.42 & 20.59  & 0.85 & 0.49 \\ 
A05-S20-nSG-R17 & 0.5   &  2.0   & 1.7  & -  & no & - &  - & 21.57 & 19.20  & 0.87 & 0.54 \\
A05-S10-SG-R12 &0.5   &  1.0   & 1.2  & yes & no & - &  - & 31.11 & 25.50  & 0.66 & 0.18 \\ 
A05-S10-SG-R17 & 0.5   &  1.0   & 1.7  & yes & no & - &  - & 31.29 & 25.61  & 0.67 & 0.20 \\ 
A05-S14-SG-R10 & 0.5   &  1.4   & 1.02  & yes  & no & - &  - & 31.00 &  25.90 & 0.77 & 0.29 \\ 
A05-S20-SG-R12 & 0.5   &  2.0   & 1.2  & yes & no & - &  - & 23.49 & 19.71  & 0.85 & 0.46 \\ 
A05-S20-SG-R17 &0.5   &  2.0   & 1.7  & yes & no & - &  - & 23.55 & 19.72  & 0.83 & 0.48 \\ 
\hline
A06-S10-nSG-R12 &0.6   &  1.0   & 1.2  & -  & no & - &  - & 33.49 & 28.11  & 0.69 & 0.20 \\ 
A06-S10-nSG-R17 & 0.6   &  1.0   & 1.7  & -  & no & - &  - & 33.48 & 28.11  & 0.71  & 0.21 \\ 
A06-S20-nSG-R12 &0.6   &  2.0   & 1.2  & -  & no & - &  - & 22.39 & 19.82  & 0.85 & 0.50 \\ 
A06-S10-nSG-R17 &0.6   &  2.0   & 1.7  & -  & no & - &  - & 21.57 &  19.20 & 0.86  & 0.53 \\ 
A06-S10-SG-R12 & 0.6   &  1.0   & 1.2  & yes & no & - &  - & 30.66 & 25.73  & 0.69 & 0.19 \\ 
A06-S10-SG-R17 &0.6   &  1.0   & 1.7  & yes & no & - &  - & 30.60 & 25.68  & 0.71 & 0.20 \\ 
A06-S20-SG-R12 & 0.6   &  2.0   & 1.2  & yes & no & - &  - & 22.73 & 19.50  & 0.88 & 0.47 \\ 
A06-S20-SG-R17 & 0.6   &  2.0   & 1.7  & yes & no & - &  - & 25.27 & 20.75  & 0.76 & 0.43 \\ 
\hline
A08-S10-nSG-R12 & 0.85   &  1.0   & 1.2  & -  & no & - &  - & 33.49 &  28.11 & 0.85  & 0.18\\ 
A08-S10-nSG-R17 & 0.85   &  1.0   & 1.7  & -  & no & - &  - & 33.48 &  28.10  & 0.85 & 0.20 \\ 
A08-S14-nSG-R12 &0.85   &  1.4   & 1.2  & -  & no & - &  - & 32.43 &  26.89 & 0.84 & 0.27 \\ 
A08-S20-nSG-R12 &0.85   &  2.0   & 1.2  & -  & no & - &  - & 23.55 & 20.68  & 0.87 & 0.45 \\ 
A08-S20-nSG-R17 & 0.85   &  2.0   & 1.7  & -  & no & - &  - & 21.29 &  18.99 & 0.88 & 0.51 \\ 
A08-S10-SG-R12 & 0.85   &  1.0   & 1.2  & yes & no & - &  - & 30.79 & 25.82  & 0.85 & 0.18 \\ 
A08-S10-SG-R13 &  0.85   &  1.0   & 1.3  & yes & no & - &  - & 30.78 & 25.81  & 0.85 & 0.19 \\
A08-S10-SG-R17 &  0.85   &  1.0   & 1.7  & yes & no & - &  - & 30.72 & 25.76  & 0.85 & 0.19 \\
A08-S14-SG-R12 & 0.85   &  1.4   & 1.2  & yes  & no & - &  - & 31.29 &  26.14 & 0.82 & 0.24 \\ 
A08-S20-SG-R12 & 0.85   &  2.0   & 1.2  & yes & no & - &  - & 23.61 & 20.01  & 0.88 & 0.42\\ 
A08-S20-SG-R13 &  0.85   &  2.0   & 1.3  & yes & no & - &  - &23.89 & 20.17  & 0.88 & 0.42 \\ 
A08-S20-SG-R17 & 0.85   &  2.0   & 1.7  & yes & no & - &  - & 23.54 & 19.81  & 0.91 & 0.46 \\ 
\hline
A03-S10-nSG-sMF & 0.3   &  1.0   & 1.2  & -  & vertical & - &  10 & 33.49 & 28.11  & 0.61 & 0.21 \\ 
A03-S10-nSG-wMF &0.3   &  1.0   & 1.2  & -  & vertical & - &  100 & 33.49 & 28.11  & 0.61 & 0.21 \\ 
A03-S20-nSG-sMF &  0.3   &  2.0   & 1.2  & -  & vertical & - &  10 & 20.18 & 18.17  & 0.84 & 0.58 \\ 
A03-S20-nSG-wMF &0.3   &  2.0   & 1.2  & -  & vertical & - &  100 & 21.14 & 18.88  & 0.85 & 0.56 \\ 
A03-S10- SG-sMF & 0.3   &  1.0   & 1.2  & yes  & vertical & - &  10 & 30.32 &  25.48 & 0.62 & 0.20 \\ 
A03-S10-SG-wMF & 0.3   &  1.0   & 1.2  & yes  & vertical & - &  100 & 30.43 & 25.56  & 0.62 & 0.21 \\ 
A03-S20-SG-sMF &0.3   &  2.0   & 1.2  & yes  & vertical & - &  10 & 23.98 & 20.11  & 0.81 & 0.48 \\ 
A03-S20-SG-wMF & 0.3   &  2.0   & 1.2  & yes  & vertical & - &  100 & 24.25 & 20.27 & 0.85 & 0.48 \\ 
\hline
A05-S10-nSG-sMF &0.5   &  1.0   & 1.2  & -  & vertical & - &  10 & 33.49 & 28.11  & 0.66 & 0.20 \\ 
A05-S10-nSG-wMF & 0.5   &  1.0   & 1.2  & -  & vertical & - &  100 & 33.49 & 28.11  & 0.66 & 0.20 \\ 
A05-S20-nSG-sMF &  0.5   &  2.0   & 1.2  & -  & vertical & - &  10 & 21.64 &  19.26 & 0.83 & 0.52 \\ 
A05-S20-nSG-wMF & 0.5   &  2.0   & 1.2  & -  & vertical & - &  100 & 23.22 & 20.44  & 0.85 & 0.49 \\ 
A05-S10-SG-sMF & 0.5   &  1.0   & 1.2  & yes  & vertical & - &  10 & 30.37 & 25.51  & 0.66 & 0.19 \\ 
A05-S10-SG-wMF &0.5   &  1.0   & 1.2  & yes  & vertical & - &  100 & 30.55 & 25.64  & 0.66 & 0.19\\ 
A05-S20-SG-sMF &0.5   &  2.0   & 1.2  & yes  & vertical & - &  10 & 22.98 & 19.59  & 0.88 & 0.48 \\ 
A05-S20-SG-wMF &0.5   &  2.0   & 1.2  & yes  & vertical & - &  100 & 24.06   & 20.21 & 0.80 & 0.46 \\ 
\hline
A06-S10-nSG-sMF & 0.6   &  1.0   & 1.2  & -  & vertical & - &  10 & 33.49 & 28.11  &  0.69 & 0.20 \\ 
A06-S10-nSG-wMF & 0.6   &  1.0   & 1.2  & -  & vertical & - &  100 & 33.49 & 28.11  &  0.69 & 0.20 \\ 
A06-S20-nSG-sMF &0.6   &  2.0   & 1.2  & -  & vertical & - &  10 & 20.12 & 18.12  & 0.82 & 0.53 \\ 
A06-S20-nSG-wMF &0.6   &  2.0   & 1.2  & -  & vertical & - &  100 & 23.84 & 20.90  & 0.86 & 0.48 \\ 
A06-S10-SG-sMF &0.6   &  1.0   & 1.2  & yes  & vertical & - &  10 & 30.44 & 25.56  & 0.69 & 0.19 \\ 
A06-S10-SG-wMF &0.6   &  1.0   & 1.2  & yes  & vertical & - &  100 & 30.65 & 25.72  & 0.69 & 0.19 \\ 
A06-S20-SG-sMF & 0.6   &  2.0   & 1.2  & yes  & vertical & - &  10 & 23.26   & 19.81 & 0.87 & 0.46 \\ 
A06-S20-SG-wMF & 0.6   &  2.0   & 1.2  & yes  & vertical & - &  100 & 23.20  & 19.75  &0.88  & 0.47 \\ 
\hline
A08-S10-nSG-sMF & 0.85   &  1.0   & 1.2  & -  & vertical & - &  10 & 33.49 & 28.11  & 0.85 & 0.18 \\ 
A08-S10-nSG-wMF & 0.85   &  1.0   & 1.2  & -  & vertical & - &  100 & 33.49 & 28.11  & 0.85 & 0.18 \\ 
A08-S20-nSG-sMF &0.85   &  2.0   & 1.2  & -  & vertical & - &  10 & 20.16 & 18.16  & 0.85 & 0.50 \\ 
A08-S20-nSG-wMF &0.85   &  2.0   & 1.2  & -  & vertical & - &  100 & 24.23 & 21.19  & 0.87 & 0.44 \\ 
A08-S10-SG-sMF &0.85   &  1.0   & 1.2  & yes  & vertical & - &  10 & 30.62 & 25.69  & 0.85 & 0.18 \\ 
A08-S10-SG-wMF & 0.85   &  1.0   & 1.2  & yes  & vertical & - &  100 & 30.65 & 25.72  & 0.85 & 0.18 \\ 
A08-S20-SG-sMF & 0.85   &  2.0   & 1.2  & yes  & vertical & - &  10 & 24.44  & 20.54 & 0.87 & 0.41 \\ 
A08-S20-SG-wMF & 0.85   &  2.0   & 1.2  & yes  & vertical & - &  100 & 24.00 & 20.24  & 0.88 & 0.42 \\ 
\hline
\hline
 D08-S10-nSG-b01 & 0.85   &  1.0   & 1.2  & -  & dipole & - &  0.1  & 33.75 & 28.09  & 0.85 & 0.20 \\ 
D08-S10-nSG-b05 & 0.85   &  1.0   & 1.2  & -  & dipole & - &  0.5  & 33.75  & 28.09 & 0.85 & 0.20 \\ 
D08-S10-nSG-b10 & 0.85   &  1.0   & 1.2  & -  & dipole & - &  1.0  & 33.75 & 28.09 &  0.85 &  0.20 \\
D08-S10-nSG-b50 & 0.85   &  1.0   & 1.2  & -  & dipole & - &  50.0 & 33.75 & 28.09 & 0.85  &  0.20 \\
D08-S10-nSG-s01 & 0.85   &  1.0   & 1.2  & -  & dipole & 0.1 &  -  & 33.45 & 28.09 &  0.85 &  0.20 \\
D08-S10-nSG-s10 & 0.85   &  1.0   & 1.2  & -  & dipole & 1.0 &  - & 33.45 & 28.16 & 0.85  &  0.20 \\
D08-S10-SG-b01 & 0.85   &  1.0   & 1.2  & yes  & dipole & - &  0.1  & 34.23 & 28.27  & 0.85 &  0.21 \\ 
 D08-S10-SG-b05 & 0.85   &  1.0   & 1.2  & yes  & dipole & - &  0.5  & 34.23  & 28.27 & 0.85 &  0.21 \\
D08-S10-SG-b10 & 0.85   &  1.0   & 1.2  & yes& dipole & - &  1.0  & 34.23 & 28.26 & 0.85  & 0.21 \\
D08-S10-SG-b50 & 0.85   &  1.0   & 1.2  & yes& dipole & - &  50.0 & 34.23 & 28.26 & 0.85  & 0.21 \\
D08-S10-SG-s01 & 0.85   &  1.0   & 1.2  & yes& dipole & 0.1 &  -  & 33.88 & 28.27 & 0.85 &  0.21 \\
D08-S10-SG-s10 & 0.85   &  1.0   & 1.2  & yes& dipole & 1.0 &  - & 33.88 & 28.27 &  0.85 &  0.21 \\
D08-S20-nSG-b01 & 0.85   &  2.0   & 1.2  & -  & dipole & - &  0.1  & 25.62 & 21.18  & 0.87 & 0.42 \\ 
D08-S20-nSG-b05 & 0.85   &  2.0   & 1.2  & -  & dipole & - &  0.5  & 26.03  & 22.48 & 0.87 & 0.42 \\ 
D08-S20-nSG-b10 & 0.85   &  2.0   & 1.2  & -  & dipole & - &  1.0  & 25.78 & 22.30 &  0.87 &  0.42 \\
D08-S20-nSG-b50 & 0.85   &  2.0   & 1.2  & -  & dipole & - &  50.0 & 25.78 & 22.25 & 0.87  &  0.42 \\
D08-S20-nSG-s01 & 0.85   &  2.0   & 1.2  & -  & dipole & 0.1 &  - & 25.59 & 22.21  & 0.87  &  0.42 \\
D08-S20-nSG-s10 & 0.85   &  2.0   & 1.2  & -  & dipole & 1.0 &  - & 19.99 & 18.02  & 0.87 & 0.50  \\
D08-S20-SG-b01 &  0.85   &  2.0   & 1.2  & yes  & dipole & - &  0.1  & 34.85 & 28.30 &  0.85 & 0.40 \\
D08-S20-SG-b05 & 0.85   &  2.0   & 1.2  & yes  & dipole & - &  0.5 &36.79  & 28.30 & 0.85  &  0.40 \\
D08-S20-SG-b10  & 0.85   &  2.0   & 1.2  & yes  & dipole & - &  1.0  & 34.72 & 28.29 &  0.85 & 0.40 \\
 D08-S20-SG-b50 & 0.85   &  2.0   & 1.2  & yes  & dipole & - &  50.0 &34.76  & 28.29 & 0.85  &  0.40 \\
D08-S20-SG-s01 & 0.85   &  2.0   & 1.2  & yes  & dipole & 0.1 &  - & 33.76 & 28.30  & 0.85  &  0.41 \\
D08-S20-SG-s10 & 0.85   &  2.0   & 1.2  & yes  & dipole & 1.0 &  - & 33.88 & 28.33  & 0.85 & 0.41  \\
\hline
\hline
 W08-S10-nSG-b50 & 0.85   &  1.0   & 1.2  & -   & Wald & -   &  50.0 & 0.092$^{*}$  &  3.0 & 0.85  & 0.76  \\
 W08-S10-SG-b50 & 0.85   &  1.0   & 1.2  & yes & Wald & -   &  50.0 & 0.087$^{*}$  & 3.0 & 0.85  & 0.75 \\
 W08-S20-nSG-b50& 0.85   &  2.0   & 1.2  & -   & Wald & -   &  50.0 & 0.116$^{*}$  & 3.0  & 0.85  & 0.71 \\
W08-S20-SG-b50 & 0.85   &  2.0   & 1.2  & yes & Wald & -   &  50.0 & 0.165$^{*}$  & 3.0 & 0.85  & 0.66  \\
  \end{longtable}
  \end{center}
\clearpage
\twocolumn


\section{Discussion}
\label{sec:diss}

In this work, we present for the first time the new version of our time-dependent code, based on HARM scheme but supplemented
with the dynamically evolving space-time metric coefficients, as described and 
developed originally by \citet{Janiuk2018} and later explored in \citet{Dominika2021}.
The current significant modification aims to numerically account for the Kerr metric perturbation due to the gravitational self-force, which is acting on the matter inside the orbit of a given fluid element, and changing with the distance from the black hole. This dynamical treatment of the metric perturbation is not providing a method of solving the full set of Einstein field equations,
such as is possible in the Einstein Toolkit\footnote{https://einsteintoolkit.org/} framework. Nevertheless, we argue that it provides a good approximation to the collapsar problem and allows to compute the stellar collapse with a wide range of black hole spin parameters and dynamical evolution of the black hole mass and spin, while the mass of the self-gravitating envelope which imposes the perturbation in the Kerr metric is non-negligibly large. Such a calculations is, to the best of our knowledge, currently not possible with other methods.

In this paper, we compared the new results with the cases when the self-gravity perturbation was neglected, and we found dramatic differences between these two cases, mainly in the early phase of the collapsing star time evolution.
We then focused on the potential role of the gravitational instability in collapsing stellar cores, across the broad range of values of the initial spins of the black hole. 
We studied both non-magnetized and magnetized models. In the latter case, we explored in most detail the self-gravitating collapsing stars embedded in an initially vertical magnetic field, in order to be able to compare them quantitatively with our previous results, where only such a configuration was used. In addition, we adopted an alternative configuration of a dipole magnetic field, which is presumably more adequate for large scale field initialization in collapsing stars. Nevertheless, we found out that this configuration itself does not lead to the magnetic flux being brought into the black hole horizon, which would be large enough to account for magnetically arrested state formation, MAD \citep{JaniukJames2022}. The most promising way to produce bi-polar jets which would be able to emerge from the collapsar envelope, while being also able to be powered by magnetized accretion originally seeded in the stellar magnetic field, is therefore a hybrid configuration, such as the one proposed in \cite{Gottlieb2022}. Exploring such models should be done in a 3-dimensional setup, which is more demanding computationally. We notice that when one accounts for self-gravity, apart from the Kerr metric changes because of black hole growth, two additional equations for perturbation of mass and angular momentum have to be solved and integrated over volume in each grid point. Both cases were studied so far in 2D only. We postpone the 3D task to our future work.

In our self-gravitating models we assumed the accreting black hole of initial mass of $3 M_{\odot}$, which is slightly larger than possible maximum mass of a neutron star, and ensures that our core collapsed directly to a black hole. After that, the compact object increases its mass due to the fallback of stellar envelope. Our adopted model assumes the envelope mass is fixed and equal to 25 $M_{\odot}$.
This mass is smaller than a typical core-collapse supernova may have, while it is adequate for the massive star that has been already stripped off from its Hydrogen envelope \citep{2003Podsiadlowski}. Therefore, as for the black hole growth, we allow for a change of its mass up to this possible value of about $28 M_{\odot}$, which is found to be in the lower end of the mass distribution of black holes detected by gravitational wave interferometers (10-85 $M_{\odot}$, \cite{2020Abbott}).

The initial spin of the newly formed black hole is our model parameter and ranges from 0.3 to 0.85.
We do not start from a non-spinning black hole, as we did in our previous work (e.g. \cite{Murguia2020}), because in general what is of interest for this study is an electromagnetic counterpart of the collapse, in the form of a GRB event that is presumably powered by a spinning black hole.
The spin changes during the collapse, due to accretion of the envelope where some content of the angular momentum was already available. As
the stellar rotation is parameterized in our models
by the ratio between given specific angular momentum and critical rotation speed at the ISCO, while on the other hand, a moderately rotating black hole was already seeded in the core, the ultimate outcome of the process naturally depends on the two-parameter space, namely $S$ and $A_{0}$. This parameter space was explored in our set of simulations, while we also aimed to break the degeneracy between these parameters by introducing magnetic field in some of the models.

\begin{figure*}[ht]
      \centering
   \includegraphics[scale=0.59]{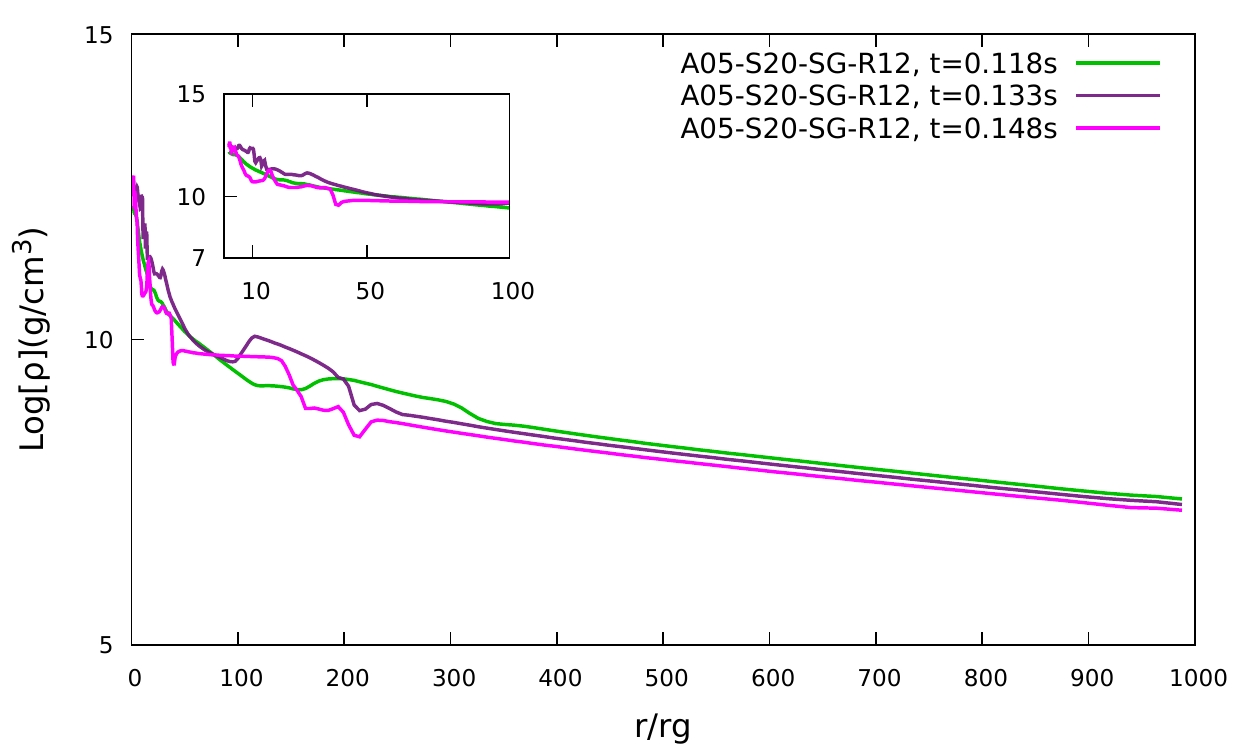}
   \includegraphics[scale=0.59]{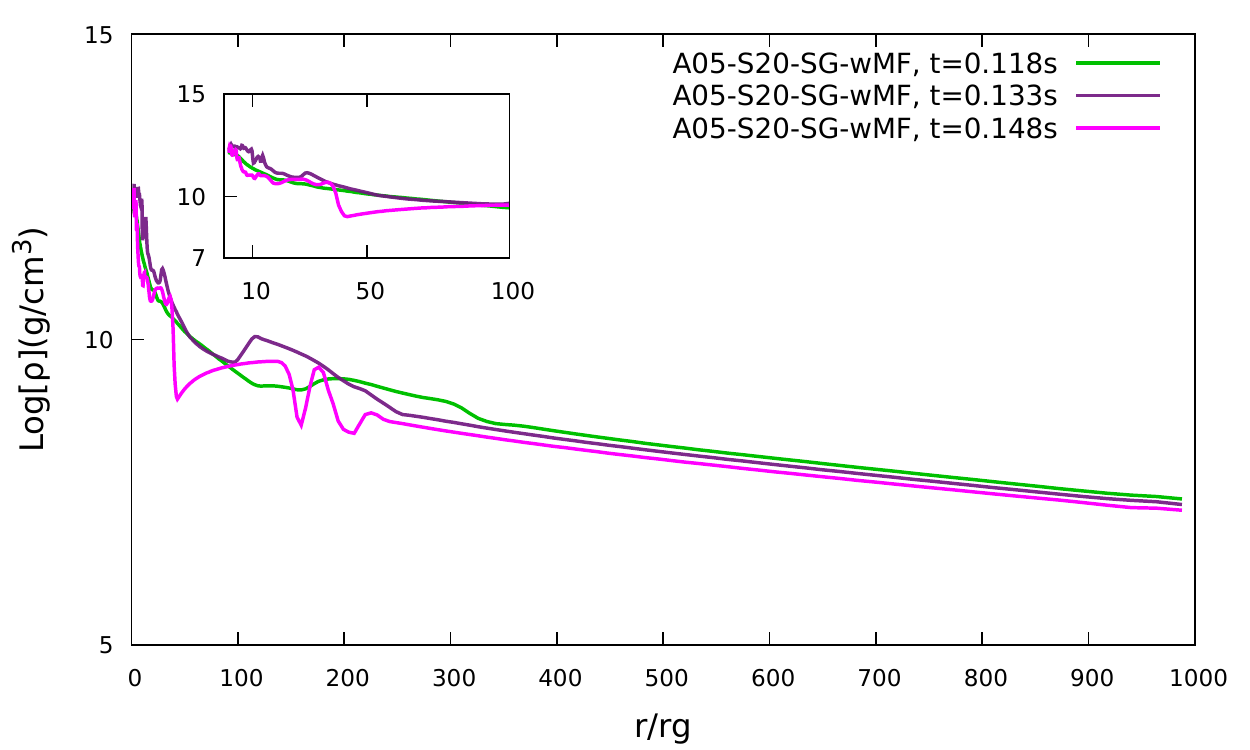}
   \includegraphics[scale=0.59]{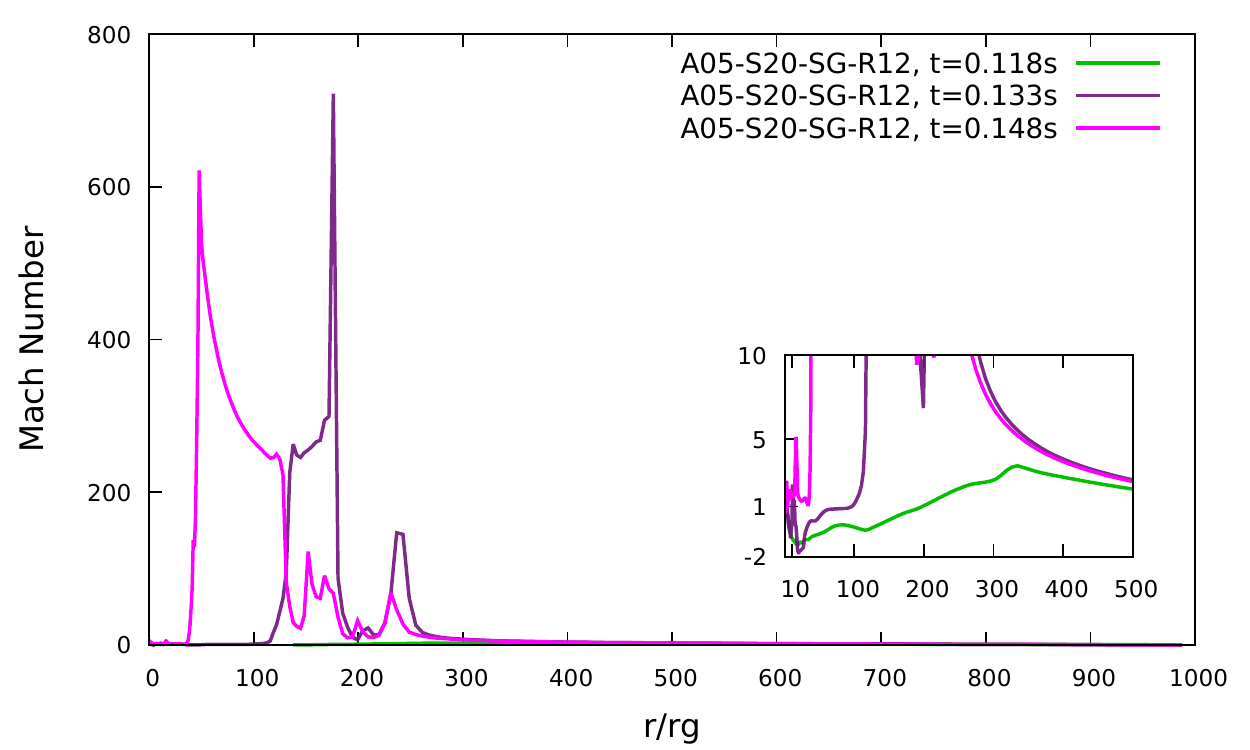}
   \includegraphics[scale=0.59]{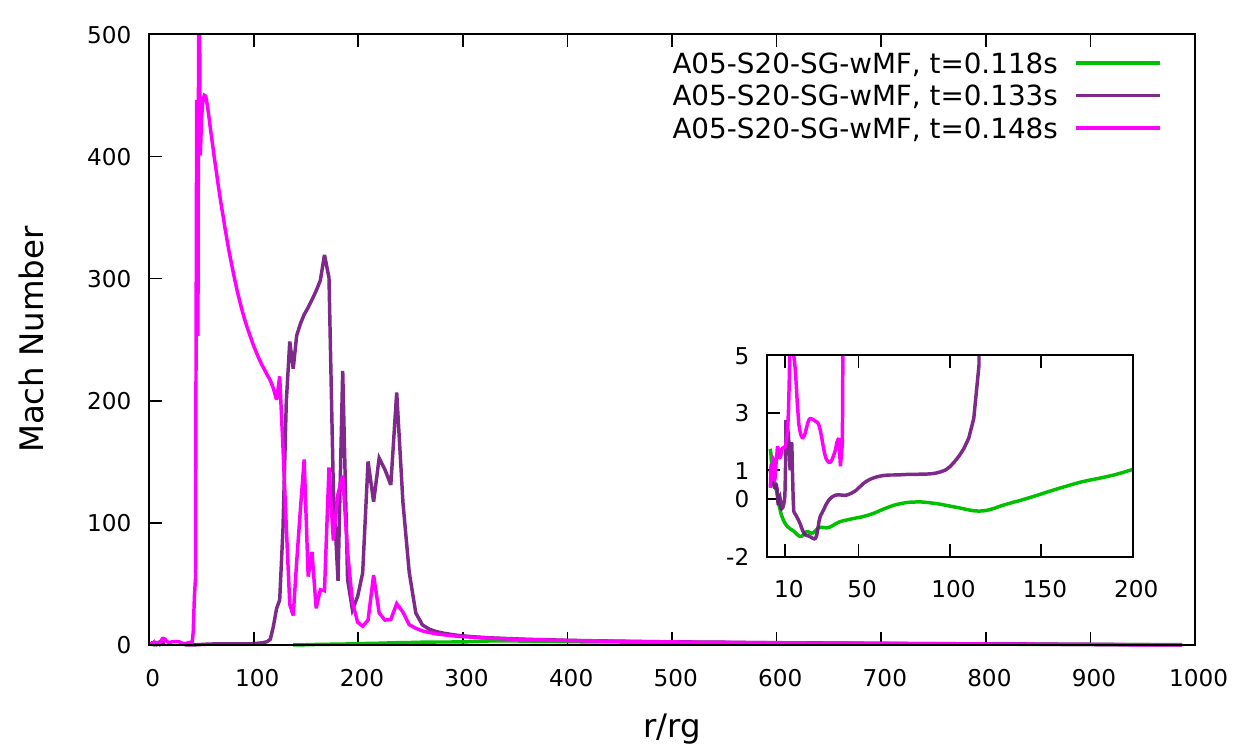}
   \caption{Radial profiles of Mach number and density at some specific time snapshots, or models with $S=2$ and $A_{0}=0.5$.
     The left panels show the curves in the self-gravitating case, while the plots in the right panels show those of self-gravitating magnetized case. For the sake of more visibility, we provide zommed-in inset panels representing the inner regions. Models are labeled in the all panels with symbols referring to Tab \ref{tab:modele}. }
    \label{shock}
\end{figure*}

Our simulations are started with the smooth transonic solution, when the gas radial velocity is supersonic within inner 80 gravitational radii.
This is different from a multi-transonic shock solution, introduced in the simulations of \citep{Sukova2015}, which would naturally lead to
shock oscillations.
Here we observe the sonic front expansion, and also some transient shock formation during the collapse.
At early times, the small transonic shocks appear located around 100 $r_{g}$. They are presenting a moderate density contrast (pre-shock to post-shock density ratio $R=\rho_{1}/\rho_{2}\sim 10$). Such shocks appear also at later times during collapsar evolution. We find that their formation is enhanced by the self-gravity effects. 
Regarding the magnetic field impact, we found that it does not make any significant difference in these time scales, although it may influence the strength of the transient shock which appeared in the case of high rotation, $S=2$. This comparison can be confirmed by the Figure \ref{shock} which indicates the radial evolution of density and Mach number through early stage of the simulation. The left hand plots are associated with the self-gravitating case while the right ones also contain the effects of magnetic field. This impact of the magnetic field on shock strength is consistent with previous studies.

  We notice therefore that magnetic field does not significantly change the properties of inhomogeneous region.
The latter conclusion about magnetized shocks is consistent with the Fermi acceleration process studies made for the pulsar wind nebulae. As found by 
\cite{SironiSpitkovsky}, the particle acceleration in magnetized shocks leads to smaller speed and energy for larger magnetisation, $\sigma$. Moreover \cite{1999MNRAS.308.1069K} showed that even for moderately magnetized plasma the formation of the strong shocks is different than in purely hydrodynamical case. In our models, the typical magnetisation is weak, $\sigma<1$, but still in the early times of collapse, the shocks which form in the innermost regions of the star have larger density contrast in non-magnetized cases.

Several scenarios have been proposed to explain the temporal variability of the GRB light curves. Such fluctuations are detected in both prompt emission of Gamma rays and flaring activity in X-ray or optical bands.
The internal shock model \citep{piran1993hydrodynamics, katz1994two}, turbulence or magnetic reconnections in the jet \citep{narayan2009turbulent, beloborodov1998self,beloborodov2000power, amati2018theseus}, viscous and thermal instabilities in hyperaccretion disk, as the GRB central engine, \citep{janiuk2007instabilities, lei2009magnetically,  kawanaka2012possible, kawanaka2013discovery} are among the studies to model this erratic behavior.
As suggested already by \cite{Perna2006}, the phenomenology of short and long gamma ray bursts indicates that the gravitational instability in their engines may lead to the flaring activity \citep{Margutti2010}.
In particular, the self-gravity can lead to a clumpy structure (see also \citet{shahamat2020viscous} for the case of flaring activity, \citet{shahamat2021grb} for the prompt emission variability, and \citet{coughlin2020variability} regarding either cases of the flares and prompt emission's oscillatory behavior) or spiral density waves \citep{masada2007dead} in the central engine.
This activity perturbation may however be further modulated by the jet breakout from the star \citep{Petropoulou2020}.
What is then measured in the data is the amplitude of the the count rate, for the variability observed in jets of GRBs, and that depends strongly on the selected energy band.
About 30\% of GRBs present X-ray flares whose origin is now a subject of discussion.
The longest of them, lasting few hundred seconds, are attributed to the reverse shock emission, while the following plateau phase, seen in Optical and X-ray bands, may be related to the late central engine activity in GRB180205 \citep{Becerra2019}.

Prompt gamma-ray emission exhibits typically a stochastic variability, described by a Poisson noise.
The variability time scale is estimated to be of the order of $<10s$ down to $10 ms$ \citep{bhat2011temporal, golkhou2014uncovering}.
It naturally gives rise to clusterization, i.e., time intervals characterized by an intense activity with a high rate of peaks, are interspersed with quiescent periods, during
which the rate drops significantly \citep{Guidorzi2015}.
This high energy variability and late X-ray flares are related to each other, and presumably they just represent different fragments of the GRB central engine, being accreted at the beginning and at the end, respectively, and are likely following the same the same mechanism.
In our calculations, the central engine is represented by the innermost parts of the collapsing star, enclosed within the computational domain. 
The very first fragments of the self-gravitating star will lead to the prompt emission variability on the scale of 0.1-0.2 seconds, while the 
black hole spin is also changing on time.
The remaining parts of the envelope, additionally broken into further fragments and rings over larger inhomogeneities, should lead to longer timescale variability. In this phase, the black hole spin should reach a final value, and the process of energy extraction to the jets will not be affected by the spin changes. Thus, the extended quiescent period of the prompt emission, can be followed by several subsequent pulses, when the matter clumps incoming form the outermost regions of the engine will eventually fall onto the black hole, likely delayed by the viscous spreading \citep{DallOsso2017}.
Flares occur therefore $100-1000 s$ after the prompt emission, while the light curves are generally of around $>10s$ length.

Our study identified several key factors that may influence accretion rate variability, and potentially explain the oscillations in prompt emission. First, the barrier between gravitational torque of the black hole and the centrifugal force, which pushes matter outwards especially in the cases of higher rotations (i.e., $S=1.4,~2$) and near the equator, causes some fluctuations in the size of outflow, while this region shrinks due to the dominance of the central gravity. It can produce a pulse with a duration of around $\lesssim0.05~s$. In cases without self-gravity, on the other hand, the outflow zone moves outward, and shrinks in a more stable manner during a longer period of time. Consequently, the accretion rate is much smoother, with a variability of the order of a few $10^{-1}~s$ (see the right hand side panel in the Figure \ref{M_S_A}). We identify this factor as the only responsible for the smooth variability of the accretion rate, in the absence of self-gravity.
Second, the inhomogeneous structure of the inner stellar core due to SGI instability, in addition to the creation of transient shock (seen in the earlier time steps of the case $S=2$) leads to a variable accretion rate. We estimate that the former generates a pulse of the width less than $\sim 0.074~s$ (regarding the period of time during which the specific angular momentum, through the inner regions, encounters a drop as demonstrated in Figure \ref{lspec}) and the latter provides a variability of the order of $\sim10^{-2}~s$.
  On the other hand, shocks are considered as another factor that can provide a short-term variability, and also a high efficiency of energy conversion, in the astrophysical accreting systems such as active galactic nuclei \citep{meszaros1983shocks}.  
  In conclusion, we are of the opinion that the short-term variability of the GRB's prompt emission can be well-explained in terms of these mechanisms through our model.

\section{Conclusions}
\label{sec:conclusions}

We studied collapsing stellar core models accounting for the changing black hole mass and spin, the retaletd coefficients of the Kerr space-time metric,
and, for the first time, the self-gravity of the star. In the main part of this analysis, we
compared our new results to the cases without the self-gravity terms. We also analyzed the impact of SGI instability on the properties of the collapsing flow.
In addition, some of our models were embedded in a magnetic field of a various strength and couple typical configurations.

 The main findings of this study are the following:
 \begin{itemize}
     \item We show that evolution of spin and mass of the black hole are quantitatively and qualitatively affected by the self-gravitation of the envelope.
    \item We show that accretion rate variability at early times is much stronger in self-gravitating collapsing stars and may lead to detectable signals in long GRB prompt emission.
    \item We find that self-gravity effects provide mechanism of the transport of angular momentum, and that final black hole mass and spin are reached much earlier during the collapse.
      \item We see a weak and non-linear dependence of the black hole evolution on its initial spin,  manifested mainly when the magnetic field present in super-critically rotating envelopes. 
      \item We detect formation of transient shocks, with moderate density contrast, also in magnetized models. 
   \item At early times of simulation, the density contrast in transonic shocks seems to be higher in non-magnetized cases.
 \end{itemize}

\begin{acknowledgements}
We thank Petra Sukova and Ishika Palit for helpful discussions.
The project was partially supported by grant 2019/35/B/ST9/04000 from Polish National Science Center. We made use of computational resources of the PL-Grid infrastructure, under grant \textit{plggrb5}. Additionally D.~{\L}.~K. was supported by the Polish National Science Center DEC-2019/35/O/ST9/04054. We hereby acknowledge Sci-HPC center of Ferdowsi University of Mashhad, Iran, where some part of this research was performed.
 AJ acknowledges the Czech-Polish mobility program (M\v{S}MT 8J20PL037
and PPN/BCZ/2019/1/00069)
\end{acknowledgements}

\bibliographystyle{aa}
\bibliography{aselfg}

\begin{thebibliography}{77}
\expandafter\ifx\csname natexlab\endcsname\relax\def\natexlab#1{#1}\fi

\bibitem[{{Abbott} {et~al.}(2020){Abbott}, {LIGO Scientific Collaboration}, \&
  {Virgo Collaboration}}]{2020Abbott}
{Abbott}, R., {LIGO Scientific Collaboration}, \& {Virgo Collaboration}. 2020,
  \apjl, 900, L13

\bibitem[{Amati {et~al.}(2018)Amati, O’Brien, G{\"o}tz, Bozzo, Tenzer,
  Frontera, Ghirlanda, Labanti, Osborne, Stratta, {et~al.}}]{amati2018theseus}
Amati, L., O’Brien, P., G{\"o}tz, D., {et~al.} 2018, Advances in Space
  Research, 62, 191

\bibitem[{{Armitage}(2011)}]{2011ARA&A..49..195A}
{Armitage}, P.~J. 2011, \araa, 49, 195

\bibitem[{Balbus(2003)}]{balbus2003enhanced}
Balbus, S.~A. 2003, Annual Review of Astronomy and Astrophysics, 41, 555

\bibitem[{{Becerra} {et~al.}(2019){Becerra}, {De Colle}, {Watson}, {Fraija},
  {Butler}, {Lee}, {Rom{\'a}n-Z{\'u}{\~n}iga}, {Bloom}, {Gonz{\'a}lez},
  {Kutyrev}, {Prochaska}, {Ramirez-Ruiz}, {Richer}, \& {Troja}}]{Becerra2019}
{Becerra}, R.~L., {De Colle}, F., {Watson}, A.~M., {et~al.} 2019, \apj, 887,
  254

\bibitem[{Beloborodov {et~al.}(1998)Beloborodov, Stern, \&
  Svensson}]{beloborodov1998self}
Beloborodov, A.~M., Stern, B.~E., \& Svensson, R. 1998, The Astrophysical
  Journal, 508, L25

\bibitem[{Beloborodov {et~al.}(2000)Beloborodov, Stern, \&
  Svensson}]{beloborodov2000power}
Beloborodov, A.~M., Stern, B.~E., \& Svensson, R. 2000, The Astrophysical
  Journal, 535, 158

\bibitem[{Bhat {et~al.}(2011)Bhat, Briggs, Connaughton, Kouveliotou, van~der
  Horst, Paciesas, Meegan, Bissaldi, Burgess, Chaplin,
  {et~al.}}]{bhat2011temporal}
Bhat, P., Briggs, M.~S., Connaughton, V., {et~al.} 2011, The Astrophysical
  Journal, 744, 141

\bibitem[{{Christodoulou} \& {Narayan}(1992)}]{christo1992}
{Christodoulou}, D.~M. \& {Narayan}, R. 1992, \apj, 388, 451

\bibitem[{{Chrzanowski}(1975)}]{Chrzanowski1975}
{Chrzanowski}, P.~L. 1975, \prd, 11, 2042

\bibitem[{{Cohen} \& {Kegeles}(1974)}]{Cohen1974}
{Cohen}, J.~M. \& {Kegeles}, L.~S. 1974, \prd, 10, 1070

\bibitem[{Coughlin {et~al.}(2020)Coughlin, Nixon, Barnes, Metzger, \&
  Margutti}]{coughlin2020variability}
Coughlin, E.~R., Nixon, C., Barnes, J., Metzger, B.~D., \& Margutti, R. 2020,
  The Astrophysical Journal Letters, 896, L38

\bibitem[{{Dall'Osso} {et~al.}(2017){Dall'Osso}, {Perna}, {Tanaka}, \&
  {Margutti}}]{DallOsso2017}
{Dall'Osso}, S., {Perna}, R., {Tanaka}, T.~L., \& {Margutti}, R. 2017, \mnras,
  464, 4399

\bibitem[{{Eichler} \& {Cheng}(1989)}]{Eichler1989}
{Eichler}, D. \& {Cheng}, A.~F. 1989, \apj, 336, 360

\bibitem[{{Gammie} {et~al.}(2003){Gammie}, {McKinney}, \&
  {T{\'o}th}}]{Gammie2003}
{Gammie}, C.~F., {McKinney}, J.~C., \& {T{\'o}th}, G. 2003, \apj, 589, 444

\bibitem[{Golkhou \& Butler(2014)}]{golkhou2014uncovering}
Golkhou, V.~Z. \& Butler, N.~R. 2014, The Astrophysical Journal, 787, 90

\bibitem[{{Gottlieb} {et~al.}(2022){Gottlieb}, {Lalakos}, {Bromberg}, {Liska},
  \& {Tchekhovskoy}}]{Gottlieb2022}
{Gottlieb}, O., {Lalakos}, A., {Bromberg}, O., {Liska}, M., \& {Tchekhovskoy},
  A. 2022, \mnras, 510, 4962

\bibitem[{{Guidorzi} {et~al.}(2015){Guidorzi}, {Dichiara}, {Frontera},
  {Margutti}, {Baldeschi}, \& {Amati}}]{Guidorzi2015}
{Guidorzi}, C., {Dichiara}, S., {Frontera}, F., {et~al.} 2015, \apj, 801, 57

\bibitem[{{Hachisu} {et~al.}(1987){Hachisu}, {Tohline}, \&
  {Eriguchi}}]{Hachisu1987}
{Hachisu}, I., {Tohline}, J.~E., \& {Eriguchi}, Y. 1987, \apj, 323, 592

\bibitem[{Harten {et~al.}(1983)Harten, Lax, \& Leer}]{Harten1983}
Harten, A., Lax, P.~D., \& Leer, B.~v. 1983, SIAM Review, 25, 35

\bibitem[{{Hjorth} {et~al.}(2003){Hjorth}, {Sollerman}, {M{\o}ller}, {Fynbo},
  {Woosley}, {Kouveliotou}, {Tanvir}, {Greiner}, {Andersen}, {Castro-Tirado},
  {Castro Cer{\'o}n}, {Fruchter}, {Gorosabel}, {Jakobsson}, {Kaper}, {Klose},
  {Masetti}, {Pedersen}, {Pedersen}, {Pian}, {Palazzi}, {Rhoads}, {Rol}, {van
  den Heuvel}, {Vreeswijk}, {Watson}, \& {Wijers}}]{Hjorth2003}
{Hjorth}, J., {Sollerman}, J., {M{\o}ller}, P., {et~al.} 2003, \nat, 423, 847

\bibitem[{Hueckstaedt {et~al.}(2005)Hueckstaedt, Peterson, \&
  Hunter~Jr}]{hueckstaedt2005parameter}
Hueckstaedt, R., Peterson, A., \& Hunter~Jr, J. 2005, Monthly Notices of the
  Royal Astronomical Society: Letters, 361, L35

\bibitem[{Hunter~Jr {et~al.}(1997)Hunter~Jr, Whitaker, \&
  Lovelace}]{hunter1997kelvin}
Hunter~Jr, J.~H., Whitaker, R.~W., \& Lovelace, R.~V. 1997, The Astrophysical
  Journal, 482, 852

\bibitem[{Hunter~Jr {et~al.}(1998)Hunter~Jr, Whitaker, \&
  Lovelace}]{hunter1998stability}
Hunter~Jr, J.~H., Whitaker, R.~W., \& Lovelace, R.~V. 1998, The Astrophysical
  Journal, 508, 680

\bibitem[{{Janiuk}(2022{\natexlab{a}})}]{JaniukJames2022}
{Janiuk}, A. 2022{\natexlab{a}}, in Black-hole activity feedback from
  Bondi-radius to galaxy-cluster scales, Vol. 2022, 29

\bibitem[{{Janiuk}(2022{\natexlab{b}})}]{Janiukproc}
{Janiuk}, A. 2022{\natexlab{b}}, in XL Polish Astronomical Society Meeting, ed.
  E.~{Szuszkiewicz}, A.~{Majczyna}, K.~{Ma{\l}ek}, M.~{Ratajczak},
  E.~{Niemczura}, U.~{B{\k{a}}k-St{\k{e}}{\'s}licka}, R.~{Poleski},
  M.~{Bilicki}, \& {\L}.~{Wyrzykowski}, Vol.~12, 221--224

\bibitem[{{Janiuk} {et~al.}(2008){Janiuk}, {Moderski}, \& {Proga}}]{JMP2008}
{Janiuk}, A., {Moderski}, R., \& {Proga}, D. 2008, \apj, 687, 433

\bibitem[{{Janiuk} \& {Proga}(2008)}]{JP2008}
{Janiuk}, A. \& {Proga}, D. 2008, \apj, 675, 519

\bibitem[{{Janiuk} {et~al.}(2018){Janiuk}, {Sukova}, \& {Palit}}]{Janiuk2018}
{Janiuk}, A., {Sukova}, P., \& {Palit}, I. 2018, \apj, 868, 68

\bibitem[{Janiuk {et~al.}(2007)Janiuk, Yuan, Perna, \&
  Di~Matteo}]{janiuk2007instabilities}
Janiuk, A., Yuan, Y., Perna, R., \& Di~Matteo, T. 2007, The Astrophysical
  Journal, 664, 1011

\bibitem[{{Janka} {et~al.}(2007){Janka}, {Langanke}, {Marek},
  {Mart{\'\i}nez-Pinedo}, \& {M{\"u}ller}}]{Janaka2007}
{Janka}, H.~T., {Langanke}, K., {Marek}, A., {Mart{\'\i}nez-Pinedo}, G., \&
  {M{\"u}ller}, B. 2007, \physrep, 442, 38

\bibitem[{{Karas} {et~al.}(2020){Karas}, {Sapountzis}, \& {Janiuk}}]{Karas2020}
{Karas}, V., {Sapountzis}, K., \& {Janiuk}, A. 2020, arXiv e-prints,
  arXiv:2012.15105

\bibitem[{Katz(1994)}]{katz1994two}
Katz, J. 1994, The Astrophysical Journal, 422, 248

\bibitem[{Kawanaka \& Kohri(2012)}]{kawanaka2012possible}
Kawanaka, N. \& Kohri, K. 2012, Monthly Notices of the Royal Astronomical
  Society, 419, 713

\bibitem[{Kawanaka {et~al.}(2013)Kawanaka, Mineshige, \&
  Piran}]{kawanaka2013discovery}
Kawanaka, N., Mineshige, S., \& Piran, T. 2013, The Astrophysical Journal
  Letters, 777, L15

\bibitem[{Kifonidis {et~al.}(2003)Kifonidis, Plewa, Janka, \&
  M{\"u}ller}]{kifonidis2003non}
Kifonidis, K., Plewa, T., Janka, H.-T., \& M{\"u}ller, E. 2003, Astronomy \&
  Astrophysics, 408, 621

\bibitem[{{Komissarov}(1999)}]{1999MNRAS.308.1069K}
{Komissarov}, S.~S. 1999, \mnras, 308, 1069

\bibitem[{{Kr{\'o}l} \& {Janiuk}(2021)}]{Dominika2021}
{Kr{\'o}l}, D.~{\L}. \& {Janiuk}, A. 2021, \apj, 912, 132

\bibitem[{{Kumar} \& {Zhang}(2015)}]{Kumar2015}
{Kumar}, P. \& {Zhang}, B. 2015, \physrep, 561, 1

\bibitem[{Lei {et~al.}(2009)Lei, Wang, Zhang, Gan, Zou, \&
  Xie}]{lei2009magnetically}
Lei, W., Wang, D., Zhang, L., {et~al.} 2009, The Astrophysical Journal, 700,
  1970

\bibitem[{{Lodato}(2008)}]{Lodato2008}
{Lodato}, G. 2008, \nar, 52, 21

\bibitem[{{Margutti} {et~al.}(2010){Margutti}, {Guidorzi}, {Chincarini},
  {Bernardini}, {Genet}, {Mao}, \& {Pasotti}}]{Margutti2010}
{Margutti}, R., {Guidorzi}, C., {Chincarini}, G., {et~al.} 2010, \mnras, 406,
  2149

\bibitem[{Masada {et~al.}(2007)Masada, Kawanaka, Sano, \&
  Shibata}]{masada2007dead}
Masada, Y., Kawanaka, N., Sano, T., \& Shibata, K. 2007, The Astrophysical
  Journal, 663, 437

\bibitem[{Meszaros \& Ostriker(1983)}]{meszaros1983shocks}
Meszaros, P. \& Ostriker, J. 1983, The Astrophysical Journal, 273, L59

\bibitem[{{Murguia-Berthier} {et~al.}(2020){Murguia-Berthier}, {Batta},
  {Janiuk}, {Ramirez-Ruiz}, {Mandel}, {Noble}, \& {Everson}}]{Murguia2020}
{Murguia-Berthier}, A., {Batta}, A., {Janiuk}, A., {et~al.} 2020, \apjl, 901,
  L24

\bibitem[{Narayan \& Kumar(2009)}]{narayan2009turbulent}
Narayan, R. \& Kumar, P. 2009, Monthly Notices of the Royal Astronomical
  Society, 394, L117

\bibitem[{{Narayan} {et~al.}(1992){Narayan}, {Paczynski}, \&
  {Piran}}]{Narayan1992}
{Narayan}, R., {Paczynski}, B., \& {Piran}, T. 1992, \apjl, 395, L83

\bibitem[{{Noble} {et~al.}(2006){Noble}, {Gammie}, {McKinney}, \& {Del
  Zanna}}]{Noble2006}
{Noble}, S.~C., {Gammie}, C.~F., {McKinney}, J.~C., \& {Del Zanna}, L. 2006,
  \apj, 641, 626

\bibitem[{{Obergaulinger} {et~al.}(2009){Obergaulinger}, {Cerd{\'a}-Dur{\'a}n},
  {M{\"u}ller}, \& {Aloy}}]{Obergaulinger2009}
{Obergaulinger}, M., {Cerd{\'a}-Dur{\'a}n}, P., {M{\"u}ller}, E., \& {Aloy},
  M.~A. 2009, \aap, 498, 241

\bibitem[{{Palit} {et~al.}(2019){Palit}, {Janiuk}, \&
  {Sukova}}]{2019MNRAS.487..755P}
{Palit}, I., {Janiuk}, A., \& {Sukova}, P. 2019, \mnras, 487, 755

\bibitem[{{Perna} {et~al.}(2006){Perna}, {Armitage}, \& {Zhang}}]{Perna2006}
{Perna}, R., {Armitage}, P.~J., \& {Zhang}, B. 2006, \apjl, 636, L29

\bibitem[{{Petit} {et~al.}(2019){Petit}, {Wade}, {Schneider}, {Fossati},
  {Kamp}, {Neiner}, {David-Uraz}, {Alecian}, \& {MiMeS
  Collaboration}}]{Petit2019}
{Petit}, V., {Wade}, G.~A., {Schneider}, F.~R.~N., {et~al.} 2019, \mnras, 489,
  5669

\bibitem[{{Petropoulou} {et~al.}(2020){Petropoulou}, {Beniamini},
  {Vasilopoulos}, {Giannios}, \& {Barniol Duran}}]{Petropoulou2020}
{Petropoulou}, M., {Beniamini}, P., {Vasilopoulos}, G., {Giannios}, D., \&
  {Barniol Duran}, R. 2020, \mnras, 496, 2910

\bibitem[{{Piran}(2004)}]{Piran2004}
{Piran}, T. 2004, Reviews of Modern Physics, 76, 1143

\bibitem[{Piran {et~al.}(1993)Piran, Shemi, \&
  Narayan}]{piran1993hydrodynamics}
Piran, T., Shemi, A., \& Narayan, R. 1993, Monthly Notices of the Royal
  Astronomical Society, 263, 861

\bibitem[{{Podsiadlowski} {et~al.}(2003){Podsiadlowski}, {Rappaport}, \&
  {Han}}]{2003Podsiadlowski}
{Podsiadlowski}, P., {Rappaport}, S., \& {Han}, Z. 2003, \mnras, 341, 385

\bibitem[{{Pontzen} {et~al.}(2010){Pontzen}, {Deason}, {Governato}, {Pettini},
  {Wadsley}, {Quinn}, {Brooks}, {Bellovary}, \& {Fynbo}}]{Pontzen2010}
{Pontzen}, A., {Deason}, A., {Governato}, F., {et~al.} 2010, \mnras, 402, 1523

\bibitem[{{Reichert} {et~al.}(2023){Reichert}, {Obergaulinger}, {Aloy},
  {Gabler}, {Arcones}, \& {Thielemann}}]{Reichert2023}
{Reichert}, M., {Obergaulinger}, M., {Aloy}, M.~{\'A}., {et~al.} 2023, \mnras,
  518, 1557

\bibitem[{{Ryu} {et~al.}(2020){Ryu}, {Krolik}, {Piran}, \& {Noble}}]{Ryu2020}
{Ryu}, T., {Krolik}, J., {Piran}, T., \& {Noble}, S.~C. 2020, \apj, 904, 99

\bibitem[{Shahamat \& Abbassi(2020)}]{shahamat2020viscous}
Shahamat, N. \& Abbassi, S. 2020, The Astrophysical Journal, 888, 64

\bibitem[{Shahamat {et~al.}(2021)Shahamat, Abbassi, \& Liu}]{shahamat2021grb}
Shahamat, N., Abbassi, S., \& Liu, T. 2021, Monthly Notices of the Royal
  Astronomical Society, 508, 6068

\bibitem[{{Shapiro} \& {Teukolsky}(1986)}]{1986bhwd.book.....S}
{Shapiro}, S.~L. \& {Teukolsky}, S.~A. 1986, {Black Holes, White Dwarfs and
  Neutron Stars: The Physics of Compact Objects}

\bibitem[{{Sironi} \& {Spitkovsky}(2011)}]{SironiSpitkovsky}
{Sironi}, L. \& {Spitkovsky}, A. 2011, \apj, 741, 39

\bibitem[{{Spruit}(2002)}]{Spruit2002}
{Spruit}, H.~C. 2002, \aap, 381, 923

\bibitem[{{Sukov{\'a}} {et~al.}(2017){Sukov{\'a}}, {Charzy{\'n}ski}, \&
  {Janiuk}}]{2017MNRAS.472.4327S}
{Sukov{\'a}}, P., {Charzy{\'n}ski}, S., \& {Janiuk}, A. 2017, \mnras, 472, 4327

\bibitem[{{Sukov{\'a}} \& {Janiuk}(2015)}]{Sukova2015}
{Sukov{\'a}}, P. \& {Janiuk}, A. 2015, \mnras, 447, 1565

\bibitem[{{Teukolsky}(1972)}]{Teukolsky1972}
{Teukolsky}, S.~A. 1972, \prl, 29, 1114

\bibitem[{{Toomre}(1964)}]{Toomre1964}
{Toomre}, A. 1964, \apj, 139, 1217

\bibitem[{{van de Meent}(2017)}]{Meent2017}
{van de Meent}, M. 2017, Classical and Quantum Gravity, 34, 124003

\bibitem[{{Wald}(1974)}]{Wald1974}
{Wald}, R.~M. 1974, \prd, 10, 1680

\bibitem[{{Wald}(1978)}]{Wald1978}
{Wald}, R.~M. 1978, \prl, 41, 203

\bibitem[{{White} {et~al.}(2022){White}, {Burrows}, {Coleman}, \&
  {Vartanyan}}]{Burrows2022}
{White}, C.~J., {Burrows}, A., {Coleman}, M. S.~B., \& {Vartanyan}, D. 2022,
  \apj, 926, 111

\bibitem[{{Woosley}(1993)}]{Woosley1993}
{Woosley}, S.~E. 1993, \apj, 405, 273

\bibitem[{{Woosley} \& {Bloom}(2006)}]{Woosley2006}
{Woosley}, S.~E. \& {Bloom}, J.~S. 2006, \araa, 44, 507

\bibitem[{{Woosley} \& {Heger}(2006)}]{WoosleyHeger2006}
{Woosley}, S.~E. \& {Heger}, A. 2006, \apj, 637, 914

\bibitem[{{Woosley} \& {Heger}(2015)}]{Woosley2015}
{Woosley}, S.~E. \& {Heger}, A. 2015, \apj, 810, 34

\bibitem[{{Zhang} {et~al.}(2003){Zhang}, {Woosley}, \& {MacFadyen}}]{Zhang2003}
{Zhang}, W., {Woosley}, S.~E., \& {MacFadyen}, A.~I. 2003, \apj, 586, 356

\end{thebibliography}

\end{document}